\documentclass[%
 reprint,
 amsmath,amssymb,
 aps,
]{revtex4-2}
\usepackage{graphicx}
\usepackage{dcolumn}
\usepackage{bm}
\usepackage{color} 
\usepackage{cancel}
\usepackage{array}
\usepackage{tabularx}
\usepackage{graphicx}
\usepackage{dcolumn}
\usepackage{braket}

\begin{document}

\title{Two-dimensional Weyl points and nodal lines in pentagonal materials and their optical response}

\author{Sergio Bravo}
\email{sergio.bravo.14@sansano.usm.cl}
\address{Departamento de F\' isica, Universidad T\'ecnica Federico Santa Mar\' ia, Valpara\' iso, Chile}
\author{M. Pacheco}
\email{monica.pacheco@usm.cl}
\address{Departamento de F\' isica, Universidad T\'ecnica Federico Santa Mar\' ia, Valpara\' iso, Chile}

\author{V. Nuñez}%
\address{Departamento de F\' isica, Universidad T\'ecnica Federico Santa Mar\' ia, Valpara\' iso, Chile}
\author{J.D. Correa}%
\address{Facultad de Ciencias  B\'asicas, Universidad de Medell\' \i n, 
Medell\' \i n, Colombia}
\author{Leonor Chico}%
\address{Departamento de Física de Materiales, 
Facultad de Ciencias Físicas, 
Universidad Complutense de Madrid, 
28040 Madrid, Spain}


\date{\today}

\begin{abstract}

Two-dimensional pentagonal structures based on the Cairo tiling are the basis of a family of layered
materials with appealing physical properties. 
In this work we present a theoretical study of 
the symmetry-based electronic and optical properties of these pentagonal materials. 
We provide a complete classification of the space groups that support 
pentagonal structures for binary and ternary systems. 
By means of first-principles 
calculations, their electronic band structures and the local spin textures in momentum space
are analyzed. Our results show that pentagonal structures can be realized in chiral and achiral
lattices with Weyl nodes pinned at high-symmetry points and nodal lines along the Brillouin zone boundary;  
these degeneracies are protected by the combined action of crystalline and time-reversal symmetries. Additionally, we discuss the linear and nonlinear optical features of some penta-materials, such as the shift current, which shows an enhancement due to the presence of nodal lines and points, 
and their possible applications.         

\end{abstract}


\maketitle

\section{\label{sec:level1} Introduction \protect\\} 

Low-dimensional materials have provided multiple possibilities for
the development of technological applications as well as the 
exploration of novel physical phenomena and states of matter. 
Within this class of materials, layered structures and two-dimensional (2D) 
systems \cite{Pere2014}, have stimulated great interest: graphene, hexagonal 
BN, transition metal dichalcogenides and few-layer stacking of van der Waals
materials show unexpected characteristics which are being intensely explored
\cite{Novoselov_2016,Das_2015_ARCM}. 
Furthermore, from the technological viewpoint, 
2D materials are highly adaptable for their use on devices, within
fields like electronics, spintronics and valleytronics, thanks 
to their planar geometry which is desirable for gate-controlled
designs \cite{Fiori_2014, Wang_2012}. 

The upsurge of interest in these materials has motivated many computational
high-throughput searches. This has resulted in the generation of several databases 
\cite{jain_commentary_2013,choudhary_high-throughput_2017,haastrup_computational_2018,mounet_two-dimensional_2018,choudhary_jarvis_2020,talirz_materials_2020}, 
with a great amount of prospective 2D materials with energetically 
favorable phases. In some of them, interlayer van der Waals interactions imply 
that they might be obtained by 
exfoliation \cite{mounet_two-dimensional_2018}. 

Among the most interesting candidates are the so-called
\textit{pentagonal materials}, for which the first experimental
example has been synthesized recently, in the form of 
PdSe$_2$ \cite{oyedele_pdse_2017}. The origin of pentagonal
materials can be traced to the theoretical prediction of penta-graphene 
\cite{2014_tang,2015_Zhang_pentagraphene},  
and their hallmark is the geometry of their 
planar projection, the so-called Cairo tiling. After these 
seminal works, many pentagonal systems have been proposed by 
means of first-principles calculations 
\cite{2016_TiC2,ma_room_2016,li_half-metallicity_2016,zhao_elastic_2017,pang_mechanical_2017,naseri_investigation_2018,naseri_penta-p2x_2018,naseri_penta-sic_2018,liu_single-layer_2018,zhao_penta-bcn_2020}, 
with remarkable characteristics such as excellent thermoelectrical 
properties \cite{2016Liu_thermoPG}, potential application 
for anode materials \cite{2016_anodePG} and auxetic 
behavior \cite{PG_auxetic_2019}, among others. 
Also, recent experimental studies have shown that the pentagonal 
material PdSe$_2$ presents good performance as an ohmic contact
\cite{oyedele_defect-mediated_2019,nguyen_atomically_2020}. 

\begin{figure}[ht]
\centering
\includegraphics[width=0.8\columnwidth,clip]{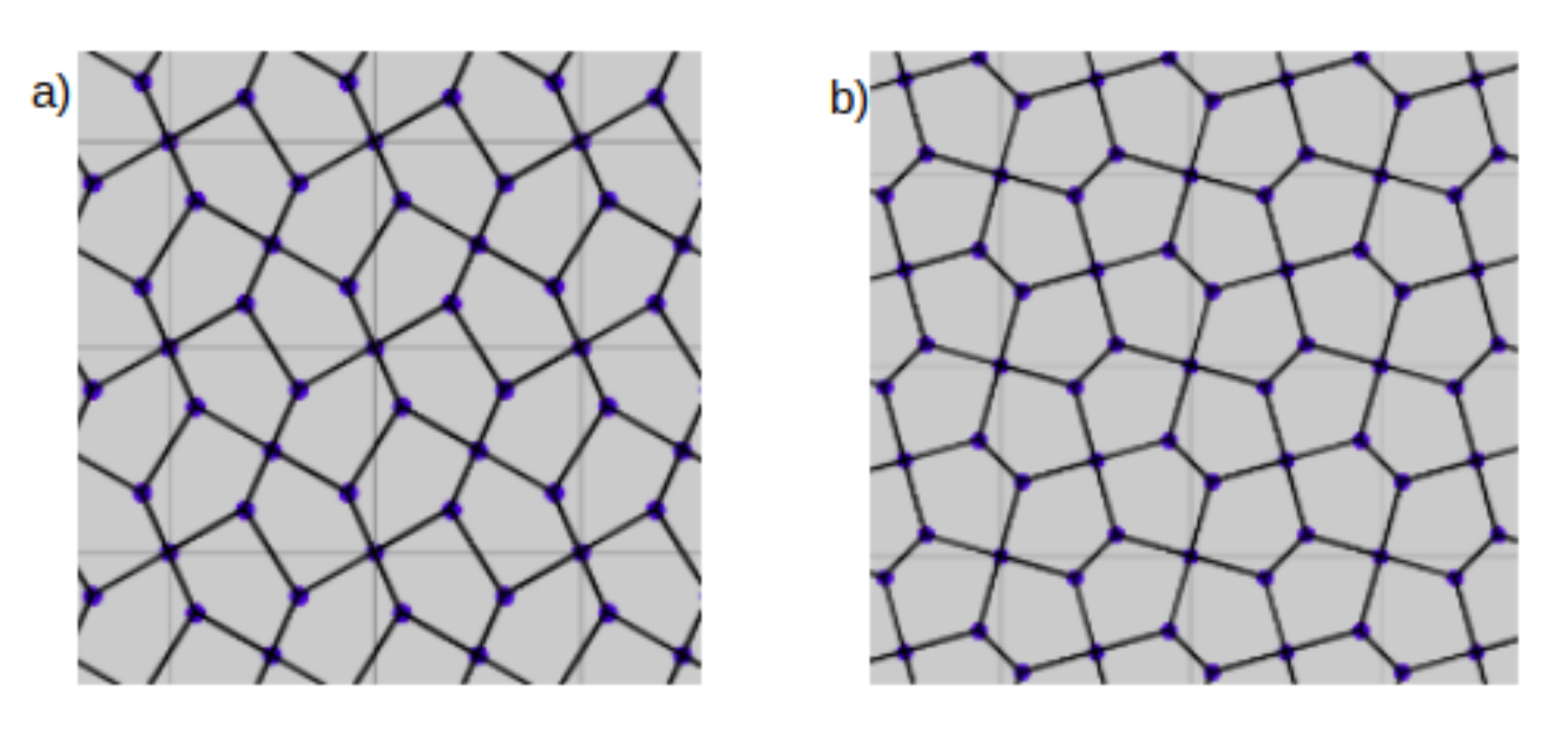}
 \caption{Wallpaper tessellations with a) type-2 and b) type-4 pentagons.}
  \label{tessellation} 
\end{figure}

Such findings have compelled us to classify systematically these 
pentagonal materials and their symmetry-based properties. 
Previous work along this line was done by H. Zhuang \textit{et al.} 
in a series of articles \cite{PhysRevMat_liu2018,ZHUANG2019448,Lui_Zhuang2019_ES}. 
They studied all possible pentagonal tessellations of the plane 
(with fifteen different types of pentagons \cite{2017Rao}) and their 
possible realizations in monolayer materials by first-principles 
calculations. They showed that the pentagonal lattice structure  
is feasible with only two configurations composed of two kinds of
pentagons, namely, type-2 and type-4 pentagons \cite{ZHUANG2019448}. 
An example of the planar projection of these two structures 
is presented in Fig. \ref{tessellation}. 
Despite of this result, no further studies have been done in order 
to relate these two possible structures with spatial groups, which
would be greatly advantageous in the search for particular electronic 
properties linked to symmetry protection. 

Thus, in this work, we classify all (layer) space groups (SG)
that support unstrained pentagonal structures with up to ternary 
composition. We also provide a partial classification for few-layer
structures with specific stackings. Further, we analyze the electronic
properties constrained by crystalline symmetries. 
This allows us to identify nodal points and nodal lines in 
noncentrosymmetric SG hosting pentagonal lattices. In such cases, if 
a significant spin-orbit coupling (SOC) is present, it can lead to 
topologically nontrivial Weyl nodes pinned at high-symmetry points, 
as well as to nodal lines at the Brillouin zone (BZ) boundary. We present 
first-principles calculations for several examples of pentagonal 
materials supporting the group-theoretical predictions. Also, with
the aim to complement our findings, we provide calculations of the
linear optical conductivity, natural optical activity and for the
nonlinear shift current for all such examples. 

This article is organized as follows. In Section \ref{II} we provide
a brief description of the conditions under which the first-principles 
calculations were performed. Also the specific formulas for the optical 
responses are sketched. In Section \ref{result} our main results are 
presented. Firstly, the procedure to find all SG allowing for pentagonal
structures is described. A more detailed discussion for particularly
appealing SG with noncentrosymmetric character follows.
Subsequently, the electronic properties of several pentagonal materials 
belonging to such SG is presented, providing their band structures 
and spin textures. To close this Section, optical calculations based
on the first-principles results are presented. Specifically, the linear 
optical conductivity, the natural rotatory power and the nonlinear shift current
are reviewed, with discussing their features and possible relations to the 
electronic structures. Finally, in Section \ref{end} we state our conclusions
and provide an outlook pinpointing some issues that we consider worth 
to be further developed in future works. Part of our results are left 
as Supplementary Material.  

\section{Computational details}\label{II}

Electronic properties were computed by first-principles calculations using the QUANTUM ESPRESSO package \cite{QE_2020}. The GGA approximation with the PBE functional were used throughout the work with the addition of the the 2D cutoff for layered structures. A cutoff in energy of 90 Ry and a $17 \times 17\times 1$ Monkhorst-Pack grid were selected, ensuring the convergence of the total energy. 
For relaxation purposes the 
force tolerance was set to $10^{-5}$  [eV/\AA]  
In the case of optical calculations, Wannier interpolated
bands were constructed, using the WANNIER90 code \cite{mostofi_updated_2014}. 
After the Wannier model is obtained, calculations with very dense momentum
space grids can be performed. 
We used grids ranging from $301\times 301 \times 1$ to $901\times 901\times 1$
in order to check convergence. Additional post processing of the band 
structure calculations were carried with the WannierTools package 
\cite{WU2017}.
Optical calculations were performed also with WANNIER90 (postw90). 
In particular, the optical conductivity is calculated with the Kubo formula
in terms of the Berry connection, given as a function of the photon
energy $\hbar\omega$ by \cite{mostofi_updated_2014}

\begin{equation}
\sigma _{\alpha \beta }(\hbar\omega) =\frac{ie^{2}}{\hbar N_{k} V}\sum _{\boldsymbol{k},n,m}
f_{mn}E_{mn}\frac{A^{\alpha}_{nm} A^{\beta}_{nm}}{E_{mn}-(\hbar \omega +i \delta )},
\end{equation}
where $A^{\alpha}_{nm}
=\bra{u_{n}}i\nabla _{k_\alpha}\ket{u_{m}}$ is the Berry connection and  
$n$ and $m$ are band indices. Also, $V$ is the cell volume, N$_k$ is the 
number of k-points in the BZ, $E_{mn}=E_{m}-E_{n}$ ($E_{n}=E_{n\boldsymbol{k}}$
is the band dispersion), $f_{mn}=f(E_{m\boldsymbol{k}})-f(E_{n\boldsymbol{k}})$, 
where $f(E_{m\boldsymbol{k}})$ is the Fermi-Dirac function, and $\delta$ is
a parameter with units of energy. 

The rotatory power was calculated using the expression \cite{Malashevich_2010}: 
\begin{equation}
\rho ( \omega ) =\frac{\omega ^{2}}{2c^{2}} \ {\rm Re}[ \gamma _{xyz}( \omega )]
\end{equation}
where the tensor $\gamma_{abc}$ is defined in terms of the band properties as ($\partial_{c}=\partial/\partial_{k_{c}}$)
\cite{Malashevich_2010}

\begin{equation}
\begin{aligned}
{\rm Re}[ \gamma _{abc}( \omega )] = {} & \\
& \frac{e^{2}}{\epsilon _{0} \hbar ^{2}}\int [ d\boldsymbol{k}]\sum ^{occ,emp}_{n,l}\left[\frac{{\rm Re}\left(
A^{b}_{ln} B^{ac}_{nl} -A^{a}_{ln} B^{bc}_{nl}\right)} {\omega ^{2}_{ln} -\omega ^{2}}\right] \\
& -\frac{3\omega ^{2}_{ln} -\omega ^{2}}{\left( \omega ^{2}_{ln} -\ \omega ^{2}\right)^{2}} \partial _{c}( E_{l} +E_{n}) \ {\rm Im}\left(A^{a}_{nl} A^{b}_{nl}\right).
\end{aligned}
\end{equation}
Here $[d\boldsymbol{k}]=d^{d}k/(2\pi)^d $(with d as the dimension), $\omega_{ln}=(E_l-E_n)/\hbar$ and $B^{ac}_{nl}$ ( also valid for $B^{bc}_{nl}$) is given by 

\begin{equation*}
B^{ac}_{nl}=B^{ac \ ORB}_{nl}+B^{ac \ SPIN}_{nl},
\end{equation*}
where $B^{ac \ ORB}_{nl}$ is the orbital contribution and $B^{ac \ SPIN}_{nl}$ is
the spin part which are respectively expressed as
\begin{equation*}
B^{ac\ ( ORB)}_{ln} =\bra{u_{n}}\partial _{c}H \ket{\partial _{c} u_{l}} -\bra{\partial _{c}u_{n}}\partial _{c} H \ket{u_{l}}
\end{equation*}
and
\begin{equation*}
B^{ac\ (SPIN)}_{ln} =-\frac{i\hbar ^{2}}{m_{e}} \epsilon _{abc}\bra{u_{n}}\sigma _{b} \ket{u_{l}} .
\end{equation*}

Finally, we compute the interband contribution to the nonlinear shift current.
It relates the appearance of a current density $J_{a}$ in the system due 
to an electric field $\boldsymbol{E}$ to second 
order, such that $J_{a}=2\sigma_{abc}{\rm Re}[E^{*}_{b}E_{c}]$, where 
$\sigma_{abc}$ is the third rank shift current tensor \cite{Sype_2000}. 
This is obtained using the following formula:

\begin{equation}
\begin{aligned}
\sigma_{a b c}(0 ; \omega,-\omega)
=-\frac{i \pi e^{3}}{2 \hbar^{2}} \int[d \boldsymbol{k}] \sum_{n, m} f_{n m}\\
\times \left(r_{m n}^{b} r_{n m ; a}^{c}
+r_{m n}^{c} r_{n m; a}^{b}\right)
&\delta\left(\omega_{m n}-\omega\right),
\end{aligned}\label{sct}
\end{equation}

 where $r_{nm ; a}^{b}=\partial_{k_{a}}r_{n m}^{b}-i\left(A_{nn}^{a}-A_{m m}^{a}\right) r_{n m}^{b}$ and $r_{nm}^{a}=\left(1-\delta_{n m}\right) A_{nm}^{a}$. 
 The rest of the quantities are defined as in the previous formulas. 

\section{Results and discussion}\label{result}

\subsection{Space groups supporting pentagonal structures}\label{R1} 

Pentagonal materials have the Cairo tiling as their two dimensional projection. 
The essential feature of this pattern is that it possesses two glide planes, 
as it can be visually appreciated in Fig. \ref{tessellation}. This greatly 
constrains the spatial groups of a pentagonal structure.
Therefore, we start identifying the largest tetragonal (layer) SG with this  
characteristic. 
In Table \ref{tab1} we find that SG \#127, which
possesses 16 symmetry operations, is the greatest layer group with this 
feature \cite{bilbao_doubleSG}; note that SG\#129 also has two glide 
planes, but it does not have the adequate Wyckoff positions (WP), as explained below. 
Within this group, 
we can search for all the subgroups that 
preserve this two-glide-plane condition. 
Since we are only using the lattice symmetry, we are implicitly
assuming that all atomic positions are equal. However, this condition
can be relaxed by considering that the spatial symmetry classifies the atomic 
sites in different Wyckoff positions \cite{bradley_group}. 

Thus, in order to preserve a SG we must only care about having equal 
elements within each WP. As it is well-known, the basic pentagonal structure comprises
two WPs, namely, a 4-fold WP and a 2-fold WP, giving six atoms per unit cell \cite{bravo_symmetry-protected_2019}.
We have checked 
that for binary
compounds with different atomic species at each WP, 
the possible SG does not change, being the same as for 
monatomic pentamaterials. We present all possible SG supporting a monolayer pentagonal
structure composed of one or two elements in Table \ref{tab1}. 

This procedure can be extended to ternary penta-monolayer compounds. 
This is done by either 
replacing one atom in the 2-fold WP or by replacing two atoms of the same type 
in the 4-fold WP (replacing one atom is also possible, but it generally gives the trivial space
group P1 as a result). 
 The list of SG for ternary monolayer structures
is given in Table \ref{tab2}, obtained by choosing one SG in Table \ref{tab1} and replacing
the atoms in the WP as described, which yields a new SG with lower symmetry. 
Further increasing the number of elements is not
pursued here since, in general, it will give similar low-symmetry SG as the ones listed
in Tables \ref{tab1} and \ref{tab2}.

Another route to explore is the stacking of pentagonal
layers. We have studied translational (the so-called slip configurations,
no rotation implied) stackings of few-layer pentagonal structures. 
The translational vectors were selected to comprise typical stackings 
such as AA, AB, AC and other, more exotic stackings.
These stackings are amenable to automated SG calculation;  we have
obtained the SG for binary structures for two and three layers 
(composed of the same monolayer). The results are presented in 
the Supplementary Material. 

Let us remark two features related to the obtained SG. \\
(i) There are planar and non-planar structures, which provide a means to 
study the effects of buckling in the electronic properties. For 
example, SG \#127 and \#55 can describe planar (atomically thin)
structures \cite{zhao_2dplanar_2019}, but they can also correspond to 
non-planar pentagonal structures with eight-coordinated atoms where an eight-fold WP
is used to describe the site symmetry, such as in \cite{yuan_prediction_2017}.
\\
(ii)
Restricting ourselves to monatomic and binary systems, SG\#113, 
SG\#100, SG\#90, SG\#32, SG\#18 and SG\#8 are noncentrosymmetric. 
This is one of the conditions needed for the appearance of nontrivial
degeneracy points and lines in the electronic structure
\cite{chang_topological_2018} (see part \ref{R2}). 
These noncentrosymmetric SG can also be subdivided in chiral and
achiral groups depending on the presence of mirror and roto-inversion
symmetry operations \cite{Nye_physical_1985}.
Thus, SG\#90  and SG\#18 are chiral and SG\#113, SG\#100, SG\#32 
and SG\#8 are achiral. 

We center the upcoming discussion in the  
noncentrosymmetric SG, as we aim to find novel examples of 
pentagonal systems with Weyl points in their electronic band structure. 
In the following we 
discuss the nature of degeneracy points in the band structure 
for some specific materials, based on the characteristics of their SG.         

\begin{table}[ht]
\centering{}%
\begin{tabular}{lll}
\hline
\hline
\textbf{Group Label (\#)} & \textbf{Point group} & \textbf{Wyckoff positions}  \\
\hline
\hline
$P4/mbm \: (\#127)$ & 4/mmm  & (4e,2a)  \\
$ P\overline{4}b2  \: (\#117)$ & 4/mmm  & (4e,2a) \\
$P\overline{4}21m \: (\#113) $ & $\overline{4}2m$ & (4d,2a) \\
$P4bm \:  (\#100) $ & 4mm    & (4c,2a)   \\
$P42_{1}2 \: (\#90)$  & 422 & (4d,2b) \\
$ Pbam \: (\#55)$  & mmm    & (4e,2a)  \\
$ Pba2 \: (\#32)$  & mm2    & (4c,2a)  \\
$ P2_{1}2_{1}2 \: (\#18)$   & 222     & (4c,2a) \\
$ P2_{1}/c\: (\#14)$ & 2/m  & (4e,2a) \\
$ Cm\: (\#8)$ & m  & (4b,2a) \\ 
\hline
\hline
\end{tabular}
\caption{\label{tab1} Space groups that can support pentagonal structures with one and two
elements. The Wyckoff positions used to conform the structure are given in the third column.}
\end{table}

\begin{table}[ht]
\centering{}%
\begin{tabular}{llll}
\hline
\hline
\textbf{Parent group} & \textbf{WP change} & \textbf{Ternary group} & \textbf{Point Group} \\
\hline
\hline
\#127* & 2a $\rightarrow$ 1a,1b   & $P4/m$ (\#83)  & $4/m$  \\
\#127 & 4g $\rightarrow$ 2i,2g   & $Cmmm$ (\#65)  & $mmm$ \\
\#127 & 4g $\rightarrow$ 2a,2a    & $Pmc2_{1}$ (\#26)  & $mm2$ \\
\#113 & 2b $\rightarrow$ 1a,1b    & $P\overline{4}$ (\#81) & $\overline{4}$ \\
\#113 & 4e $\rightarrow$ 2e,2d   & $Cmm2$ (\#35)  & $mm2$ \\
\#113 & 4e $\rightarrow$ 2a,2a       & $P2_{1}$ (\#4) & $2$ \\
\#100 & 2a $\rightarrow$ 1a,1b    & $P4$ (\#75)    & $4$ \\
\#100 & 4c $\rightarrow$ 2e,2d   & $Cmm2$ (\#35)  & $mm2$ \\
\#100 & 4c $\rightarrow$ 2a,2a       & $Pc$ (\#7)     & $m$ \\
\#90 & 2c $\rightarrow$ 1a,1b     & $P4$ (\#75)      & $4$ \\
\#90 & 4d $\rightarrow$ 2e,2g    & $C222$ (\#21)     & $222$ \\
\#90 & 4d $\rightarrow$ 2a,2a        & $P2_{1}$ (\#4)     & $2$ \\
\#55 & 2a $\rightarrow$ 1a,1e     & $P2/m$ (\#10)     & $2/m$ \\
\#55 & 4e $\rightarrow$ 2m,2n     & $P2/m$ (\#10)     & $2/m$ \\
\#55 & 4e $\rightarrow$ 2a,2b     & $Pmc2_{1}$ (\#26)     & $2$ \\
\#32 & 2a $\rightarrow$ 1a,1a     & $P2$ (\#3)     & $2$ \\
\#32 & 4c $\rightarrow$ 2e,2e     & $P2$ (\#3)     & $2$ \\
\#32 & 4c $\rightarrow$ 2a,2a     & $Pc$ (\#7)     & $m$ \\
\#18 & 2a $\rightarrow$ 1a,1a     & $P2$ (\#3)     & $2$ \\
\#18 & 4d $\rightarrow$ 2e,2e     & $P2$ (\#3)     & $2$ \\
\#18 & 4d $\rightarrow$ 2a,2e     & $P2_{1}$ (\#4)     & $2$ \\
\#14 & 2a $\rightarrow$ 1a,1a     & $P\overline{1}$ (\#2)     & $\overline{1}$ \\
\#14 & 4e $\rightarrow$ 2a,2a     & $P2_{1}$ (\#4)     & $2$ \\
\#14 & 4e $\rightarrow$ 2a,2a     & $Pc$ (\#7)     & $2$ \\
\#8 & 2a $\rightarrow$ 1a,1b   & $Pm$ (\#6)     & $m$ \\
\#8 & 4b $\rightarrow$ 2a,2a   & $P1$ (\#1)     & $1$ \\
\#8 & 4b $\rightarrow$ 2a,2b   & $Pm$ (\#6)     & $m$ \\
\hline
\hline
\end{tabular}
\caption{\label{tab2} Ternary space groups supporting the pentagonal structure, 
arising from site substitutions in the SG of Table \ref{tab1}. Information was retrieved
from the Bilbao Crystallographic Server \cite{bilbao_serverII}. *The SG \#117 gives the same ternary 
groups as the same WP coordinates are used for both groups.}
\end{table}

\begin{figure}[ht]
\includegraphics[width=1.0\columnwidth,clip]{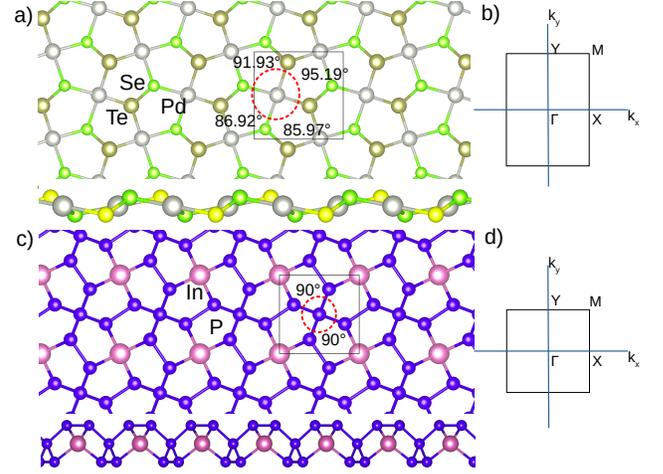}
\caption{Lattice structure for (a) SG \#4 and (c) SG \#81. Brillouin zone for (b) SG \#4 and (d) SG \#81. The angles given are measured considering only the planar projection and not the entire three-dimensional lattice structure.}
\label{lattBZ} 
\end{figure}

\subsection{Electronic structure of noncentrosymmetric pentagonal materials}\label{R2}

Since the proposal of penta-graphene, several pentagonal materials have been put forward. 
Most of the materials with one and two elements belong to SG \#14, \#113 and \#127.
To our knowledge, there are not reported materials for the rest of the groups in Table \ref{tab1}. 
Those from SG \#14 are the most numerous; several databases for 2D materials report them   
\cite{haastrup_computational_2018,mounet_two-dimensional_2018}. 
In fact, PdSe$_2$, the only pentagonal 2D material synthesized to date, 
pertains to this SG \cite{oyedele_pdse_2017}. From these three groups, 
only SG \#113 is noncentrosymmetric. 
We focus on materials with noncentrosymmetric SG because they can host 
nodal points or lines, Weyl points and non-trivial topological features when SOC is relevant. 
In order to find more examples of those, we propose two possible routes. 
First, we search for ternary materials with noncentrosymmetric SG (Table \ref{tab2}).
This is the case of SG \#4; a list of ternary materials has already 
been presented in \cite{liu_computational_2019}.
We take an alternative route for SG \#4 and start with highly
stable SG \#14 materials \cite{haastrup_computational_2018} to derive 
new ternary materials via atomic substitution, following the information of 
Table \ref{tab2}. We propose two systems, PdSeS and PdSeTe, which are formed 
from PdSe$_2$.

Another example is  SG \#81, for which dynamically stable  
materials with formula XP$_5$ (X=Al, Ga, In) have also 
been proposed \cite{naseri_investigation_2018}. 
Note that although XP$_5$ materials are compositionally binary, from the group-theoretical viewpoint
it can be considered ternary, given that one P atom is located in a WP different to the other four 
P atoms, breaking the two-glide plane symmetries present in the parent SG \# 113.   
We select the InP$_5$ system for SG \#81;  although it was studied previously, SOC was not considered 
\cite{naseri_investigation_2018}. We include this effect in order to explore the spin texture and 
the appearance of Weyl points.  

A second possible route is related to bilayers. Specifically, bilayers composed of 
SG \#113 monolayers with AB stacking give rise to noncentrosymmetric achiral
SG \#111. However, we do not follow this approach in the present work.

To summarize, we have chosen three noncentrosymmetric SG for pentagonal 
2D materials, namely, \#4, \#81, and \#113. We provide examples of 
the lattice structure for SG \#4 and \#81 in Fig. \ref{lattBZ} a) and c), respectively.
Notice that the SG \#113 lattice is similar to the SG \#81, but the 4-coordinated 
atoms are equal for the \#113 case.
Our selection for these particular SG is guided by two factors: on one hand, their 
proximity to recently reported materials, such as PdSe$_2$, and on the other hand, 
the attractive chiral/achiral relationship of their
nodal points and lines, as discussed below.

\begin{figure}[ht]
\centerline{
\includegraphics[width=1.0\columnwidth,clip]{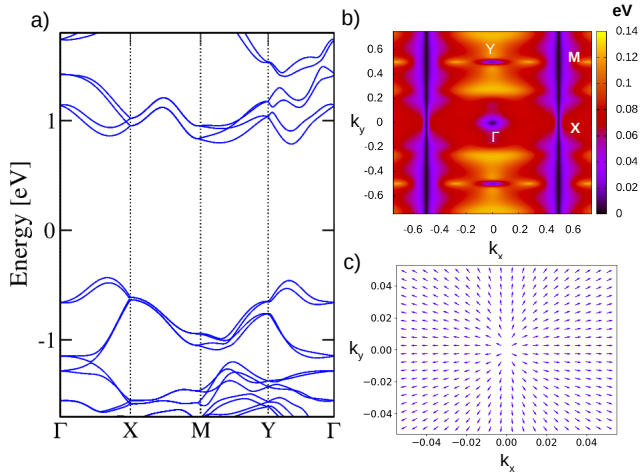}}
 \caption{a) First-principles electronic band structure for PdSeTe. b) A color map 
 for the energy difference between the lowest pair of conduction bands mapped over 
 the entire BZ for PdSeTe. c) Spin texture for the fourth valence band in the vicinity of
 the $\Gamma$ point for PdSeTe.}
  \label{PdSeTe_QE} 
\end{figure}

The first example we present is from SG \#4 (P2$_1$), with formula PdSeTe, which has been
derived by atomic substitution of two Se atoms by Te atoms at a 4-fold WP. The electronic 
band structure including SOC for PdSeTe is presented in Fig. \ref{PdSeTe_QE} a), where
the typical \textit{stick-together} bands along the 
high symmetry line X-M can be observed, due to the nonsymmorphic character of 
the SG  \cite{dresselhaus_group_2008}. Further, around the $\Gamma$ point
we can observe the spin splitting of bands, which show a 2-fold degeneracy 
just at $\Gamma$. 

In order to analyze the character of this nodal point we have numerically calculated 
the Berry phase \cite{vanderbilt_2018} along a circular path around the $\Gamma$ point, 
which yields a value of $\pi$, indicating that it corresponds to a Weyl point 
with topological charge $|C|=1$ \cite{Hu_2019_ARCM,nagaosa_transport_2020}. 
Additionally, we present in Fig. \ref{PdSeTe_QE} c) the spin texture of the fourth
uppermost valence band to gain further insight into the nature of the Weyl 
node. For this particular band the spin texture shows a radial
pattern near $\Gamma$, suggesting that these Weyl points are indeed a two-dimensional 
version of Kramers-Weyl points, as it was recently realized in other systems
\cite{chang_topological_2018}. Kramers-Weyl points are present in every chiral SG. 
They are pinned to high-symmetry points as long as no nodal lines are
connected to them \cite{xie_kramers_2020}. In view of this, the Y point at 
the corner of the BZ can also host a Kramers-Weyl node, as it can been seen
from Fig. \ref{PdSeTe_QE} a). 
From a direct analysis of the physically irreducible representations (irreps) of this 
SG \cite{bilbao_doubleSG}, it can be seen that the $\Gamma$ and Y points only have 
two-dimensional irreps, indicating that only Weyl nodes can be formed. Additionally, along
the X-M high-symmetry line, only two dimensional irreps are possible, thus implying the
formation of a Weyl nodal line. 

This nodal line also has a little group that only has a two-dimensional
irrep if time-reversal holds. This irrep can be expressed as $e^{i \pi u}i\sigma_z$,
where $u$ is the fractional coordinate along the nodal line \cite{bilbao_doubleSG}. 
Following the results in \cite{2012_MultiWeyl_PRL,2017_Multiweyl_PRB}, 
the ratio $\alpha_{v}/\alpha_{\mu}$, where $\alpha_{v/\mu}$ are the irrep eigenvalues, 
indicates the type of Weyl point present along a symmetry line. In this case
$\alpha_{v}/\alpha_{\mu}=-1$, meaning that a symmetry-protected Weyl node is 
produced at every point along the BZ boundary, with Chern number of magnitude $|C|=1$.

These features should be present in every material with the SG \#4 and time-reversal symmetry. 
These point nodes and nodal lines are protected by the combined action of time-reversal 
$T$ and $2_1$ symmetries $(TC_{2}|1/2,0,0)$ \cite{bradley_group,el-batanouny_wooten_2008}. 
Similar results are obtained for PdSeS, 
which are presented in the Supplementary Material. The Weyl points and nodal line can be
clearly identified in a color map plot of the energy difference between two adjacent bands, 
as shown in Fig. \ref{PdSeTe_QE} b), for the lowest pair of conduction bands,
where the Weyl nodes are clearly spotted and the 
nodal line appears along all the X-M line. 

\begin{figure}[ht]
\centering
\includegraphics[width=1.0\columnwidth,clip]{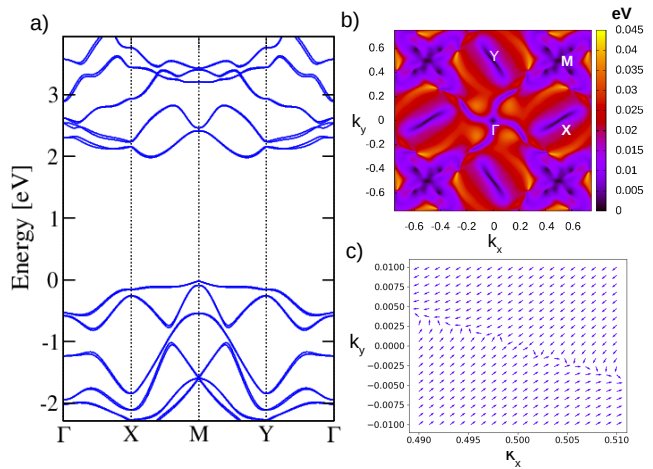}
 \caption{a) Electronic band structure for InP$_5$. b) Color map 
 for the energy difference between the third and fourth uppermost valence bands, 
 mapped over the entire BZ for InP$_5$. Darker zones away from high-symmetry points
 are not degenerate, but very close in energy. c) The spin texture for the first
 conduction band for a region near the X point for InP$_5$.}
  \label{InP5_bands}
\end{figure}

Next, we turn to the SG \#81 (P$\bar{4}$), presenting the first-principles band structure with SOC 
for InP$_5$ in Fig. \ref{InP5_bands} a). The spin splitting can also be observed, being larger  
for the lowest conduction bands near the $\Gamma$ point. We also have computed the Berry
phase for this material, obtaining a value of $\pi$ for all the high-symmetry points of 
interest, mentioned below. In this case the lattice structure is achiral, due to the
presence of roto-inversion $S_4$ and mirror planes. Still, as mentioned 
in \cite{chang_topological_2018}, there exists the possibility 
to have a reminiscence of chirality, so to say, at momentum space. 
This is because certain little groups (at high-symmetry points) of 
these achiral groups could be chiral \cite{bilbao_serverII}. 
This is the case for SG \#81, 
where the X(Y) point located at [1/2,0] ([0,1/2]) has a chiral little 
group \cite{bradley_group}. To explore this assertion, the spin texture for 
the first conduction band is presented in Fig. \ref{InP5_bands} c), in the
vicinity of the X point. It can be observed that the radial pattern is partially
present, which is related to the competition of the spin-orbit
interactions that can be present in the neighborhood of X, due to the $C_{2}$
little group. 
This competition distorts the spin texture and masks the chiral 
signatures of the node. 

It can be mentioned that at the $\Gamma$ point (and also at the M point), there exists a Weyl 
point that is hosted by an achiral little group, and therefore no Kramers-Weyl node is
allowed \cite{sanchez_topological_2019}. One of the interesting characteristics of these points
($\Gamma$ and M) is that they can be considered as two-dimensional projections of 
Kramers-Weyl nodal lines \cite{xie_kramers_2020}.
That is to say, they correspond separately to the end points of two nodal lines that
would exist in a three-dimensional BZ for the same SG along $\Gamma$-A and
M-R \cite{xie_kramers_2020}. Notice also that there are no nodal lines, since
only one-dimensional irreps are present away from the high-symmetry 
points \cite{bilbao_serverII}. All these facts can be better appreciated
in the color map of the spin splitting (energy difference) for the second pair
of adjacent valence bands in Fig. \ref{InP5_bands} b). 
It can be seen that the Weyl nodes appear exactly at the high-symmetry
points, along with zones of relative high and low spin splitting. Similar 
conclusions could be attained for the SG \#111 ($P\bar{4}2m$), where the X(Y) 
point also hosts Kramers-Weyl nodes. The reason is the same as for SG \#81. 
As was mentioned, this SG can be formed by a bilayer composed of two (equal) 
SG \#113 monolayers with AB stacking. We leave the detailed analysis of 
multilayer pentagonal structures for a later work.

\begin{figure}[ht]
\centering
\includegraphics[width=1.0\columnwidth,clip]{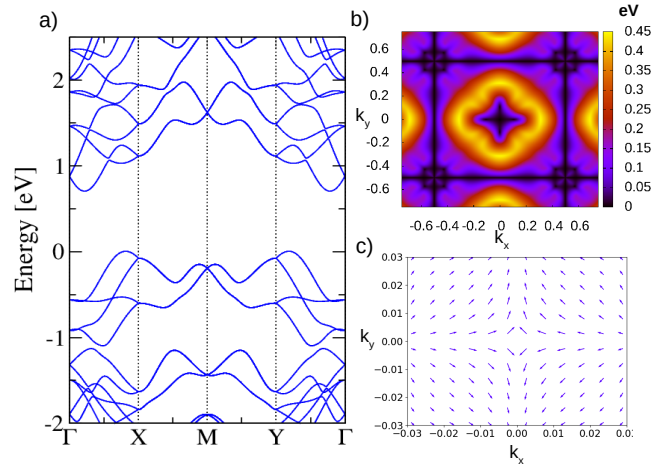}
 \caption{a) First-principles electronic band structure for GeBi$_2$. b) Color map 
 for the energy difference between the two uppermost valence bands mapped over the 
 entire BZ for GeBi$_2$. c) Spin texture for the uppermost valence band near the 
 $ \Gamma$ point for GeBi$_2$.}
  \label{GeBi2_bands}
\end{figure}

Finally, we propose a stable binary SG \#113 (P$\overline{4}21m $) material, GeBi$_2$. Its band
structure with SOC is presented in Fig. \ref{GeBi2_bands} a). In this
achiral SG, the M point little group protects a fourfold degenerate node 
\cite{flicker_chiral_2018,sanchez_linear2019}. 
This can be clearly observed in the band structure of Fig. \ref{GeBi2_bands}, 
as for example in the upper group of valence bands, where the fourfold 
fermion locates at an energy of $-0.20$ eV. 
This fermion is not chiral, since the little group of the M point 
is isomorphic to $D_{2d}$ \cite{bilbao_serverII}. The rest of the 
high-symmetry points have two-fold Weyl points with no chiral character,
namely, the $\Gamma$, X and Y points. 
However, only the $\Gamma$ point comprises an isolated node, since the X and 
Y points are part of a closed twofold nodal line that goes around the whole
BZ boundary, as was reported in our previous work \cite{bravo_symmetry-protected_2019}. 
The analysis for this nodal line is analogous to that of the SG \#4 case. 
The ratio for the 2D irrep in the little group
is $\alpha_v/\alpha_\mu =-1$, giving again a line of Weyl points all
along the BZ boundary, with $|C|=1$ \cite{2017_Multiweyl_PRB}.
We should mention that this twofold nodal line is formed by the 
splitting of a fourfold nodal line present in the case without SOC. The
two-dimensional character of this nodal line, and that occurring in 
SG \#4, 
is an interesting feature that has attracted the attention of several groups 
\cite{nodal_Nanoscale_2017,nodal_PRL_2018,nodal_PRB_2018,nodal_Nanoscale_2019,nodal_PRM_2019}.

To recapitulate, 
in this SG we encounter the peculiar coexistence of a 
fourfold fermion, a twofold Weyl node and a closed twofold nodal line, 
which gives several possibilities to study the interaction between degenerate fermions. 
Additional insight can be acquired by looking at Fig. \ref{GeBi2_bands} b), where all the nodal points 
and the closed nodal line mentioned above are clearly shown in the color map for 
the upper pair of valence bands. We complete this characterization by presenting   
the spin texture for the top valence band in the vicinity of the $\Gamma$ point in Fig. \ref{GeBi2_bands} c). It shows a 
Dresselhaus-like spin texture, as allowed by the point group associated 
to SG \#113 \cite{bradley_group}.
The fourfold M point also presents this type of spin-orbit texture,
as it can be observed in Fig. S.21 in the Supplementary Material.

The diverse patterns arising in the spin textures can be better explained 
by resorting to an effective model that we present in the Appendix.
There, it is shown that the spin texture patterns depend on the relative 
magnitude of the coefficients accompanying the momentum components in 
the $\boldsymbol{k}\cdot \boldsymbol{p}$ Hamiltonian. 
One of the most appealing features of the SG \#4 materials is that the
spin texture depends not only on the spin-orbit interaction but also on 
the anisotropy due to the low spatial symmetry.  Likewise, the distortion
of the spin texture due to the competition between SOC and anisotropy 
found in group \#81 is further analyzed in the Appendix. Different spin
textures are realized for each band in these materials; several examples 
are depicted in the Appendix illustrating the situations found in the 
three groups under study.

We have chosen several penta-materials to illustrate the characteristics associated 
with  different SG. We would like to emphasize that, to the best of our knowledge, 
there are not any reported materials for the SG \#90 to this date. Nonetheless, 
this group is of great interest, since it can host chiral multifold 
fermions \cite{bradlyn_beyond_2016}. In particular, at the M point this SG has
a chiral fourfold degenerate fermion, with a double spin-1/2 
representation \cite{bradlyn_beyond_2016,flicker_chiral_2018}. 
Also a twofold chiral fermion is present at $\Gamma$, and  
twofold nodal lines along the whole BZ boundary are present, similar to the
SG \#113 but with chiral nature. This implies that the interplay between 
chiral multifold and chiral twofold fermions could be studied in this particular SG. 
Finding a material realization of this SG would certainly be appealing.

\subsection{Optical properties}\label{optics}

The existence of Weyl points is related to enhanced physical responses, compared 
to materials without topological nodes. For example, a large 
magnetoresistance \cite{shekhar_extremely_2015,Huang_2015_PRX} in
magnetic materials, transport characteristics 
such as the chiral magnetic effect, the nonlinear 
anomalous Hall effect, the kinetic Faraday effect and the gyrotropic
magnetic effect, also called kinetic magnetoelectric effect 
\cite{Wang_2017_AdvPhysX,Tsirkin_2018,Hu_2019_ARCM}. 
In addition, optical effects are among the most promising responses due to
the momentum space selection rules, that could give characteristic features  
at certain frequencies \cite{chang_topological_2018,ibanez-azpiroz_ab_2018}. 
Here we report optical first-principles calculations in the Wannier basis for the  materials previously analyzed. 
Specifically, we have computed the optical conductivity, the 
rotatory power in terms of the natural optical activity (NOA) and the interband
contribution to the nonlinear shift current (for details see Section \ref{II}). 
In what follows we describe some of the main aspects of the optical properties.

First, all calculations were carried out using Wannier-interpolated models for each
material. In particular, for PdSeTe it includes the outermost valence 
orbitals from each atom; i.e., $d$ orbitals for Pd and $p$ orbitals for Se and 
Te; the same type of orbitals were employed for PdSeS. For InP$_5$, $p$ valence 
orbitals were used for P and $sp^3$ hybrid orbitals for In. 
In the case of GeBi$_2$, $p$ orbitals were used for Ge and hybrid $sp^{3}$ 
orbitals for Bi. These choices yield a good agreement for the low-energy bands. 
The Wannier band structures for each material are presented in the 
Supplementary Material.

\begin{figure}[ht]
\centering
\includegraphics[width=1.0\columnwidth,clip]{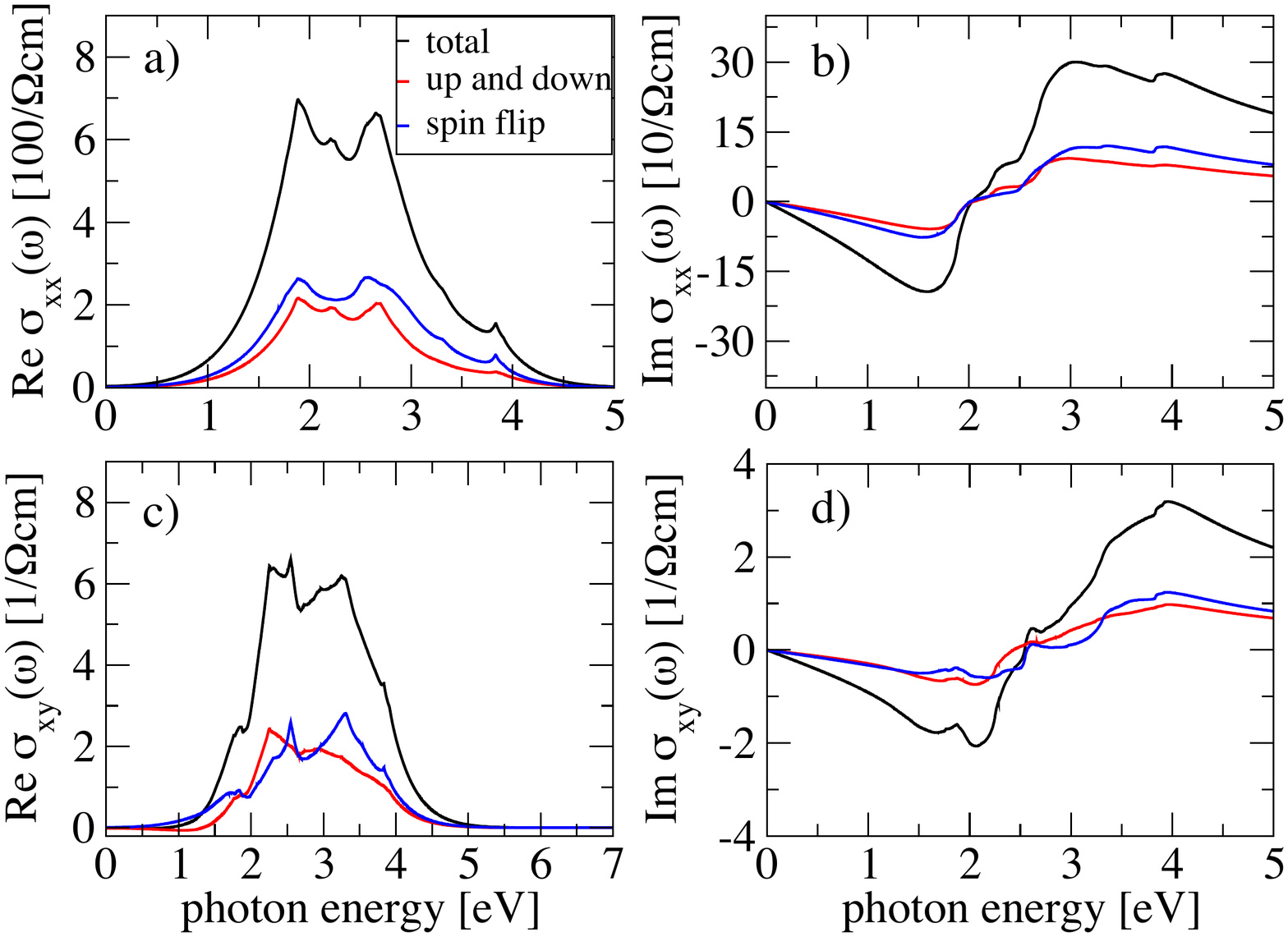}
\caption{Optical conductivity for PdSeTe with spin components along the $z$ direction. 
(a) Real part for $\sigma_{xx}$. (b) Imaginary part of $\sigma_{xx}$. 
(c) Real part for $\sigma_{xy}$. (d) Imaginary part of $\sigma_{xy}$.}
\label{kubo_PdSeTe} 
\end{figure}

We begin by reporting the optical conductivity for PdSeTe, projected in the 
spin $z$ direction, which is presented in Fig. \ref{kubo_PdSeTe}. It can be seen
that no spin polarization is present in this system (other spin directions give 
similar results). Due to the low symmetry of this SG, there are nonzero off-diagonal 
(Hall-like) components of the optical conductivity, which are two orders
of magnitude smaller than the longitudinal part. Still, this could give
rise to photoinduced currents in the transverse directions and also to optical 
rotation. The low symmetry also implies that there are few selection rules
for interband transitions in momentum space. In consequence, contributions from
all the BZ make the distinction of peaks rather opaque.

\begin{figure}[ht]
\centering
\includegraphics[width=1.0\columnwidth,clip]{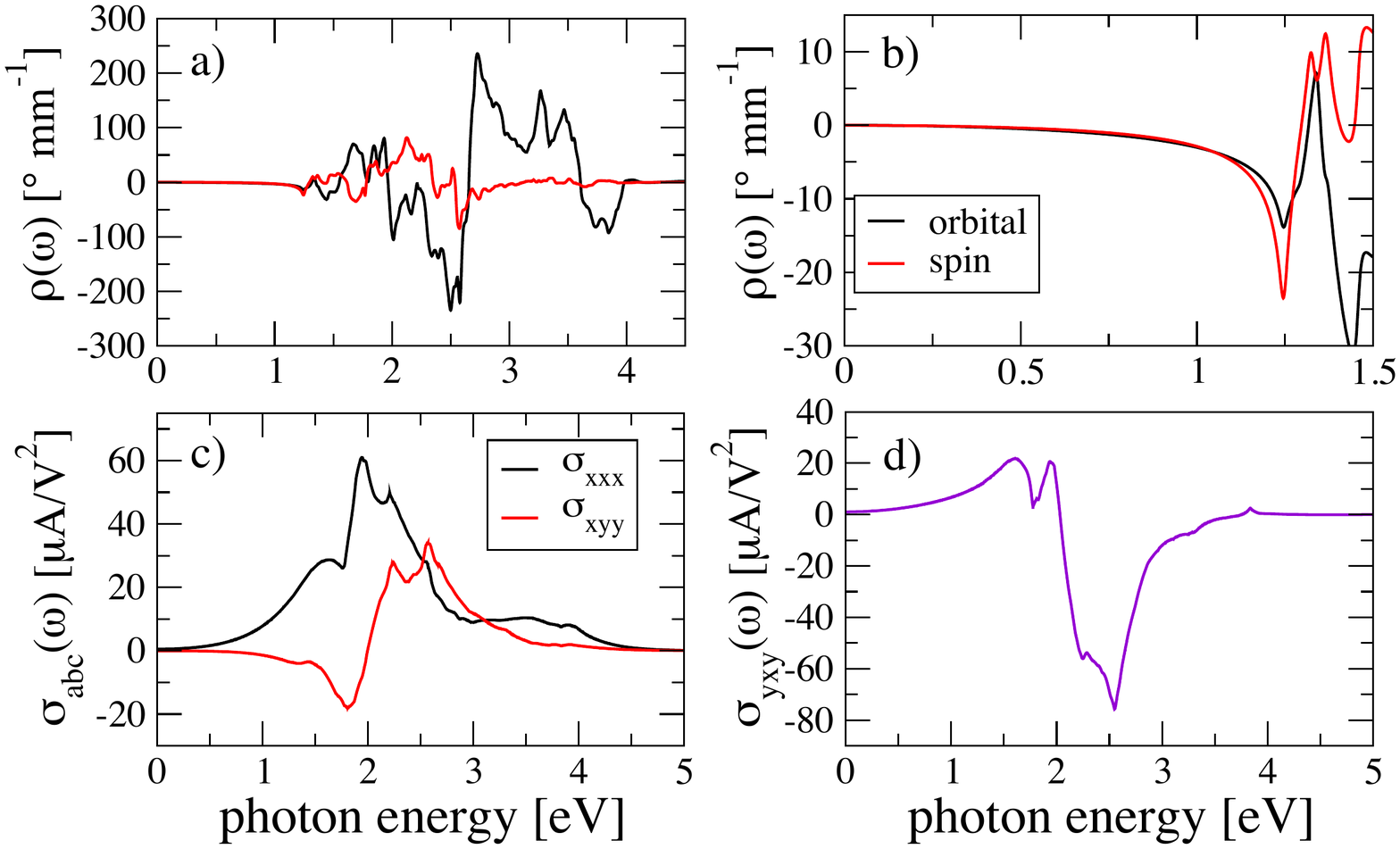}
 \caption{a) Rotatory power from NOA for PdSeTe with orbital and spin contribution. 
 (b) Zoom-in from (a) along the region below the optical direct bandgap.
 (c) Shift current along the $x$ direction with in-plane polarization for PdSeTe.
 (d) Shift current along the $y$ direction with in-plane polarization for PdSeTe.}
  \label{PdSeTe_NOA_SC} 
\end{figure}

In order to gain insight about the optical response we calculated 
the interband contribution to the rotatory power, obtained from 
the NOA tensor, as shown in Fig. \ref{PdSeTe_NOA_SC} (a) for PdSeTe. 
There exists a nonzero response in a wide range of frequencies, which 
gives to this material a great potential for optoelectronic applications
\cite{pospischil_optoelectronic_2016}. 
We hypothesize that such response could be augmented by using multilayers,
since the chiral nature is enhanced by increasing thickness
\cite{long_chiral-perovskite_2020}. Even a higher outcome could be
obtained if rotated layers are employed, as previous works on 
twisted materials have reported \cite{Suarez_Chico_2017,slip_twist_PRB_2019}. 
Note that the orbital contribution is generally 
greater than the spin contribution, with the exception of the 
low-frequency region, including the zone below the absorbing edge ($\sim$ 1.2 [eV]).
This is shown in Fig. \ref{PdSeTe_NOA_SC} (b), where it is evident that
the spin contribution is of the same order as the orbital contribution,
even surpassing it close to the gap. This last behavior
is encountered also in the NOA response of GeBi$_2$ along the whole 
frequency range, as shown in Fig. S.23 in the Supplementary Material. This is
in fair contrast with other materials such as elemental tellurium, where the 
spin contribution was found to be marginal \cite{Tsirkin_2018}. 

Nonlinear optical effects occurring in these materials with chiral nodes 
have special relevance for potential applications  
\cite{tan_shift_2016,xu_comprehensive_2020,le_ab_2020,ni_linear_2020}.
In order to explore this issue, we computed the interband contribution to the nonlinear shift 
current $\sigma_{abc}$, which arises due to the charge center shift associated to the
effects of a nonzero interband Berry 
connection \cite{cook_design_2017,tan_shift_2016}. 
The effect depends
on the type of (linear) polarization of light and the symmetry
character of the $\sigma_{abc}$ tensor, given in Eq. \ref{sct}. 
For PdSeTe(S) (SG\#4) the tensor has 4 independent 
nonzero components, while for InP$_{5}$ (SG \#81) the tensor has 2 nonzero
independent components \cite{Gallego_tensor}.
For the purpose of yielding more reliable results, the shift current of 
the layered system
is rescaled by a numerical factor taking into account the slab
geometry, as implemented in \cite{Rangel_GeS,ibanez-azpiroz_ab_2018}. 
This factor is defined as the ratio between the length of the 
vacuum region $d_{vac}$ and the monolayer thickness $d_{m}$, such
that the final shift current tensor $\sigma_{abc}$ is 
given by \cite{ibanez-azpiroz_ab_2018}

\begin{equation}
\sigma_{abc}=(\frac{d_{vac}}{d_{m}})\sigma^{layer}_{abc}.
\end{equation}

\begin{figure}[ht]
\centering
\includegraphics[width=1.0\columnwidth,clip]{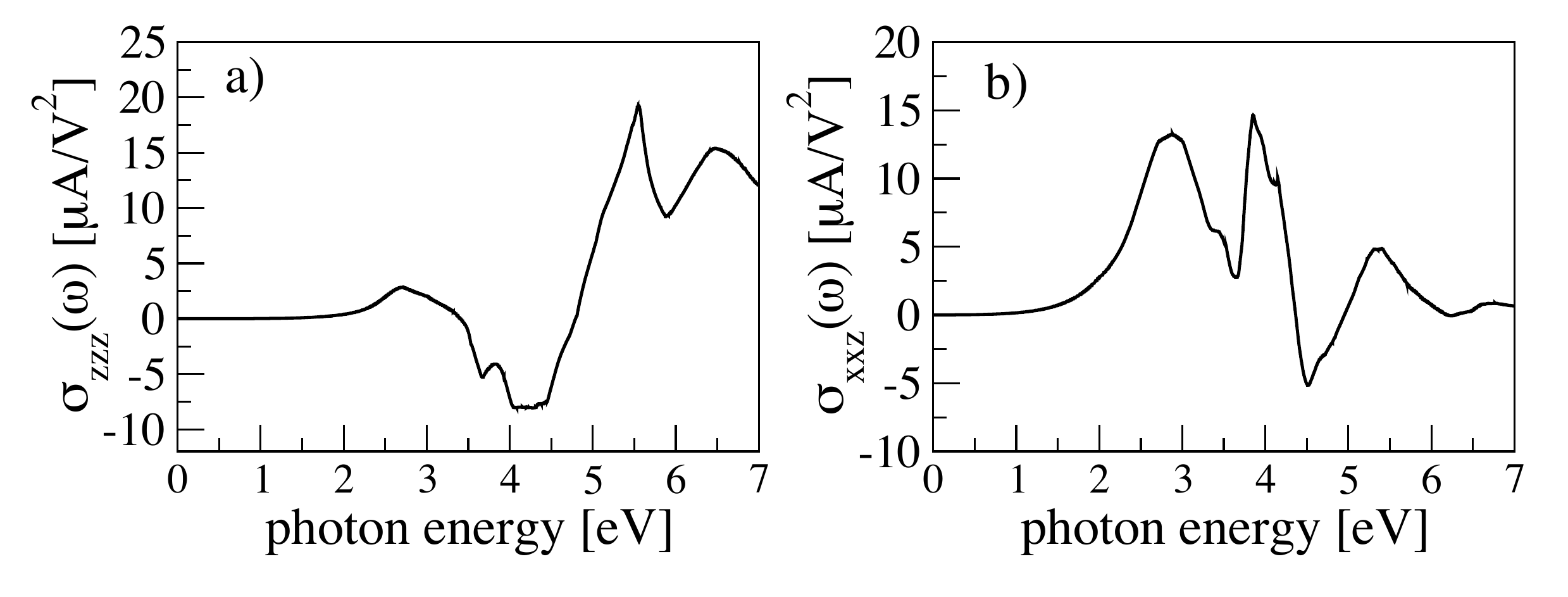}
 \caption{Components of the shift current tensor $\sigma_{abc} (\omega)$  for InP$_5$.}
  \label{shift_InP5} 
\end{figure}

We present the shift current tensor for PdSeTe along the $x$ and $y$ directions in 
Fig. \ref{PdSeTe_NOA_SC} (c) and (d).   
The rest of the components are not shown because they have a negligible value in comparison. 
Additionally, in Fig. \ref{shift_InP5} the shift current for InP$_5$ 
is shown, with values in other polarization directions,  
specifically, in the $z$ direction. 

This current is more sensitive to specific regions in momentum space, 
at variance with other optical responses \cite{zhang_photogalvanic_2018}. 
Peaks in these currents are slightly related to peaks in the optical
conductivity contributions since two-band and three-band transitions
are the most important weights 
\cite{flicker_chiral_2018,zhang_photogalvanic_2018,zhang_strong_2019}. 
Points where Weyl nodes are located ($\Gamma$ and Y) have a large contribution,
as for example in $\sigma_{xyy}$ for PdSeTe, depicted in  
Fig. \ref{PdSeTe_NOA_SC} (c). The peaks are localized within regions 
which coincide with the energy difference between two Weyl points. 
Unfortunately, this contribution from Weyl points is not exclusive 
for this particular energy range, due to the low symmetry of the 
material and to the proximity in energy of trivial bands 
along the BZ. 

To sum up, 
whereas 
Weyl points contribute significantly 
to the shift current and, in fact, certain peaks in the spectrum 
can be attributed to them, 
no intrinsic Weyl signatures
are expected in this system, such as it occurs in the 
circular photogalvanic effect \cite{zhang_photogalvanic_2018}. 
A similar conclusion is reached by inspecting the shift current 
of InP$_5$ in Fig. \ref{shift_InP5}, where achiral Weyl nodes
do not show any particular features.
Still, there are  certain aspects that can be pointed out, such
as the occurrence of anisotropy, which is observed by contrasting
different components of the $\sigma_{yxy}$ tensor. Since the 
contribution to the shift current from these components depends
on the polarization of light, anisotropy introduces selectivity 
along some particular spatial directions. 
This is highly desirable for applications in photovoltaics, 
which requires preferred directions in materials in order 
to efficiently transport the induced current to the 
electrodes \cite{cook_design_2017}.

One of the main drawbacks of this bulk photovoltaic response is the
low conversion ratio, which can hinder the applicability of 
the effect \cite{tan_shift_2016}. 
Consequently, the magnitude of the shift current is also important 
\cite{young_first_2012,young_first-principles_2012}. 
The search for novel materials with high shift currents is thus
the focus of great interest \cite{xu_comprehensive_2020}.  
In this line, we would like to emphasize the remarkable values obtained 
for the pentagonal material PdSeTe for component
$\sigma_{xxx}$, as shown in Fig. \ref{PdSeTe_NOA_SC} (c), with a peak value
of 60 [$\mu$A/V$^{2}$] at $\sim 1.9$ eV and for $\sigma_{yxy}$ 
in Fig. \ref{PdSeTe_NOA_SC} (d), with a peak value of $-76$ [$\mu$A/V$^{2}$] 
at $\sim 2.5$ eV. 
We can also observe in Fig. \ref{shift_InP5}(a) that the $\sigma_{zzz}$ 
component in InP$_5$ has a peak value of 130 [$\mu$A/V$^{2}$] at 
$\sim 5.55$ eV. Finally, $\sigma_{xxz}$ has two peaks of 89 [$\mu$ A/V$^{2}$]
and 98 [$\mu$A/V$^{2}$] at $\sim 2.88$ eV and $\sim 3.87$ eV, respectively.    
These magnitudes compare very favorably with previously reported values
for layered materials, as for example with monolayer GeS, which possesses 
a shift current of 100 [$\mu$A/$V^{2}$] \cite{Rangel_GeS}. 
Furthermore, since the values are within the visible/UV spectrum,  
they convey very promising prospects for these systems as 
photovoltaic materials \cite{zenkevich_giant_2014}. 

As it is well-known, the use of the PBE functional underestimates 
the band gaps. This also has an obvious consequence in the optical 
spectra, producing a redshift of the optical features. For those 
materials including a transition group element, we also calculated
the electronic properties with a hybrid functional, showing that 
our results are not affected by this approximation, besides the 
expected energy displacement mentioned above.

In the Supplementary Information we present additional material
related to the optical response of InP$_5$ and for GeBi$_2$ (SG \#113). 

\section{Conclusions and outlook}\label{end}

In this work we have classified all the layer groups that can support 
a pentagonal structure with up to three elements. We have also classified 
multilayer structure with slip stackings up to three layers.
We have focused on specific noncentrosymmetric SG, with
chiral and achiral point groups. The corresponding SG were described in 
detail and their degeneracies linked to Weyl nodes and Kramers-Weyl nodes 
depending on their little group. This framework was subsequently applied to several 
examples of pentagonal materials to describe their electronic structures.
Additionally, their optical properties were reported, complementing 
the description of their electronic band structures.  
Although the existence of Weyl points in these materials does 
not imply any specific signature in their physical properties, 
they can be related to the enhancement of the shift currents, 
yielding high values of this magnitude. 

We expect 
that pentagonal materials such as PdSeTe and PdSeS to be 
fairly feasible experimentally, since they are closely related with
the recently synthesized pentagonal material PdSe$_2$. 
Other routes that are worth to be explored further are the signatures
of the spin textures and their spin-orbit competition and anisotropy. 
Also, multilayers are highly attractive since little work has been 
done for pentagonal systems and, as we mentioned above, different
stackings could give emergent SG with contrasting symmetry-related properties or boost chirality-related phenomena. 
Finally, our results indicate that the optical response for other 
pentagonal materials deserves to be explored, since the interplay of Weyl nodes, 
low-symmetry and chemical composition could yield novel systems with enhanced 
photovoltaic currents.

\section*{Appendix: Effective models and spin textures for two-fold Weyl points}\label{append}
\renewcommand{\theequation}{A$.$\arabic{equation}}
\renewcommand{\thefigure}{A$.$\arabic{figure}}  
\setcounter{equation}{0}
\setcounter{figure}{0}

Since the degenerate points are local properties in momentum space, 
effective models are helpful to understand in more detail the relation
between structural chirality and spin textures. Here we analyze specific nodal points
in SG \#4, SG \#81 and SG \#113, selected from those presented in section \ref{R2}.

\textit{SG \#4 and SG \#81}. For both groups, the most interesting 
$k$ points are those with chiral little groups ($\Gamma$ and Y for SG \#4 and 
X for SG \#81). These little groups are all isomorphic to $C_2$
\cite{bilbao_serverII}, which has only one symmetry operation, 
namely, a $\pi$ rotation about the axis perpendicular to the layer plane. 
Using the theory of invariants \cite{bir1974symmetry}, 
the most general $\boldsymbol{k} \cdot \boldsymbol{p}$
effective model can be expressed up to linear order in the crystal momentum

\begin{equation}
H_{C_{2}}=\alpha_{0}+\alpha_{1xx}k_{x}\sigma_{x}-\alpha_{1xy}k_{x}\sigma_{y}
+\alpha_{1yx}k_{y}\sigma_{x}-\alpha_{1yy}k_{y}\sigma_{y},
\label{kpmodel_1_v2}
\end{equation}
where $k_{i}$ is the crystal momentum in the $i$ direction measured from 
the corresponding high-symmetry point, $\sigma_{i}$ 
are the Pauli matrices; $\alpha_{0}$ and $\alpha_{1ij}$ correspond 
to the free parameters that describe the energy bands in the vicinity
of the chiral point.  
Note that this model allows for the presence of anisotropy, given that different
parameters are associated with the $x$ and $y$ momentum components. Besides, 
depending on the sign relation of these parameters, different spin-orbit 
interactions can be realized. Thus, a competition between spin-orbit effects
could be present, depending on the relative magnitudes 
and signs of the linear parameters. The competing interplay of both 
effects, anisotropy and spin-orbit, produces in some cases spin 
textures appreciably different from the ideal radial textures which
are characteristic of Kramers-Weyl nodes \cite{chang_topological_2018}. 
These different spin textures appear by choosing different values for 
the parameters $\alpha_{1ij}$ in Eq. (\ref{kpmodel_1_v2}). 

For instance, to realize the spin textures that are observed
in the pentagonal materials of SG \#4, an additional condition is
imposed such that $\alpha_{1xy} \ll  \alpha_{1xx},\alpha_{1yy}$ 
and $\alpha_{1yx} \ll  \alpha_{1xx},\alpha_{1yy}$.
Taking this into account, the Hamiltonian in 
Eq. (\ref{kpmodel_1_v2}) can be reduced to 

\begin{equation}
H_{SG\#4}=\alpha_{0}+\alpha_{1xx}k_{x}\sigma_{x}-\alpha_{1yy}k_{y}\sigma_{y},
\label{kpmodel_SG4}
\end{equation}
 
\begin{figure}[ht]
\centering
\includegraphics[width=1.0\columnwidth,clip]{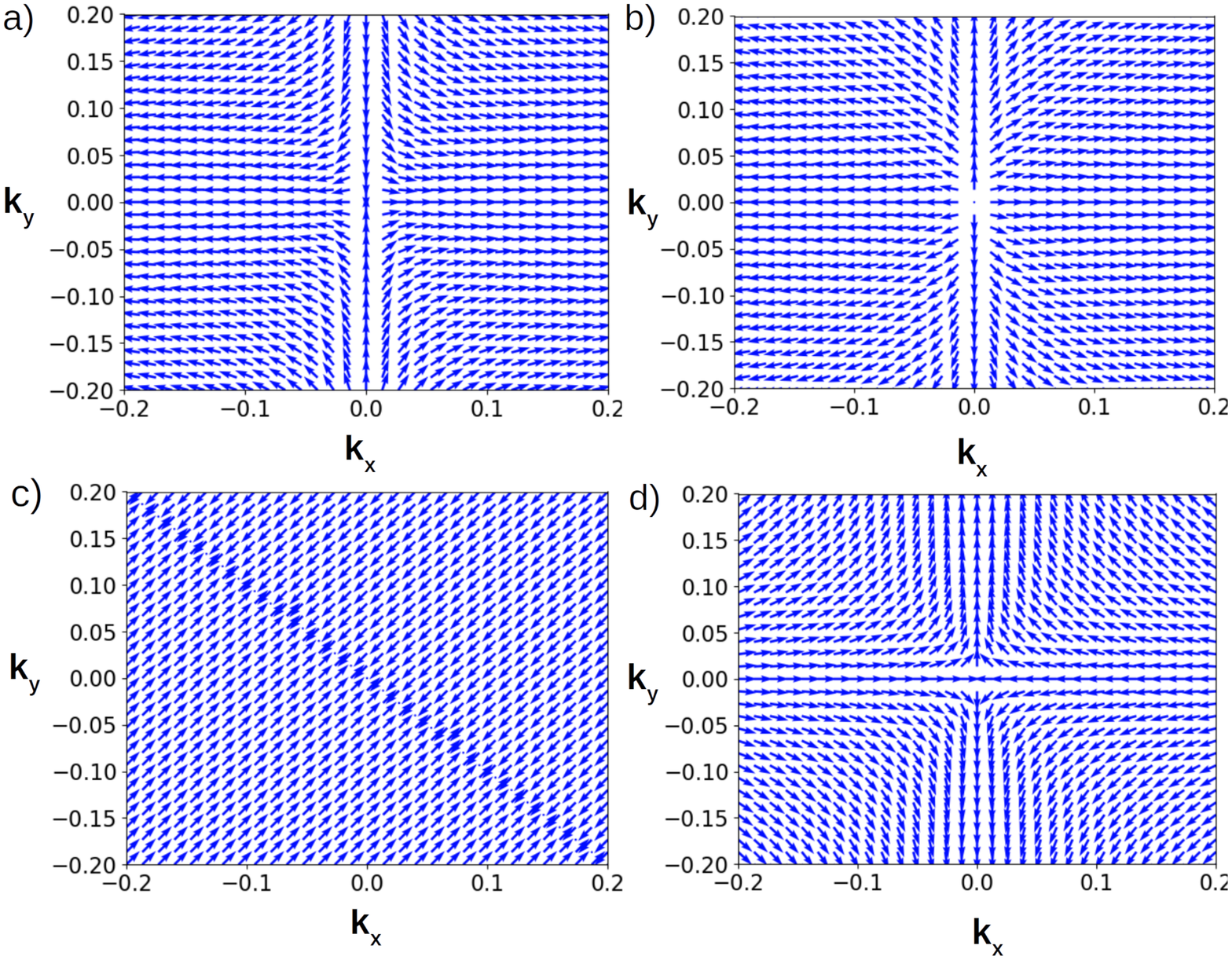}
 \caption{Spin texture from the $\boldsymbol{k} \cdot \boldsymbol{p}$ model for 
 a) SG \#4 with $\text{sign}(\alpha_{1xy}) = \text{sign}(\alpha_{1xx})$
 b) SG \#4 with $\text{sign}(\alpha_{1xy}) = -\text{sign}(\alpha_{1xx})$
 c) $X$ point in SG \#81 and d) $\Gamma$ point in SG \#113.}
 \label{spintext_kp} 
\end{figure}

For the sake of illustrating this point, Figs. \ref{spintext_kp} a) and b) depict these
cases for parameter 
values chosen without the aim of modeling a particular material. In Fig. \ref{spintext_kp} a)
a Dresselhaus-like texture is obtained by using the constraint  
$\text{sign}(\alpha_{1xy}) = \text{sign}(\alpha_{1xx})$. This texture can be appreciated in 
the uppermost valence bands of PdSeTe as shown in Fig. S12 of the 
Supplementary Information. 
In the case of Fig. \ref{spintext_kp} b) the condition 
$\text{sign}(\alpha_{1xy}) = -\text{sign}(\alpha_{1xx})$ gives a radial 
spin texture, very similar to the one presented above for PdSeTe in 
Fig. \ref{PdSeTe_QE} c). The competition and variability in 
spin textures within the same material generalizes the 
exclusive radial textures encountered in previous works and 
opens the way to explore the interplay of anisotropy and 
competing spin-orbit couplings in pentagonal systems. 

For SG \#81 at the X point, the model must include all $\alpha_{1ij}$
parameters, resulting in the interplay between all spin-orbit 
couplings, which yields diverse textures. In Fig.  \ref{spintext_kp} c)
a spin texture with a partial inward radial distribution is presented. This
is in good agreement with the spin texture near X, obtained for 
InP$_5$ in Fig. \ref{InP5_bands} c).

\textit{SG \#113}. For this SG the only nodal point that can be treated
with a two-band model is the $\Gamma$ point, which has an achiral little group isomorphic to $D_{2d}$. In this case, the effective model is isotropic, and up to linear order in crystal momentum is given by 

\begin{equation}
H_{SG\#113}=\beta_{0}+\beta_{1}(k_{x}\sigma_{x}-k_{y}\sigma_{y}),
\label{kpmodel_2_v2}
\end{equation}
where $\beta_{i}$ are free parameters. This model produces isotropic spin textures in the neighborhood of the $\Gamma$ point, as it can 
be observed in Fig. \ref{spintext_kp} d), where a Dresselhaus spin texture 
arises. This compares well with the results for GeBi$_2$ depicted in 
Fig. \ref{GeBi2_bands} c), which indicates the dominating spin-orbit
interaction is this material. 
The analysis of the fourfold fermion requires a four-band model along
with new symmetry considerations, which are outside 
the scope of this Appendix.

\section*{Acknowledgements}
This work has been partially supported by Chilean FONDECYT Grant 1201876, Spanish MCIU and AEI and the European Union under Grant No. PGC2018-097018-B-I00 (MCIU/AEI/FEDER, UE), and by Grant USM-DGIIP PI-LI 1925. V. N. acknowledges financial support from Beca Doctorado Nacional ANID No. 21170872 (CL). J. D. C. acknowledges
 Universidad Universidad de Medell\'{\i}n under grant 1120 and  the Laboratorio de Simulaci\'on y Computaci\'on Cient\'{\i}fica for computational facilities.

\bibliography{main}

\begin{thebibliography}{86}%
\makeatletter
\providecommand \@ifxundefined [1]{%
 \@ifx{#1\undefined}
}%
\providecommand \@ifnum [1]{%
 \ifnum #1\expandafter \@firstoftwo
 \else \expandafter \@secondoftwo
 \fi
}%
\providecommand \@ifx [1]{%
 \ifx #1\expandafter \@firstoftwo
 \else \expandafter \@secondoftwo
 \fi
}%
\providecommand \natexlab [1]{#1}%
\providecommand \enquote  [1]{``#1''}%
\providecommand \bibnamefont  [1]{#1}%
\providecommand \bibfnamefont [1]{#1}%
\providecommand \citenamefont [1]{#1}%
\providecommand \href@noop [0]{\@secondoftwo}%
\providecommand \href [0]{\begingroup \@sanitize@url \@href}%
\providecommand \@href[1]{\@@startlink{#1}\@@href}%
\providecommand \@@href[1]{\endgroup#1\@@endlink}%
\providecommand \@sanitize@url [0]{\catcode `\\12\catcode `\$12\catcode
  `\&12\catcode `\#12\catcode `\^12\catcode `\_12\catcode `\%12\relax}%
\providecommand \@@startlink[1]{}%
\providecommand \@@endlink[0]{}%
\providecommand \url  [0]{\begingroup\@sanitize@url \@url }%
\providecommand \@url [1]{\endgroup\@href {#1}{\urlprefix }}%
\providecommand \urlprefix  [0]{URL }%
\providecommand \Eprint [0]{\href }%
\providecommand \doibase [0]{https://doi.org/}%
\providecommand \selectlanguage [0]{\@gobble}%
\providecommand \bibinfo  [0]{\@secondoftwo}%
\providecommand \bibfield  [0]{\@secondoftwo}%
\providecommand \translation [1]{[#1]}%
\providecommand \BibitemOpen [0]{}%
\providecommand \bibitemStop [0]{}%
\providecommand \bibitemNoStop [0]{.\EOS\space}%
\providecommand \EOS [0]{\spacefactor3000\relax}%
\providecommand \BibitemShut  [1]{\csname bibitem#1\endcsname}%
\let\auto@bib@innerbib\@empty
\bibitem [{\citenamefont {Mir\'o}\ \emph {et~al.}(2014)\citenamefont {Mir\'o},
  \citenamefont {Audiffred},\ and\ \citenamefont {Heine}}]{Pere2014}%
  \BibitemOpen
  \bibfield  {author} {\bibinfo {author} {\bibfnamefont {P.}~\bibnamefont
  {Mir\'o}}, \bibinfo {author} {\bibfnamefont {M.}~\bibnamefont {Audiffred}},\
  and\ \bibinfo {author} {\bibfnamefont {T.}~\bibnamefont {Heine}},\ }\bibfield
   {title} {\bibinfo {title} {{An atlas of two-dimensional materials}},\ }\href
  {https://doi.org/10.1039/c4cs00102h} {\bibfield  {journal} {\bibinfo
  {journal} {Chem. Soc. Rev.}\ }\textbf {\bibinfo {volume} {43}},\ \bibinfo
  {pages} {6537} (\bibinfo {year} {2014})}\BibitemShut {NoStop}%
\bibitem [{\citenamefont {Novoselov}\ \emph {et~al.}(2016)\citenamefont
  {Novoselov}, \citenamefont {Mishchenko}, \citenamefont {Carvalho},\ and\
  \citenamefont {Castro~Neto}}]{Novoselov_2016}%
  \BibitemOpen
  \bibfield  {author} {\bibinfo {author} {\bibfnamefont {K.}~\bibnamefont
  {Novoselov}}, \bibinfo {author} {\bibfnamefont {A.}~\bibnamefont
  {Mishchenko}}, \bibinfo {author} {\bibfnamefont {A.}~\bibnamefont
  {Carvalho}},\ and\ \bibinfo {author} {\bibfnamefont {A.~H.}\ \bibnamefont
  {Castro~Neto}},\ }\bibfield  {title} {\bibinfo {title} {2d materials and van
  der waals heterostructures},\ }\bibfield  {journal} {\bibinfo  {journal}
  {Science}\ }\textbf {\bibinfo {volume} {353}},\ \href
  {https://doi.org/10.1126/science.aac9439} {10.1126/science.aac9439} (\bibinfo
  {year} {2016})\BibitemShut {NoStop}%
\bibitem [{\citenamefont {Saptarshi}\ \emph {et~al.}(2015)\citenamefont
  {Saptarshi}, \citenamefont {Robinson}, \citenamefont {Dubey}, \citenamefont
  {Terrones},\ and\ \citenamefont {Terrones}}]{Das_2015_ARCM}%
  \BibitemOpen
  \bibfield  {author} {\bibinfo {author} {\bibfnamefont {D.}~\bibnamefont
  {Saptarshi}}, \bibinfo {author} {\bibfnamefont {J.~A.}\ \bibnamefont
  {Robinson}}, \bibinfo {author} {\bibfnamefont {M.}~\bibnamefont {Dubey}},
  \bibinfo {author} {\bibfnamefont {H.}~\bibnamefont {Terrones}},\ and\
  \bibinfo {author} {\bibfnamefont {M.}~\bibnamefont {Terrones}},\ }\bibfield
  {title} {\bibinfo {title} {Beyond graphene: Progress in novel two-dimensional
  materials and van der waals solids},\ }\href
  {https://doi.org/10.1146/annurev-matsci-070214-021034} {\bibfield  {journal}
  {\bibinfo  {journal} {Annual Review of Materials Research}\ }\textbf
  {\bibinfo {volume} {45}},\ \bibinfo {pages} {1} (\bibinfo {year}
  {2015})}\BibitemShut {NoStop}%
\bibitem [{\citenamefont {Fiori}\ \emph {et~al.}(2014)\citenamefont {Fiori},
  \citenamefont {Bonaccorso}, \citenamefont {Iannaccone}, \citenamefont
  {Palacios}, \citenamefont {Neumaier}, \citenamefont {Seabaugh}, \citenamefont
  {Banerjee},\ and\ \citenamefont {Colombo}}]{Fiori_2014}%
  \BibitemOpen
  \bibfield  {author} {\bibinfo {author} {\bibfnamefont {G.}~\bibnamefont
  {Fiori}}, \bibinfo {author} {\bibfnamefont {F.}~\bibnamefont {Bonaccorso}},
  \bibinfo {author} {\bibfnamefont {G.}~\bibnamefont {Iannaccone}}, \bibinfo
  {author} {\bibfnamefont {T.}~\bibnamefont {Palacios}}, \bibinfo {author}
  {\bibfnamefont {D.}~\bibnamefont {Neumaier}}, \bibinfo {author}
  {\bibfnamefont {A.}~\bibnamefont {Seabaugh}}, \bibinfo {author}
  {\bibfnamefont {S.~K.}\ \bibnamefont {Banerjee}},\ and\ \bibinfo {author}
  {\bibfnamefont {L.}~\bibnamefont {Colombo}},\ }\bibfield  {title} {\bibinfo
  {title} {Electronics based on two-dimensional materials},\ }\href
  {https://doi.org/10.1038/nnano.2014.207} {\bibfield  {journal} {\bibinfo
  {journal} {Nature Nanotechnology}\ }\textbf {\bibinfo {volume} {9}},\
  \bibinfo {pages} {768} (\bibinfo {year} {2014})}\BibitemShut {NoStop}%
\bibitem [{\citenamefont {Wang}\ \emph {et~al.}(2012)\citenamefont {Wang},
  \citenamefont {Kalantar-Zadeh}, \citenamefont {Kis}, \citenamefont
  {Coleman},\ and\ \citenamefont {Strano}}]{Wang_2012}%
  \BibitemOpen
  \bibfield  {author} {\bibinfo {author} {\bibfnamefont {Q.~H.}\ \bibnamefont
  {Wang}}, \bibinfo {author} {\bibfnamefont {K.}~\bibnamefont
  {Kalantar-Zadeh}}, \bibinfo {author} {\bibfnamefont {A.}~\bibnamefont {Kis}},
  \bibinfo {author} {\bibfnamefont {J.~N.}\ \bibnamefont {Coleman}},\ and\
  \bibinfo {author} {\bibfnamefont {M.~S.}\ \bibnamefont {Strano}},\ }\bibfield
   {title} {\bibinfo {title} {Electronics and optoelectronics of
  two-dimensional transition metal dichalcogenides},\ }\href
  {https://doi.org/10.1038/nnano.2012.193} {\bibfield  {journal} {\bibinfo
  {journal} {Nature Nanotechnology}\ }\textbf {\bibinfo {volume} {7}},\
  \bibinfo {pages} {699} (\bibinfo {year} {2012})}\BibitemShut {NoStop}%
\bibitem [{\citenamefont {Jain}\ \emph {et~al.}(2013)\citenamefont {Jain},
  \citenamefont {Ong}, \citenamefont {Hautier}, \citenamefont {Chen},
  \citenamefont {Richards}, \citenamefont {Dacek}, \citenamefont {Cholia},
  \citenamefont {Gunter}, \citenamefont {Skinner}, \citenamefont {Ceder},\ and\
  \citenamefont {Persson}}]{jain_commentary_2013}%
  \BibitemOpen
  \bibfield  {author} {\bibinfo {author} {\bibfnamefont {A.}~\bibnamefont
  {Jain}}, \bibinfo {author} {\bibfnamefont {S.~P.}\ \bibnamefont {Ong}},
  \bibinfo {author} {\bibfnamefont {G.}~\bibnamefont {Hautier}}, \bibinfo
  {author} {\bibfnamefont {W.}~\bibnamefont {Chen}}, \bibinfo {author}
  {\bibfnamefont {W.~D.}\ \bibnamefont {Richards}}, \bibinfo {author}
  {\bibfnamefont {S.}~\bibnamefont {Dacek}}, \bibinfo {author} {\bibfnamefont
  {S.}~\bibnamefont {Cholia}}, \bibinfo {author} {\bibfnamefont
  {D.}~\bibnamefont {Gunter}}, \bibinfo {author} {\bibfnamefont
  {D.}~\bibnamefont {Skinner}}, \bibinfo {author} {\bibfnamefont
  {G.}~\bibnamefont {Ceder}},\ and\ \bibinfo {author} {\bibfnamefont {K.~A.}\
  \bibnamefont {Persson}},\ }\bibfield  {title} {\bibinfo {title} {Commentary:
  {The} {Materials} {Project}: {A} materials genome approach to accelerating
  materials innovation},\ }\href {https://doi.org/10.1063/1.4812323} {\bibfield
   {journal} {\bibinfo  {journal} {APL Materials}\ }\textbf {\bibinfo {volume}
  {1}},\ \bibinfo {pages} {011002} (\bibinfo {year} {2013})},\ \bibinfo {note}
  {publisher: American Institute of Physics}\BibitemShut {NoStop}%
\bibitem [{\citenamefont {Choudhary}\ \emph {et~al.}(2017)\citenamefont
  {Choudhary}, \citenamefont {Kalish}, \citenamefont {Beams},\ and\
  \citenamefont {Tavazza}}]{choudhary_high-throughput_2017}%
  \BibitemOpen
  \bibfield  {author} {\bibinfo {author} {\bibfnamefont {K.}~\bibnamefont
  {Choudhary}}, \bibinfo {author} {\bibfnamefont {I.}~\bibnamefont {Kalish}},
  \bibinfo {author} {\bibfnamefont {R.}~\bibnamefont {Beams}},\ and\ \bibinfo
  {author} {\bibfnamefont {F.}~\bibnamefont {Tavazza}},\ }\bibfield  {title}
  {\bibinfo {title} {High-throughput {Identification} and {Characterization} of
  {Two}-dimensional {Materials} using {Density} functional theory},\ }\href
  {https://doi.org/10.1038/s41598-017-05402-0} {\bibfield  {journal} {\bibinfo
  {journal} {Scientific Reports}\ }\textbf {\bibinfo {volume} {7}},\ \bibinfo
  {pages} {5179} (\bibinfo {year} {2017})},\ \bibinfo {note} {number: 1
  Publisher: Nature Publishing Group}\BibitemShut {NoStop}%
\bibitem [{\citenamefont {Haastrup}\ \emph {et~al.}(2018)\citenamefont
  {Haastrup}, \citenamefont {Strange}, \citenamefont {Pandey}, \citenamefont
  {Deilmann}, \citenamefont {Schmidt}, \citenamefont {Hinsche}, \citenamefont
  {Gjerding}, \citenamefont {Torelli}, \citenamefont {Larsen}, \citenamefont
  {Riis-Jensen}, \citenamefont {Gath}, \citenamefont {Jacobsen}, \citenamefont
  {Jørgen~Mortensen}, \citenamefont {Olsen},\ and\ \citenamefont
  {Thygesen}}]{haastrup_computational_2018}%
  \BibitemOpen
  \bibfield  {author} {\bibinfo {author} {\bibfnamefont {S.}~\bibnamefont
  {Haastrup}}, \bibinfo {author} {\bibfnamefont {M.}~\bibnamefont {Strange}},
  \bibinfo {author} {\bibfnamefont {M.}~\bibnamefont {Pandey}}, \bibinfo
  {author} {\bibfnamefont {T.}~\bibnamefont {Deilmann}}, \bibinfo {author}
  {\bibfnamefont {P.~S.}\ \bibnamefont {Schmidt}}, \bibinfo {author}
  {\bibfnamefont {N.~F.}\ \bibnamefont {Hinsche}}, \bibinfo {author}
  {\bibfnamefont {M.~N.}\ \bibnamefont {Gjerding}}, \bibinfo {author}
  {\bibfnamefont {D.}~\bibnamefont {Torelli}}, \bibinfo {author} {\bibfnamefont
  {P.~M.}\ \bibnamefont {Larsen}}, \bibinfo {author} {\bibfnamefont {A.~C.}\
  \bibnamefont {Riis-Jensen}}, \bibinfo {author} {\bibfnamefont
  {J.}~\bibnamefont {Gath}}, \bibinfo {author} {\bibfnamefont {K.~W.}\
  \bibnamefont {Jacobsen}}, \bibinfo {author} {\bibfnamefont {J.}~\bibnamefont
  {Jørgen~Mortensen}}, \bibinfo {author} {\bibfnamefont {T.}~\bibnamefont
  {Olsen}},\ and\ \bibinfo {author} {\bibfnamefont {K.~S.}\ \bibnamefont
  {Thygesen}},\ }\bibfield  {title} {\bibinfo {title} {The {Computational} {2D}
  {Materials} {Database}: high-throughput modeling and discovery of atomically
  thin crystals},\ }\href {https://doi.org/10.1088/2053-1583/aacfc1} {\bibfield
   {journal} {\bibinfo  {journal} {2D Mater.}\ }\textbf {\bibinfo {volume}
  {5}},\ \bibinfo {pages} {042002} (\bibinfo {year} {2018})}\BibitemShut
  {NoStop}%
\bibitem [{\citenamefont {Mounet}\ \emph {et~al.}(2018)\citenamefont {Mounet},
  \citenamefont {Gibertini}, \citenamefont {Schwaller}, \citenamefont {Campi},
  \citenamefont {Merkys}, \citenamefont {Marrazzo}, \citenamefont {Sohier},
  \citenamefont {Castelli}, \citenamefont {Cepellotti}, \citenamefont {Pizzi},\
  and\ \citenamefont {Marzari}}]{mounet_two-dimensional_2018}%
  \BibitemOpen
  \bibfield  {author} {\bibinfo {author} {\bibfnamefont {N.}~\bibnamefont
  {Mounet}}, \bibinfo {author} {\bibfnamefont {M.}~\bibnamefont {Gibertini}},
  \bibinfo {author} {\bibfnamefont {P.}~\bibnamefont {Schwaller}}, \bibinfo
  {author} {\bibfnamefont {D.}~\bibnamefont {Campi}}, \bibinfo {author}
  {\bibfnamefont {A.}~\bibnamefont {Merkys}}, \bibinfo {author} {\bibfnamefont
  {A.}~\bibnamefont {Marrazzo}}, \bibinfo {author} {\bibfnamefont
  {T.}~\bibnamefont {Sohier}}, \bibinfo {author} {\bibfnamefont {I.~E.}\
  \bibnamefont {Castelli}}, \bibinfo {author} {\bibfnamefont {A.}~\bibnamefont
  {Cepellotti}}, \bibinfo {author} {\bibfnamefont {G.}~\bibnamefont {Pizzi}},\
  and\ \bibinfo {author} {\bibfnamefont {N.}~\bibnamefont {Marzari}},\
  }\bibfield  {title} {\bibinfo {title} {Two-dimensional materials from
  high-throughput computational exfoliation of experimentally known
  compounds},\ }\href {https://doi.org/10.1038/s41565-017-0035-5} {\bibfield
  {journal} {\bibinfo  {journal} {Nature Nanotech}\ }\textbf {\bibinfo {volume}
  {13}},\ \bibinfo {pages} {246} (\bibinfo {year} {2018})},\ \bibinfo {note}
  {arXiv: 1611.05234}\BibitemShut {NoStop}%
\bibitem [{\citenamefont {Choudhary}\ \emph {et~al.}(2020)\citenamefont
  {Choudhary}, \citenamefont {Garrity}, \citenamefont {Reid}, \citenamefont
  {DeCost}, \citenamefont {Biacchi}, \citenamefont {Walker}, \citenamefont
  {Trautt}, \citenamefont {Hattrick-Simpers}, \citenamefont {Kusne},
  \citenamefont {Centrone}, \citenamefont {Davydov}, \citenamefont {Jiang},
  \citenamefont {Pachter}, \citenamefont {Cheon}, \citenamefont {Reed},
  \citenamefont {Agrawal}, \citenamefont {Qian}, \citenamefont {Sharma},
  \citenamefont {Zhuang}, \citenamefont {Kalinin}, \citenamefont {Sumpter},
  \citenamefont {Pilania}, \citenamefont {Acar}, \citenamefont {Mandal},
  \citenamefont {Haule}, \citenamefont {Vanderbilt}, \citenamefont {Rabe},\
  and\ \citenamefont {Tavazza}}]{choudhary_jarvis_2020}%
  \BibitemOpen
  \bibfield  {author} {\bibinfo {author} {\bibfnamefont {K.}~\bibnamefont
  {Choudhary}}, \bibinfo {author} {\bibfnamefont {K.~F.}\ \bibnamefont
  {Garrity}}, \bibinfo {author} {\bibfnamefont {A.~C.~E.}\ \bibnamefont
  {Reid}}, \bibinfo {author} {\bibfnamefont {B.}~\bibnamefont {DeCost}},
  \bibinfo {author} {\bibfnamefont {A.~J.}\ \bibnamefont {Biacchi}}, \bibinfo
  {author} {\bibfnamefont {A.~R.~H.}\ \bibnamefont {Walker}}, \bibinfo {author}
  {\bibfnamefont {Z.}~\bibnamefont {Trautt}}, \bibinfo {author} {\bibfnamefont
  {J.}~\bibnamefont {Hattrick-Simpers}}, \bibinfo {author} {\bibfnamefont
  {A.~G.}\ \bibnamefont {Kusne}}, \bibinfo {author} {\bibfnamefont
  {A.}~\bibnamefont {Centrone}}, \bibinfo {author} {\bibfnamefont
  {A.}~\bibnamefont {Davydov}}, \bibinfo {author} {\bibfnamefont
  {J.}~\bibnamefont {Jiang}}, \bibinfo {author} {\bibfnamefont
  {R.}~\bibnamefont {Pachter}}, \bibinfo {author} {\bibfnamefont
  {G.}~\bibnamefont {Cheon}}, \bibinfo {author} {\bibfnamefont
  {E.}~\bibnamefont {Reed}}, \bibinfo {author} {\bibfnamefont {A.}~\bibnamefont
  {Agrawal}}, \bibinfo {author} {\bibfnamefont {X.}~\bibnamefont {Qian}},
  \bibinfo {author} {\bibfnamefont {V.}~\bibnamefont {Sharma}}, \bibinfo
  {author} {\bibfnamefont {H.}~\bibnamefont {Zhuang}}, \bibinfo {author}
  {\bibfnamefont {S.~V.}\ \bibnamefont {Kalinin}}, \bibinfo {author}
  {\bibfnamefont {B.~G.}\ \bibnamefont {Sumpter}}, \bibinfo {author}
  {\bibfnamefont {G.}~\bibnamefont {Pilania}}, \bibinfo {author} {\bibfnamefont
  {P.}~\bibnamefont {Acar}}, \bibinfo {author} {\bibfnamefont {S.}~\bibnamefont
  {Mandal}}, \bibinfo {author} {\bibfnamefont {K.}~\bibnamefont {Haule}},
  \bibinfo {author} {\bibfnamefont {D.}~\bibnamefont {Vanderbilt}}, \bibinfo
  {author} {\bibfnamefont {K.}~\bibnamefont {Rabe}},\ and\ \bibinfo {author}
  {\bibfnamefont {F.}~\bibnamefont {Tavazza}},\ }\bibfield  {title} {\bibinfo
  {title} {{JARVIS}: {An} {Integrated} {Infrastructure} for {Data}-driven
  {Materials} {Design}},\ }\href {http://arxiv.org/abs/2007.01831} {\bibfield
  {journal} {\bibinfo  {journal} {arXiv:2007.01831 [cond-mat,
  physics:physics]}\ } (\bibinfo {year} {2020})},\ \bibinfo {note} {arXiv:
  2007.01831}\BibitemShut {NoStop}%
\bibitem [{\citenamefont {Talirz}\ \emph {et~al.}(2020)\citenamefont {Talirz},
  \citenamefont {Kumbhar}, \citenamefont {Passaro}, \citenamefont {Yakutovich},
  \citenamefont {Granata}, \citenamefont {Gargiulo}, \citenamefont {Borelli},
  \citenamefont {Uhrin}, \citenamefont {Huber}, \citenamefont {Zoupanos},
  \citenamefont {Adorf}, \citenamefont {Andersen}, \citenamefont {Schütt},
  \citenamefont {Pignedoli}, \citenamefont {Passerone}, \citenamefont
  {VandeVondele}, \citenamefont {Schulthess}, \citenamefont {Smit},
  \citenamefont {Pizzi},\ and\ \citenamefont
  {Marzari}}]{talirz_materials_2020}%
  \BibitemOpen
  \bibfield  {author} {\bibinfo {author} {\bibfnamefont {L.}~\bibnamefont
  {Talirz}}, \bibinfo {author} {\bibfnamefont {S.}~\bibnamefont {Kumbhar}},
  \bibinfo {author} {\bibfnamefont {E.}~\bibnamefont {Passaro}}, \bibinfo
  {author} {\bibfnamefont {A.~V.}\ \bibnamefont {Yakutovich}}, \bibinfo
  {author} {\bibfnamefont {V.}~\bibnamefont {Granata}}, \bibinfo {author}
  {\bibfnamefont {F.}~\bibnamefont {Gargiulo}}, \bibinfo {author}
  {\bibfnamefont {M.}~\bibnamefont {Borelli}}, \bibinfo {author} {\bibfnamefont
  {M.}~\bibnamefont {Uhrin}}, \bibinfo {author} {\bibfnamefont {S.~P.}\
  \bibnamefont {Huber}}, \bibinfo {author} {\bibfnamefont {S.}~\bibnamefont
  {Zoupanos}}, \bibinfo {author} {\bibfnamefont {C.~S.}\ \bibnamefont {Adorf}},
  \bibinfo {author} {\bibfnamefont {C.~W.}\ \bibnamefont {Andersen}}, \bibinfo
  {author} {\bibfnamefont {O.}~\bibnamefont {Schütt}}, \bibinfo {author}
  {\bibfnamefont {C.~A.}\ \bibnamefont {Pignedoli}}, \bibinfo {author}
  {\bibfnamefont {D.}~\bibnamefont {Passerone}}, \bibinfo {author}
  {\bibfnamefont {J.}~\bibnamefont {VandeVondele}}, \bibinfo {author}
  {\bibfnamefont {T.~C.}\ \bibnamefont {Schulthess}}, \bibinfo {author}
  {\bibfnamefont {B.}~\bibnamefont {Smit}}, \bibinfo {author} {\bibfnamefont
  {G.}~\bibnamefont {Pizzi}},\ and\ \bibinfo {author} {\bibfnamefont
  {N.}~\bibnamefont {Marzari}},\ }\bibfield  {title} {\bibinfo {title}
  {Materials {Cloud}, a platform for open computational science},\ }\href
  {https://doi.org/10.1038/s41597-020-00637-5} {\bibfield  {journal} {\bibinfo
  {journal} {Scientific Data}\ }\textbf {\bibinfo {volume} {7}},\ \bibinfo
  {pages} {299} (\bibinfo {year} {2020})},\ \bibinfo {note} {number: 1
  Publisher: Nature Publishing Group}\BibitemShut {NoStop}%
\bibitem [{\citenamefont {Oyedele}\ \emph {et~al.}(2017)\citenamefont
  {Oyedele}, \citenamefont {Yang}, \citenamefont {Liang}, \citenamefont
  {Puretzky}, \citenamefont {Wang}, \citenamefont {Zhang}, \citenamefont {Yu},
  \citenamefont {Pudasaini}, \citenamefont {Ghosh}, \citenamefont {Liu},
  \citenamefont {Rouleau}, \citenamefont {Sumpter}, \citenamefont {Chisholm},
  \citenamefont {Zhou}, \citenamefont {Rack}, \citenamefont {Geohegan},\ and\
  \citenamefont {Xiao}}]{oyedele_pdse_2017}%
  \BibitemOpen
  \bibfield  {author} {\bibinfo {author} {\bibfnamefont {A.~D.}\ \bibnamefont
  {Oyedele}}, \bibinfo {author} {\bibfnamefont {S.}~\bibnamefont {Yang}},
  \bibinfo {author} {\bibfnamefont {L.}~\bibnamefont {Liang}}, \bibinfo
  {author} {\bibfnamefont {A.~A.}\ \bibnamefont {Puretzky}}, \bibinfo {author}
  {\bibfnamefont {K.}~\bibnamefont {Wang}}, \bibinfo {author} {\bibfnamefont
  {J.}~\bibnamefont {Zhang}}, \bibinfo {author} {\bibfnamefont
  {P.}~\bibnamefont {Yu}}, \bibinfo {author} {\bibfnamefont {P.~R.}\
  \bibnamefont {Pudasaini}}, \bibinfo {author} {\bibfnamefont {A.~W.}\
  \bibnamefont {Ghosh}}, \bibinfo {author} {\bibfnamefont {Z.}~\bibnamefont
  {Liu}}, \bibinfo {author} {\bibfnamefont {C.~M.}\ \bibnamefont {Rouleau}},
  \bibinfo {author} {\bibfnamefont {B.~G.}\ \bibnamefont {Sumpter}}, \bibinfo
  {author} {\bibfnamefont {M.~F.}\ \bibnamefont {Chisholm}}, \bibinfo {author}
  {\bibfnamefont {W.}~\bibnamefont {Zhou}}, \bibinfo {author} {\bibfnamefont
  {P.~D.}\ \bibnamefont {Rack}}, \bibinfo {author} {\bibfnamefont {D.~B.}\
  \bibnamefont {Geohegan}},\ and\ \bibinfo {author} {\bibfnamefont
  {K.}~\bibnamefont {Xiao}},\ }\bibfield  {title} {\bibinfo {title} {{PdSe}
  $_{\textrm{2}}$ : {Pentagonal} {Two}-{Dimensional} {Layers} with {High} {Air}
  {Stability} for {Electronics}},\ }\href
  {https://doi.org/10.1021/jacs.7b04865} {\bibfield  {journal} {\bibinfo
  {journal} {J. Am. Chem. Soc.}\ }\textbf {\bibinfo {volume} {139}},\ \bibinfo
  {pages} {14090} (\bibinfo {year} {2017})}\BibitemShut {NoStop}%
\bibitem [{\citenamefont {Tang}\ \emph {et~al.}(2014)\citenamefont {Tang},
  \citenamefont {Xiong}, \citenamefont {Shi},\ and\ \citenamefont
  {Cao}}]{2014_tang}%
  \BibitemOpen
  \bibfield  {author} {\bibinfo {author} {\bibfnamefont {C.-P.}\ \bibnamefont
  {Tang}}, \bibinfo {author} {\bibfnamefont {S.-J.}\ \bibnamefont {Xiong}},
  \bibinfo {author} {\bibfnamefont {W.-J.}\ \bibnamefont {Shi}},\ and\ \bibinfo
  {author} {\bibfnamefont {J.}~\bibnamefont {Cao}},\ }\bibfield  {title}
  {\bibinfo {title} {Two-dimensional pentagonal crystals and possible
  spin-polarized dirac dispersion relations},\ }\href
  {https://doi.org/10.1063/1.4868679} {\bibfield  {journal} {\bibinfo
  {journal} {Journal of Applied Physics}\ }\textbf {\bibinfo {volume} {115}},\
  \bibinfo {pages} {113702} (\bibinfo {year} {2014})},\ \Eprint
  {https://arxiv.org/abs/https://doi.org/10.1063/1.4868679}
  {https://doi.org/10.1063/1.4868679} \BibitemShut {NoStop}%
\bibitem [{\citenamefont {Zhang}\ \emph {et~al.}(2015)\citenamefont {Zhang},
  \citenamefont {Zhou}, \citenamefont {Wang}, \citenamefont {Chen},
  \citenamefont {Kawazoe},\ and\ \citenamefont
  {Jena}}]{2015_Zhang_pentagraphene}%
  \BibitemOpen
  \bibfield  {author} {\bibinfo {author} {\bibfnamefont {S.}~\bibnamefont
  {Zhang}}, \bibinfo {author} {\bibfnamefont {J.}~\bibnamefont {Zhou}},
  \bibinfo {author} {\bibfnamefont {Q.}~\bibnamefont {Wang}}, \bibinfo {author}
  {\bibfnamefont {X.}~\bibnamefont {Chen}}, \bibinfo {author} {\bibfnamefont
  {Y.}~\bibnamefont {Kawazoe}},\ and\ \bibinfo {author} {\bibfnamefont
  {P.}~\bibnamefont {Jena}},\ }\bibfield  {title} {\bibinfo {title}
  {Penta-graphene: A new carbon allotrope},\ }\href
  {https://doi.org/10.1073/pnas.1416591112} {\bibfield  {journal} {\bibinfo
  {journal} {Proceedings of the National Academy of Sciences}\ }\textbf
  {\bibinfo {volume} {112}},\ \bibinfo {pages} {2372} (\bibinfo {year}
  {2015})},\ \Eprint
  {https://arxiv.org/abs/https://www.pnas.org/content/112/8/2372.full.pdf}
  {https://www.pnas.org/content/112/8/2372.full.pdf} \BibitemShut {NoStop}%
\bibitem [{\citenamefont {Zhao}\ \emph {et~al.}(2016)\citenamefont {Zhao},
  \citenamefont {Zhang}, \citenamefont {Guo},\ and\ \citenamefont
  {Wang}}]{2016_TiC2}%
  \BibitemOpen
  \bibfield  {author} {\bibinfo {author} {\bibfnamefont {T.}~\bibnamefont
  {Zhao}}, \bibinfo {author} {\bibfnamefont {S.}~\bibnamefont {Zhang}},
  \bibinfo {author} {\bibfnamefont {Y.}~\bibnamefont {Guo}},\ and\ \bibinfo
  {author} {\bibfnamefont {Q.}~\bibnamefont {Wang}},\ }\bibfield  {title}
  {\bibinfo {title} {Tic2: a new two-dimensional sheet beyond mxenes},\ }\href
  {https://doi.org/10.1039/C5NR04472C} {\bibfield  {journal} {\bibinfo
  {journal} {Nanoscale}\ }\textbf {\bibinfo {volume} {8}},\ \bibinfo {pages}
  {233} (\bibinfo {year} {2016})}\BibitemShut {NoStop}%
\bibitem [{\citenamefont {Ma}\ \emph {et~al.}(2016)\citenamefont {Ma},
  \citenamefont {Kou}, \citenamefont {Li}, \citenamefont {Dai},\ and\
  \citenamefont {Heine}}]{ma_room_2016}%
  \BibitemOpen
  \bibfield  {author} {\bibinfo {author} {\bibfnamefont {Y.}~\bibnamefont
  {Ma}}, \bibinfo {author} {\bibfnamefont {L.}~\bibnamefont {Kou}}, \bibinfo
  {author} {\bibfnamefont {X.}~\bibnamefont {Li}}, \bibinfo {author}
  {\bibfnamefont {Y.}~\bibnamefont {Dai}},\ and\ \bibinfo {author}
  {\bibfnamefont {T.}~\bibnamefont {Heine}},\ }\bibfield  {title} {\bibinfo
  {title} {Room temperature quantum spin {Hall} states in two-dimensional
  crystals composed of pentagonal rings and their quantum wells},\ }\bibfield
  {journal} {\bibinfo  {journal} {NPG Asia Materials}\ }\textbf {\bibinfo
  {volume} {8}},\ \href {https://doi.org/10.1038/am.2016.51}
  {10.1038/am.2016.51} (\bibinfo {year} {2016})\BibitemShut {NoStop}%
\bibitem [{\citenamefont {Li}\ \emph {et~al.}(2016)\citenamefont {Li},
  \citenamefont {Fan}, \citenamefont {Wei}, \citenamefont {Liu}, \citenamefont
  {Li}, \citenamefont {Zhao},\ and\ \citenamefont
  {Chen}}]{li_half-metallicity_2016}%
  \BibitemOpen
  \bibfield  {author} {\bibinfo {author} {\bibfnamefont {J.}~\bibnamefont
  {Li}}, \bibinfo {author} {\bibfnamefont {X.}~\bibnamefont {Fan}}, \bibinfo
  {author} {\bibfnamefont {Y.}~\bibnamefont {Wei}}, \bibinfo {author}
  {\bibfnamefont {H.}~\bibnamefont {Liu}}, \bibinfo {author} {\bibfnamefont
  {S.}~\bibnamefont {Li}}, \bibinfo {author} {\bibfnamefont {P.}~\bibnamefont
  {Zhao}},\ and\ \bibinfo {author} {\bibfnamefont {G.}~\bibnamefont {Chen}},\
  }\bibfield  {title} {\bibinfo {title} {Half-metallicity and ferromagnetism in
  penta -{AlN} 2 nanostructure},\ }\bibfield  {journal} {\bibinfo  {journal}
  {Scientific Reports}\ }\textbf {\bibinfo {volume} {6}},\ \href
  {https://doi.org/10.1038/srep33060} {10.1038/srep33060} (\bibinfo {year}
  {2016})\BibitemShut {NoStop}%
\bibitem [{\citenamefont {Zhao}\ \emph {et~al.}(2017)\citenamefont {Zhao},
  \citenamefont {Wang}, \citenamefont {Chen}, \citenamefont {Liu},
  \citenamefont {Yu}, \citenamefont {Zhang}, \citenamefont {Chen},\ and\
  \citenamefont {Wang}}]{zhao_elastic_2017}%
  \BibitemOpen
  \bibfield  {author} {\bibinfo {author} {\bibfnamefont {L.-S.}\ \bibnamefont
  {Zhao}}, \bibinfo {author} {\bibfnamefont {Y.}~\bibnamefont {Wang}}, \bibinfo
  {author} {\bibfnamefont {C.-P.}\ \bibnamefont {Chen}}, \bibinfo {author}
  {\bibfnamefont {L.-L.}\ \bibnamefont {Liu}}, \bibinfo {author} {\bibfnamefont
  {H.-X.}\ \bibnamefont {Yu}}, \bibinfo {author} {\bibfnamefont
  {Y.}~\bibnamefont {Zhang}}, \bibinfo {author} {\bibfnamefont
  {Y.}~\bibnamefont {Chen}},\ and\ \bibinfo {author} {\bibfnamefont {X.-C.}\
  \bibnamefont {Wang}},\ }\bibfield  {title} {\bibinfo {title} {Elastic,
  electronic and optical properties of stable pentagonal {ZnO} 2},\ }\href
  {https://doi.org/10.1016/j.physe.2017.03.017} {\bibfield  {journal} {\bibinfo
   {journal} {Physica E: Low-dimensional Systems and Nanostructures}\ }\textbf
  {\bibinfo {volume} {91}},\ \bibinfo {pages} {82} (\bibinfo {year}
  {2017})}\BibitemShut {NoStop}%
\bibitem [{\citenamefont {Pang}\ \emph {et~al.}(2017)\citenamefont {Pang},
  \citenamefont {Liu},\ and\ \citenamefont {Zhao}}]{pang_mechanical_2017}%
  \BibitemOpen
  \bibfield  {author} {\bibinfo {author} {\bibfnamefont {Q.}~\bibnamefont
  {Pang}}, \bibinfo {author} {\bibfnamefont {L.}~\bibnamefont {Liu}},\ and\
  \bibinfo {author} {\bibfnamefont {J.}~\bibnamefont {Zhao}},\ }\bibfield
  {title} {\bibinfo {title} {Mechanical anisotropy and strain-tailored band
  structures of pentagonal boron nitride monolayers},\ }\href
  {https://doi.org/10.1063/1.5000700} {\bibfield  {journal} {\bibinfo
  {journal} {Journal of Applied Physics}\ }\textbf {\bibinfo {volume} {122}},\
  \bibinfo {pages} {094302} (\bibinfo {year} {2017})}\BibitemShut {NoStop}%
\bibitem [{\citenamefont
  {Naseri}(2018{\natexlab{a}})}]{naseri_investigation_2018}%
  \BibitemOpen
  \bibfield  {author} {\bibinfo {author} {\bibfnamefont {M.}~\bibnamefont
  {Naseri}},\ }\bibfield  {title} {\bibinfo {title} {Investigation on the
  stability and electronic properties of {Penta}-{XP5} ({X} = {Al}, {Ga}, {In})
  monolayer semiconductors by using first principles calculations},\ }\href
  {https://doi.org/10.1016/j.cplett.2018.05.067} {\bibfield  {journal}
  {\bibinfo  {journal} {Chemical Physics Letters}\ }\textbf {\bibinfo {volume}
  {706}},\ \bibinfo {pages} {99} (\bibinfo {year}
  {2018}{\natexlab{a}})}\BibitemShut {NoStop}%
\bibitem [{\citenamefont {Naseri}\ \emph {et~al.}(2018)\citenamefont {Naseri},
  \citenamefont {Lin}, \citenamefont {Jalilian}, \citenamefont {Gu},\ and\
  \citenamefont {Chen}}]{naseri_penta-p2x_2018}%
  \BibitemOpen
  \bibfield  {author} {\bibinfo {author} {\bibfnamefont {M.}~\bibnamefont
  {Naseri}}, \bibinfo {author} {\bibfnamefont {S.}~\bibnamefont {Lin}},
  \bibinfo {author} {\bibfnamefont {J.}~\bibnamefont {Jalilian}}, \bibinfo
  {author} {\bibfnamefont {J.}~\bibnamefont {Gu}},\ and\ \bibinfo {author}
  {\bibfnamefont {Z.}~\bibnamefont {Chen}},\ }\bibfield  {title} {\bibinfo
  {title} {Penta-{P2X} ({X}={C}, {Si}) monolayers as wide-bandgap
  semiconductors: {A} first principles prediction},\ }\href
  {https://doi.org/10.1007/s11467-018-0758-2} {\bibfield  {journal} {\bibinfo
  {journal} {Front. Phys.}\ }\textbf {\bibinfo {volume} {13}},\ \bibinfo
  {pages} {138102} (\bibinfo {year} {2018})}\BibitemShut {NoStop}%
\bibitem [{\citenamefont {Naseri}(2018{\natexlab{b}})}]{naseri_penta-sic_2018}%
  \BibitemOpen
  \bibfield  {author} {\bibinfo {author} {\bibfnamefont {M.}~\bibnamefont
  {Naseri}},\ }\bibfield  {title} {\bibinfo {title} {Penta-{SiC} 5 monolayer:
  {A} novel quasi-planar indirect semiconductor with a tunable wide band gap},\
  }\href {https://doi.org/10.1016/j.physleta.2018.01.010} {\bibfield  {journal}
  {\bibinfo  {journal} {Physics Letters A}\ }\textbf {\bibinfo {volume}
  {382}},\ \bibinfo {pages} {710} (\bibinfo {year}
  {2018}{\natexlab{b}})}\BibitemShut {NoStop}%
\bibitem [{\citenamefont {Liu}\ \emph {et~al.}(2018)\citenamefont {Liu},
  \citenamefont {Kankam},\ and\ \citenamefont
  {Zhuang}}]{liu_single-layer_2018}%
  \BibitemOpen
  \bibfield  {author} {\bibinfo {author} {\bibfnamefont {L.}~\bibnamefont
  {Liu}}, \bibinfo {author} {\bibfnamefont {I.}~\bibnamefont {Kankam}},\ and\
  \bibinfo {author} {\bibfnamefont {H.~L.}\ \bibnamefont {Zhuang}},\ }\bibfield
   {title} {\bibinfo {title} {Single-layer antiferromagnetic semiconductor
  {CoS} 2 with pentagonal structure},\ }\href
  {https://doi.org/10.1103/PhysRevB.98.205425} {\bibfield  {journal} {\bibinfo
  {journal} {Phys. Rev. B}\ }\textbf {\bibinfo {volume} {98}},\ \bibinfo
  {pages} {205425} (\bibinfo {year} {2018})}\BibitemShut {NoStop}%
\bibitem [{\citenamefont {Zhao}\ \emph {et~al.}(2020)\citenamefont {Zhao},
  \citenamefont {Guo}, \citenamefont {Shen}, \citenamefont {Wang},
  \citenamefont {Kawazoe},\ and\ \citenamefont {Jena}}]{zhao_penta-bcn_2020}%
  \BibitemOpen
  \bibfield  {author} {\bibinfo {author} {\bibfnamefont {K.}~\bibnamefont
  {Zhao}}, \bibinfo {author} {\bibfnamefont {Y.}~\bibnamefont {Guo}}, \bibinfo
  {author} {\bibfnamefont {Y.}~\bibnamefont {Shen}}, \bibinfo {author}
  {\bibfnamefont {Q.}~\bibnamefont {Wang}}, \bibinfo {author} {\bibfnamefont
  {Y.}~\bibnamefont {Kawazoe}},\ and\ \bibinfo {author} {\bibfnamefont
  {P.}~\bibnamefont {Jena}},\ }\bibfield  {title} {\bibinfo {title}
  {Penta-{BCN}: {A} {New} {Ternary} {Pentagonal} {Monolayer} with {Intrinsic}
  {Piezoelectricity}},\ }\href {https://doi.org/10.1021/acs.jpclett.0c00824}
  {\bibfield  {journal} {\bibinfo  {journal} {J. Phys. Chem. Lett.}\ }\textbf
  {\bibinfo {volume} {11}},\ \bibinfo {pages} {3501} (\bibinfo {year}
  {2020})}\BibitemShut {NoStop}%
\bibitem [{\citenamefont {Liu}\ \emph {et~al.}(2016)\citenamefont {Liu},
  \citenamefont {Qin}, \citenamefont {Lin},\ and\ \citenamefont
  {Hu}}]{2016Liu_thermoPG}%
  \BibitemOpen
  \bibfield  {author} {\bibinfo {author} {\bibfnamefont {H.}~\bibnamefont
  {Liu}}, \bibinfo {author} {\bibfnamefont {G.}~\bibnamefont {Qin}}, \bibinfo
  {author} {\bibfnamefont {Y.}~\bibnamefont {Lin}},\ and\ \bibinfo {author}
  {\bibfnamefont {M.}~\bibnamefont {Hu}},\ }\bibfield  {title} {\bibinfo
  {title} {Disparate strain dependent thermal conductivity of two-dimensional
  penta-structures},\ }\href {https://doi.org/10.1021/acs.nanolett.6b01311}
  {\bibfield  {journal} {\bibinfo  {journal} {Nano Letters}\ }\textbf {\bibinfo
  {volume} {16}},\ \bibinfo {pages} {3831} (\bibinfo {year} {2016})},\ \bibinfo
  {note} {pMID: 27228130},\ \Eprint
  {https://arxiv.org/abs/https://doi.org/10.1021/acs.nanolett.6b01311}
  {https://doi.org/10.1021/acs.nanolett.6b01311} \BibitemShut {NoStop}%
\bibitem [{\citenamefont {Xiao}\ \emph {et~al.}(2016)\citenamefont {Xiao},
  \citenamefont {Li}, \citenamefont {Yu},\ and\ \citenamefont
  {Cheng}}]{2016_anodePG}%
  \BibitemOpen
  \bibfield  {author} {\bibinfo {author} {\bibfnamefont {B.}~\bibnamefont
  {Xiao}}, \bibinfo {author} {\bibfnamefont {Y.-c.}\ \bibnamefont {Li}},
  \bibinfo {author} {\bibfnamefont {X.-f.}\ \bibnamefont {Yu}},\ and\ \bibinfo
  {author} {\bibfnamefont {J.-b.}\ \bibnamefont {Cheng}},\ }\bibfield  {title}
  {\bibinfo {title} {Penta-graphene: A promising anode material as the
  li/na-ion battery with both extremely high theoretical capacity and fast
  charge/discharge rate},\ }\href {https://doi.org/10.1021/acsami.6b12727}
  {\bibfield  {journal} {\bibinfo  {journal} {ACS Applied Materials \&
  Interfaces}\ }\textbf {\bibinfo {volume} {8}},\ \bibinfo {pages} {35342}
  (\bibinfo {year} {2016})},\ \bibinfo {note} {pMID: 27977126},\ \Eprint
  {https://arxiv.org/abs/https://doi.org/10.1021/acsami.6b12727}
  {https://doi.org/10.1021/acsami.6b12727} \BibitemShut {NoStop}%
\bibitem [{\citenamefont {Winczewski}\ and\ \citenamefont
  {Rybicki}(2019)}]{PG_auxetic_2019}%
  \BibitemOpen
  \bibfield  {author} {\bibinfo {author} {\bibfnamefont {S.}~\bibnamefont
  {Winczewski}}\ and\ \bibinfo {author} {\bibfnamefont {J.}~\bibnamefont
  {Rybicki}},\ }\bibfield  {title} {\bibinfo {title} {Anisotropic mechanical
  behavior and auxeticity of penta-graphene: {Molecular} statics/molecular
  dynamics studies},\ }\href
  {https://doi.org/https://doi.org/10.1016/j.carbon.2019.02.042} {\bibfield
  {journal} {\bibinfo  {journal} {Carbon}\ }\textbf {\bibinfo {volume} {146}},\
  \bibinfo {pages} {572 } (\bibinfo {year} {2019})}\BibitemShut {NoStop}%
\bibitem [{\citenamefont {Oyedele}\ \emph {et~al.}(2019)\citenamefont
  {Oyedele}, \citenamefont {Yang}, \citenamefont {Feng}, \citenamefont
  {Haglund}, \citenamefont {Gu}, \citenamefont {Puretzky}, \citenamefont
  {Briggs}, \citenamefont {Rouleau}, \citenamefont {Chisholm}, \citenamefont
  {Unocic}, \citenamefont {Mandrus}, \citenamefont {Meyer}, \citenamefont
  {Pantelides}, \citenamefont {Geohegan},\ and\ \citenamefont
  {Xiao}}]{oyedele_defect-mediated_2019}%
  \BibitemOpen
  \bibfield  {author} {\bibinfo {author} {\bibfnamefont {A.~D.}\ \bibnamefont
  {Oyedele}}, \bibinfo {author} {\bibfnamefont {S.}~\bibnamefont {Yang}},
  \bibinfo {author} {\bibfnamefont {T.}~\bibnamefont {Feng}}, \bibinfo {author}
  {\bibfnamefont {A.~V.}\ \bibnamefont {Haglund}}, \bibinfo {author}
  {\bibfnamefont {Y.}~\bibnamefont {Gu}}, \bibinfo {author} {\bibfnamefont
  {A.~A.}\ \bibnamefont {Puretzky}}, \bibinfo {author} {\bibfnamefont
  {D.}~\bibnamefont {Briggs}}, \bibinfo {author} {\bibfnamefont {C.~M.}\
  \bibnamefont {Rouleau}}, \bibinfo {author} {\bibfnamefont {M.~F.}\
  \bibnamefont {Chisholm}}, \bibinfo {author} {\bibfnamefont {R.~R.}\
  \bibnamefont {Unocic}}, \bibinfo {author} {\bibfnamefont {D.}~\bibnamefont
  {Mandrus}}, \bibinfo {author} {\bibfnamefont {H.~M.}\ \bibnamefont {Meyer}},
  \bibinfo {author} {\bibfnamefont {S.~T.}\ \bibnamefont {Pantelides}},
  \bibinfo {author} {\bibfnamefont {D.~B.}\ \bibnamefont {Geohegan}},\ and\
  \bibinfo {author} {\bibfnamefont {K.}~\bibnamefont {Xiao}},\ }\bibfield
  {title} {\bibinfo {title} {Defect-{Mediated} {Phase} {Transformation} in
  {Anisotropic} {Two}-{Dimensional} {PdSe} $_{\textrm{2}}$ {Crystals} for
  {Seamless} {Electrical} {Contacts}},\ }\href
  {https://doi.org/10.1021/jacs.9b02593} {\bibfield  {journal} {\bibinfo
  {journal} {J. Am. Chem. Soc.}\ }\textbf {\bibinfo {volume} {141}},\ \bibinfo
  {pages} {8928} (\bibinfo {year} {2019})}\BibitemShut {NoStop}%
\bibitem [{\citenamefont {Nguyen}\ \emph {et~al.}(2020)\citenamefont {Nguyen},
  \citenamefont {Oyedele}, \citenamefont {Haglund}, \citenamefont {Ko},
  \citenamefont {Liang}, \citenamefont {Puretzky}, \citenamefont {Mandrus},
  \citenamefont {Xiao},\ and\ \citenamefont {Li}}]{nguyen_atomically_2020}%
  \BibitemOpen
  \bibfield  {author} {\bibinfo {author} {\bibfnamefont {G.~D.}\ \bibnamefont
  {Nguyen}}, \bibinfo {author} {\bibfnamefont {A.~D.}\ \bibnamefont {Oyedele}},
  \bibinfo {author} {\bibfnamefont {A.}~\bibnamefont {Haglund}}, \bibinfo
  {author} {\bibfnamefont {W.}~\bibnamefont {Ko}}, \bibinfo {author}
  {\bibfnamefont {L.}~\bibnamefont {Liang}}, \bibinfo {author} {\bibfnamefont
  {A.~A.}\ \bibnamefont {Puretzky}}, \bibinfo {author} {\bibfnamefont
  {D.}~\bibnamefont {Mandrus}}, \bibinfo {author} {\bibfnamefont
  {K.}~\bibnamefont {Xiao}},\ and\ \bibinfo {author} {\bibfnamefont {A.-P.}\
  \bibnamefont {Li}},\ }\bibfield  {title} {\bibinfo {title} {Atomically
  {Precise} {PdSe} $_{\textrm{2}}$ {Pentagonal} {Nanoribbons}},\ }\href
  {https://doi.org/10.1021/acsnano.9b08390} {\bibfield  {journal} {\bibinfo
  {journal} {ACS Nano}\ }\textbf {\bibinfo {volume} {14}},\ \bibinfo {pages}
  {1951} (\bibinfo {year} {2020})}\BibitemShut {NoStop}%
\bibitem [{\citenamefont {Liu}\ and\ \citenamefont
  {Zhuang}(2018)}]{PhysRevMat_liu2018}%
  \BibitemOpen
  \bibfield  {author} {\bibinfo {author} {\bibfnamefont {L.}~\bibnamefont
  {Liu}}\ and\ \bibinfo {author} {\bibfnamefont {H.~L.}\ \bibnamefont
  {Zhuang}},\ }\bibfield  {title} {\bibinfo {title} {${\mathrm{ptp}}_{2}$: An
  example of exploring the hidden cairo tessellation in the pyrite structure
  for discovering novel two-dimensional materials},\ }\href
  {https://doi.org/10.1103/PhysRevMaterials.2.114003} {\bibfield  {journal}
  {\bibinfo  {journal} {Phys. Rev. Materials}\ }\textbf {\bibinfo {volume}
  {2}},\ \bibinfo {pages} {114003} (\bibinfo {year} {2018})}\BibitemShut
  {NoStop}%
\bibitem [{\citenamefont {Zhuang}(2019)}]{ZHUANG2019448}%
  \BibitemOpen
  \bibfield  {author} {\bibinfo {author} {\bibfnamefont {H.~L.}\ \bibnamefont
  {Zhuang}},\ }\bibfield  {title} {\bibinfo {title} {From pentagonal geometries
  to two-dimensional materials},\ }\href
  {https://doi.org/https://doi.org/10.1016/j.commatsci.2018.12.041} {\bibfield
  {journal} {\bibinfo  {journal} {Computational Materials Science}\ }\textbf
  {\bibinfo {volume} {159}},\ \bibinfo {pages} {448 } (\bibinfo {year}
  {2019})}\BibitemShut {NoStop}%
\bibitem [{\citenamefont {{Liu}}\ \emph {et~al.}(2019)\citenamefont {{Liu}},
  \citenamefont {{Kankam}},\ and\ \citenamefont
  {{Zhuang}}}]{Lui_Zhuang2019_ES}%
  \BibitemOpen
  \bibfield  {author} {\bibinfo {author} {\bibfnamefont {L.}~\bibnamefont
  {{Liu}}}, \bibinfo {author} {\bibfnamefont {I.}~\bibnamefont {{Kankam}}},\
  and\ \bibinfo {author} {\bibfnamefont {H.~L.}\ \bibnamefont {{Zhuang}}},\
  }\bibfield  {title} {\bibinfo {title} {{Ab initio playing of pentagonal
  puzzles}},\ }\href {https://doi.org/10.1088/2516-1075/aae303} {\bibfield
  {journal} {\bibinfo  {journal} {Electronic Structure}\ }\textbf {\bibinfo
  {volume} {1}},\ \bibinfo {pages} {015004} (\bibinfo {year}
  {2019})}\BibitemShut {NoStop}%
\bibitem [{\citenamefont {{Rao}}(2017)}]{2017Rao}%
  \BibitemOpen
  \bibfield  {author} {\bibinfo {author} {\bibfnamefont {M.}~\bibnamefont
  {{Rao}}},\ }\bibfield  {title} {\bibinfo {title} {{Exhaustive search of
  convex pentagons which tile the plane}},\ }\href@noop {} {\bibfield
  {journal} {\bibinfo  {journal} {arXiv e-prints}\ ,\ \bibinfo {eid}
  {arXiv:1708.00274}} (\bibinfo {year} {2017})},\ \Eprint
  {https://arxiv.org/abs/1708.00274} {arXiv:1708.00274 [math.CO]} \BibitemShut
  {NoStop}%
\bibitem [{\citenamefont {Giannozzi}\ \emph {et~al.}(2020)\citenamefont
  {Giannozzi}, \citenamefont {Baseggio}, \citenamefont {Bonfà}, \citenamefont
  {Brunato}, \citenamefont {Car}, \citenamefont {Carnimeo}, \citenamefont
  {Cavazzoni}, \citenamefont {de~Gironcoli}, \citenamefont {Delugas},
  \citenamefont {Ferrari~Ruffino}, \citenamefont {Ferretti}, \citenamefont
  {Marzari}, \citenamefont {Timrov}, \citenamefont {Urru},\ and\ \citenamefont
  {Baroni}}]{QE_2020}%
  \BibitemOpen
  \bibfield  {author} {\bibinfo {author} {\bibfnamefont {P.}~\bibnamefont
  {Giannozzi}}, \bibinfo {author} {\bibfnamefont {O.}~\bibnamefont {Baseggio}},
  \bibinfo {author} {\bibfnamefont {P.}~\bibnamefont {Bonfà}}, \bibinfo
  {author} {\bibfnamefont {D.}~\bibnamefont {Brunato}}, \bibinfo {author}
  {\bibfnamefont {R.}~\bibnamefont {Car}}, \bibinfo {author} {\bibfnamefont
  {I.}~\bibnamefont {Carnimeo}}, \bibinfo {author} {\bibfnamefont
  {C.}~\bibnamefont {Cavazzoni}}, \bibinfo {author} {\bibfnamefont
  {S.}~\bibnamefont {de~Gironcoli}}, \bibinfo {author} {\bibfnamefont
  {P.}~\bibnamefont {Delugas}}, \bibinfo {author} {\bibfnamefont
  {F.}~\bibnamefont {Ferrari~Ruffino}}, \bibinfo {author} {\bibfnamefont
  {A.}~\bibnamefont {Ferretti}}, \bibinfo {author} {\bibfnamefont
  {N.}~\bibnamefont {Marzari}}, \bibinfo {author} {\bibfnamefont
  {I.}~\bibnamefont {Timrov}}, \bibinfo {author} {\bibfnamefont
  {A.}~\bibnamefont {Urru}},\ and\ \bibinfo {author} {\bibfnamefont
  {S.}~\bibnamefont {Baroni}},\ }\bibfield  {title} {\bibinfo {title} {Quantum
  espresso toward the exascale},\ }\href {https://doi.org/10.1063/5.0005082}
  {\bibfield  {journal} {\bibinfo  {journal} {The Journal of Chemical Physics}\
  }\textbf {\bibinfo {volume} {152}},\ \bibinfo {pages} {154105} (\bibinfo
  {year} {2020})}\BibitemShut {NoStop}%
\bibitem [{\citenamefont {Mostofi}\ \emph {et~al.}(2014)\citenamefont
  {Mostofi}, \citenamefont {Yates}, \citenamefont {Pizzi}, \citenamefont {Lee},
  \citenamefont {Souza}, \citenamefont {Vanderbilt},\ and\ \citenamefont
  {Marzari}}]{mostofi_updated_2014}%
  \BibitemOpen
  \bibfield  {author} {\bibinfo {author} {\bibfnamefont {A.~A.}\ \bibnamefont
  {Mostofi}}, \bibinfo {author} {\bibfnamefont {J.~R.}\ \bibnamefont {Yates}},
  \bibinfo {author} {\bibfnamefont {G.}~\bibnamefont {Pizzi}}, \bibinfo
  {author} {\bibfnamefont {Y.-S.}\ \bibnamefont {Lee}}, \bibinfo {author}
  {\bibfnamefont {I.}~\bibnamefont {Souza}}, \bibinfo {author} {\bibfnamefont
  {D.}~\bibnamefont {Vanderbilt}},\ and\ \bibinfo {author} {\bibfnamefont
  {N.}~\bibnamefont {Marzari}},\ }\bibfield  {title} {\bibinfo {title} {An
  updated version of wannier90: {A} tool for obtaining maximally-localised
  {Wannier} functions},\ }\href
  {https://doi.org/https://doi.org/10.1016/j.cpc.2014.05.003} {\bibfield
  {journal} {\bibinfo  {journal} {Computer Physics Communications}\ }\textbf
  {\bibinfo {volume} {185}},\ \bibinfo {pages} {2309 } (\bibinfo {year}
  {2014})}\BibitemShut {NoStop}%
\bibitem [{\citenamefont {Wu}\ \emph {et~al.}(2018)\citenamefont {Wu},
  \citenamefont {Zhang}, \citenamefont {Song}, \citenamefont {Troyer},\ and\
  \citenamefont {Soluyanov}}]{WU2017}%
  \BibitemOpen
  \bibfield  {author} {\bibinfo {author} {\bibfnamefont {Q.}~\bibnamefont
  {Wu}}, \bibinfo {author} {\bibfnamefont {S.}~\bibnamefont {Zhang}}, \bibinfo
  {author} {\bibfnamefont {H.-F.}\ \bibnamefont {Song}}, \bibinfo {author}
  {\bibfnamefont {M.}~\bibnamefont {Troyer}},\ and\ \bibinfo {author}
  {\bibfnamefont {A.~A.}\ \bibnamefont {Soluyanov}},\ }\bibfield  {title}
  {\bibinfo {title} {Wanniertools : An open-source software package for novel
  topological materials},\ }\href
  {https://doi.org/https://doi.org/10.1016/j.cpc.2017.09.033} {\bibfield
  {journal} {\bibinfo  {journal} {Computer Physics Communications}\ }\textbf
  {\bibinfo {volume} {224}},\ \bibinfo {pages} {405 } (\bibinfo {year}
  {2018})}\BibitemShut {NoStop}%
\bibitem [{\citenamefont {Malashevich}\ and\ \citenamefont
  {Souza}(2010)}]{Malashevich_2010}%
  \BibitemOpen
  \bibfield  {author} {\bibinfo {author} {\bibfnamefont {A.}~\bibnamefont
  {Malashevich}}\ and\ \bibinfo {author} {\bibfnamefont {I.}~\bibnamefont
  {Souza}},\ }\bibfield  {title} {\bibinfo {title} {Band theory of spatial
  dispersion in magnetoelectrics},\ }\href
  {https://doi.org/10.1103/PhysRevB.82.245118} {\bibfield  {journal} {\bibinfo
  {journal} {Phys. Rev. B}\ }\textbf {\bibinfo {volume} {82}},\ \bibinfo
  {pages} {245118} (\bibinfo {year} {2010})}\BibitemShut {NoStop}%
\bibitem [{\citenamefont {Sipe}\ and\ \citenamefont
  {Shkrebtii}(2000)}]{Sype_2000}%
  \BibitemOpen
  \bibfield  {author} {\bibinfo {author} {\bibfnamefont {J.~E.}\ \bibnamefont
  {Sipe}}\ and\ \bibinfo {author} {\bibfnamefont {A.~I.}\ \bibnamefont
  {Shkrebtii}},\ }\bibfield  {title} {\bibinfo {title} {Second-order optical
  response in semiconductors},\ }\href
  {https://doi.org/10.1103/PhysRevB.61.5337} {\bibfield  {journal} {\bibinfo
  {journal} {Phys. Rev. B}\ }\textbf {\bibinfo {volume} {61}},\ \bibinfo
  {pages} {5337} (\bibinfo {year} {2000})}\BibitemShut {NoStop}%
\bibitem [{\citenamefont {Elcoro}\ \emph {et~al.}(2017)\citenamefont {Elcoro},
  \citenamefont {Bradlyn}, \citenamefont {Wang}, \citenamefont {Vergniory},
  \citenamefont {Cano}, \citenamefont {Felser}, \citenamefont {Bernevig},
  \citenamefont {Orobengoa}, \citenamefont {de~la Flor},\ and\ \citenamefont
  {Aroyo}}]{bilbao_doubleSG}%
  \BibitemOpen
  \bibfield  {author} {\bibinfo {author} {\bibfnamefont {L.}~\bibnamefont
  {Elcoro}}, \bibinfo {author} {\bibfnamefont {B.}~\bibnamefont {Bradlyn}},
  \bibinfo {author} {\bibfnamefont {Z.}~\bibnamefont {Wang}}, \bibinfo {author}
  {\bibfnamefont {M.~G.}\ \bibnamefont {Vergniory}}, \bibinfo {author}
  {\bibfnamefont {J.}~\bibnamefont {Cano}}, \bibinfo {author} {\bibfnamefont
  {C.}~\bibnamefont {Felser}}, \bibinfo {author} {\bibfnamefont {B.~A.}\
  \bibnamefont {Bernevig}}, \bibinfo {author} {\bibfnamefont {D.}~\bibnamefont
  {Orobengoa}}, \bibinfo {author} {\bibfnamefont {G.}~\bibnamefont {de~la
  Flor}},\ and\ \bibinfo {author} {\bibfnamefont {M.~I.}\ \bibnamefont
  {Aroyo}},\ }\bibfield  {title} {\bibinfo {title} {{Double crystallographic
  groups and their representations on the Bilbao Crystallographic Server}},\
  }\href {https://doi.org/10.1107/S1600576717011712} {\bibfield  {journal}
  {\bibinfo  {journal} {Journal of Applied Crystallography}\ }\textbf {\bibinfo
  {volume} {50}},\ \bibinfo {pages} {1457} (\bibinfo {year}
  {2017})}\BibitemShut {NoStop}%
\bibitem [{\citenamefont {Christopher~Bradley}(2010)}]{bradley_group}%
  \BibitemOpen
  \bibfield  {author} {\bibinfo {author} {\bibfnamefont {A.~P.~C.}\
  \bibnamefont {Christopher~Bradley}},\ }\href@noop {} {\emph {\bibinfo {title}
  {The {Mathematical} {Theory} of {Symmetry} in {Solids}: {Representation}
  {Theory} for {Point} {Groups} and {Space} {Groups}}}},\ Oxford {Classic}
  {Texts} in the {Physical} {Sciences}\ (\bibinfo  {publisher} {Oxford
  University Press},\ \bibinfo {address} {Oxford, New York},\ \bibinfo {year}
  {2010})\BibitemShut {NoStop}%
\bibitem [{\citenamefont {Bravo}\ \emph {et~al.}(2019)\citenamefont {Bravo},
  \citenamefont {Correa}, \citenamefont {Chico},\ and\ \citenamefont
  {Pacheco}}]{bravo_symmetry-protected_2019}%
  \BibitemOpen
  \bibfield  {author} {\bibinfo {author} {\bibfnamefont {S.}~\bibnamefont
  {Bravo}}, \bibinfo {author} {\bibfnamefont {J.}~\bibnamefont {Correa}},
  \bibinfo {author} {\bibfnamefont {L.}~\bibnamefont {Chico}},\ and\ \bibinfo
  {author} {\bibfnamefont {M.}~\bibnamefont {Pacheco}},\ }\bibfield  {title}
  {\bibinfo {title} {Symmetry-protected metallic and topological phases in
  penta-materials},\ }\href {https://doi.org/10.1038/s41598-019-49187-w}
  {\bibfield  {journal} {\bibinfo  {journal} {Scientific Reports}\ }\textbf
  {\bibinfo {volume} {9}},\ \bibinfo {pages} {12754} (\bibinfo {year}
  {2019})}\BibitemShut {NoStop}%
\bibitem [{\citenamefont {Zhao}\ \emph {et~al.}(2019)\citenamefont {Zhao},
  \citenamefont {Li}, \citenamefont {Wang},\ and\ \citenamefont
  {Wang}}]{zhao_2dplanar_2019}%
  \BibitemOpen
  \bibfield  {author} {\bibinfo {author} {\bibfnamefont {K.}~\bibnamefont
  {Zhao}}, \bibinfo {author} {\bibfnamefont {X.}~\bibnamefont {Li}}, \bibinfo
  {author} {\bibfnamefont {S.}~\bibnamefont {Wang}},\ and\ \bibinfo {author}
  {\bibfnamefont {Q.}~\bibnamefont {Wang}},\ }\bibfield  {title} {\bibinfo
  {title} {{2D} planar penta-{MN} $_{\textrm{2}}$ ({M} = {Pd}, {Pt}) sheets
  identified through structure search},\ }\href
  {https://doi.org/10.1039/C8CP04851G} {\bibfield  {journal} {\bibinfo
  {journal} {Phys. Chem. Chem. Phys.}\ }\textbf {\bibinfo {volume} {21}},\
  \bibinfo {pages} {246} (\bibinfo {year} {2019})}\BibitemShut {NoStop}%
\bibitem [{\citenamefont {Yuan}\ \emph {et~al.}(2017)\citenamefont {Yuan},
  \citenamefont {Zhou}, \citenamefont {Wu}, \citenamefont {Zhang},
  \citenamefont {Chen}, \citenamefont {Hou},\ and\ \citenamefont
  {Wang}}]{yuan_prediction_2017}%
  \BibitemOpen
  \bibfield  {author} {\bibinfo {author} {\bibfnamefont {S.}~\bibnamefont
  {Yuan}}, \bibinfo {author} {\bibfnamefont {Q.}~\bibnamefont {Zhou}}, \bibinfo
  {author} {\bibfnamefont {Q.}~\bibnamefont {Wu}}, \bibinfo {author}
  {\bibfnamefont {Y.}~\bibnamefont {Zhang}}, \bibinfo {author} {\bibfnamefont
  {Q.}~\bibnamefont {Chen}}, \bibinfo {author} {\bibfnamefont {J.-M.}\
  \bibnamefont {Hou}},\ and\ \bibinfo {author} {\bibfnamefont {J.}~\bibnamefont
  {Wang}},\ }\bibfield  {title} {\bibinfo {title} {Prediction of a
  room-temperature eight-coordinate two-dimensional topological insulator:
  penta-{RuS4} monolayer},\ }\href {https://doi.org/10.1038/s41699-017-0032-4}
  {\bibfield  {journal} {\bibinfo  {journal} {npj 2D Mater Appl}\ }\textbf
  {\bibinfo {volume} {1}},\ \bibinfo {pages} {29} (\bibinfo {year}
  {2017})}\BibitemShut {NoStop}%
\bibitem [{\citenamefont {Chang}\ \emph {et~al.}(2018)\citenamefont {Chang},
  \citenamefont {Wieder}, \citenamefont {Schindler}, \citenamefont {Sanchez},
  \citenamefont {Belopolski}, \citenamefont {Huang}, \citenamefont {Singh},
  \citenamefont {Wu}, \citenamefont {Chang}, \citenamefont {Neupert},
  \citenamefont {Xu}, \citenamefont {Lin},\ and\ \citenamefont
  {Hasan}}]{chang_topological_2018}%
  \BibitemOpen
  \bibfield  {author} {\bibinfo {author} {\bibfnamefont {G.}~\bibnamefont
  {Chang}}, \bibinfo {author} {\bibfnamefont {B.~J.}\ \bibnamefont {Wieder}},
  \bibinfo {author} {\bibfnamefont {F.}~\bibnamefont {Schindler}}, \bibinfo
  {author} {\bibfnamefont {D.~S.}\ \bibnamefont {Sanchez}}, \bibinfo {author}
  {\bibfnamefont {I.}~\bibnamefont {Belopolski}}, \bibinfo {author}
  {\bibfnamefont {S.-M.}\ \bibnamefont {Huang}}, \bibinfo {author}
  {\bibfnamefont {B.}~\bibnamefont {Singh}}, \bibinfo {author} {\bibfnamefont
  {D.}~\bibnamefont {Wu}}, \bibinfo {author} {\bibfnamefont {T.-R.}\
  \bibnamefont {Chang}}, \bibinfo {author} {\bibfnamefont {T.}~\bibnamefont
  {Neupert}}, \bibinfo {author} {\bibfnamefont {S.-Y.}\ \bibnamefont {Xu}},
  \bibinfo {author} {\bibfnamefont {H.}~\bibnamefont {Lin}},\ and\ \bibinfo
  {author} {\bibfnamefont {M.~Z.}\ \bibnamefont {Hasan}},\ }\bibfield  {title}
  {\bibinfo {title} {Topological quantum properties of chiral crystals},\
  }\href {https://doi.org/10.1038/s41563-018-0169-3} {\bibfield  {journal}
  {\bibinfo  {journal} {Nature Mater}\ }\textbf {\bibinfo {volume} {17}},\
  \bibinfo {pages} {978} (\bibinfo {year} {2018})}\BibitemShut {NoStop}%
\bibitem [{\citenamefont {Nye}(1985)}]{Nye_physical_1985}%
  \BibitemOpen
  \bibfield  {author} {\bibinfo {author} {\bibfnamefont {J.~F.}\ \bibnamefont
  {Nye}},\ }\href@noop {} {\emph {\bibinfo {title} {Physical Properties of
  Crystals: Their Representation by Tensors and Matrices}}}\ (\bibinfo
  {publisher} {Oxford University Press},\ \bibinfo {address} {Oxford, New
  York},\ \bibinfo {year} {1985})\BibitemShut {NoStop}%
\bibitem [{\citenamefont {Aroyo}\ \emph {et~al.}(2006)\citenamefont {Aroyo},
  \citenamefont {Kirov}, \citenamefont {Capillas}, \citenamefont {Perez-Mato},\
  and\ \citenamefont {Wondratschek}}]{bilbao_serverII}%
  \BibitemOpen
  \bibfield  {author} {\bibinfo {author} {\bibfnamefont {M.~I.}\ \bibnamefont
  {Aroyo}}, \bibinfo {author} {\bibfnamefont {A.}~\bibnamefont {Kirov}},
  \bibinfo {author} {\bibfnamefont {C.}~\bibnamefont {Capillas}}, \bibinfo
  {author} {\bibfnamefont {J.~M.}\ \bibnamefont {Perez-Mato}},\ and\ \bibinfo
  {author} {\bibfnamefont {H.}~\bibnamefont {Wondratschek}},\ }\bibfield
  {title} {\bibinfo {title} {{Bilbao Crystallographic Server. II.
  Representations of crystallographic point groups and space groups}},\ }\href
  {https://doi.org/10.1107/S0108767305040286} {\bibfield  {journal} {\bibinfo
  {journal} {Acta Crystallographica Section A}\ }\textbf {\bibinfo {volume}
  {62}},\ \bibinfo {pages} {115} (\bibinfo {year} {2006})}\BibitemShut
  {NoStop}%
\bibitem [{\citenamefont {Liu}\ and\ \citenamefont
  {Zhuang}(2019)}]{liu_computational_2019}%
  \BibitemOpen
  \bibfield  {author} {\bibinfo {author} {\bibfnamefont {L.}~\bibnamefont
  {Liu}}\ and\ \bibinfo {author} {\bibfnamefont {H.~L.}\ \bibnamefont
  {Zhuang}},\ }\bibfield  {title} {\bibinfo {title} {Computational prediction
  and characterization of two-dimensional pentagonal arsenopyrite {FeAsS}},\
  }\href {https://doi.org/10.1016/j.commatsci.2019.04.040} {\bibfield
  {journal} {\bibinfo  {journal} {Computational Materials Science}\ }\textbf
  {\bibinfo {volume} {166}},\ \bibinfo {pages} {105} (\bibinfo {year}
  {2019})}\BibitemShut {NoStop}%
\bibitem [{\citenamefont {Dresselhaus}\ \emph {et~al.}(2008)\citenamefont
  {Dresselhaus}, \citenamefont {Dresselhaus},\ and\ \citenamefont
  {Jorio}}]{dresselhaus_group_2008}%
  \BibitemOpen
  \bibfield  {author} {\bibinfo {author} {\bibfnamefont {M.~S.}\ \bibnamefont
  {Dresselhaus}}, \bibinfo {author} {\bibfnamefont {G.}~\bibnamefont
  {Dresselhaus}},\ and\ \bibinfo {author} {\bibfnamefont {A.}~\bibnamefont
  {Jorio}},\ }\href {https://doi.org/10.1007/978-3-540-32899-5} {\emph
  {\bibinfo {title} {Group {Theory}: {Application} to the {Physics} of
  {Condensed} {Matter}}}}\ (\bibinfo  {publisher} {Springer-Verlag},\ \bibinfo
  {address} {Berlin Heidelberg},\ \bibinfo {year} {2008})\BibitemShut {NoStop}%
\bibitem [{\citenamefont {Vanderbilt}(2018)}]{vanderbilt_2018}%
  \BibitemOpen
  \bibfield  {author} {\bibinfo {author} {\bibfnamefont {D.}~\bibnamefont
  {Vanderbilt}},\ }\href {https://doi.org/10.1017/9781316662205} {\emph
  {\bibinfo {title} {Berry Phases in Electronic Structure Theory: Electric
  Polarization, Orbital Magnetization and Topological Insulators}}}\ (\bibinfo
  {publisher} {Cambridge University Press},\ \bibinfo {year}
  {2018})\BibitemShut {NoStop}%
\bibitem [{\citenamefont {Hu}\ \emph {et~al.}(2019)\citenamefont {Hu},
  \citenamefont {Xu}, \citenamefont {Ni},\ and\ \citenamefont
  {Mao}}]{Hu_2019_ARCM}%
  \BibitemOpen
  \bibfield  {author} {\bibinfo {author} {\bibfnamefont {J.}~\bibnamefont
  {Hu}}, \bibinfo {author} {\bibfnamefont {S.-Y.}\ \bibnamefont {Xu}}, \bibinfo
  {author} {\bibfnamefont {N.}~\bibnamefont {Ni}},\ and\ \bibinfo {author}
  {\bibfnamefont {Z.}~\bibnamefont {Mao}},\ }\bibfield  {title} {\bibinfo
  {title} {Transport of topological semimetals},\ }\href
  {https://doi.org/10.1146/annurev-matsci-070218-010023} {\bibfield  {journal}
  {\bibinfo  {journal} {Annual Review of Materials Research}\ }\textbf
  {\bibinfo {volume} {49}},\ \bibinfo {pages} {207} (\bibinfo {year}
  {2019})}\BibitemShut {NoStop}%
\bibitem [{\citenamefont {Nagaosa}\ \emph {et~al.}(2020)\citenamefont
  {Nagaosa}, \citenamefont {Morimoto},\ and\ \citenamefont
  {Tokura}}]{nagaosa_transport_2020}%
  \BibitemOpen
  \bibfield  {author} {\bibinfo {author} {\bibfnamefont {N.}~\bibnamefont
  {Nagaosa}}, \bibinfo {author} {\bibfnamefont {T.}~\bibnamefont {Morimoto}},\
  and\ \bibinfo {author} {\bibfnamefont {Y.}~\bibnamefont {Tokura}},\
  }\bibfield  {title} {\bibinfo {title} {Transport, magnetic and optical
  properties of {Weyl} materials},\ }\href
  {https://doi.org/10.1038/s41578-020-0208-y} {\bibfield  {journal} {\bibinfo
  {journal} {Nat Rev Mater}\ }\textbf {\bibinfo {volume} {5}},\ \bibinfo
  {pages} {621} (\bibinfo {year} {2020})}\BibitemShut {NoStop}%
\bibitem [{\citenamefont {Xie}\ \emph {et~al.}(2020)\citenamefont {Xie},
  \citenamefont {Gao}, \citenamefont {Xu}, \citenamefont {Zhang}, \citenamefont
  {Hu},\ and\ \citenamefont {Law}}]{xie_kramers_2020}%
  \BibitemOpen
  \bibfield  {author} {\bibinfo {author} {\bibfnamefont {Y.-M.}\ \bibnamefont
  {Xie}}, \bibinfo {author} {\bibfnamefont {X.-J.}\ \bibnamefont {Gao}},
  \bibinfo {author} {\bibfnamefont {X.~Y.}\ \bibnamefont {Xu}}, \bibinfo
  {author} {\bibfnamefont {C.-P.}\ \bibnamefont {Zhang}}, \bibinfo {author}
  {\bibfnamefont {J.-X.}\ \bibnamefont {Hu}},\ and\ \bibinfo {author}
  {\bibfnamefont {K.~T.}\ \bibnamefont {Law}},\ }\bibfield  {title} {\bibinfo
  {title} {Kramers {Nodal} {Line} {Metals}},\ }\href
  {http://arxiv.org/abs/2008.03967} {\bibfield  {journal} {\bibinfo  {journal}
  {arXiv:2008.03967 [cond-mat]}\ } (\bibinfo {year} {2020})},\ \bibinfo {note}
  {arXiv: 2008.03967}\BibitemShut {NoStop}%
\bibitem [{\citenamefont {Fang}\ \emph {et~al.}(2012)\citenamefont {Fang},
  \citenamefont {Gilbert}, \citenamefont {Dai},\ and\ \citenamefont
  {Bernevig}}]{2012_MultiWeyl_PRL}%
  \BibitemOpen
  \bibfield  {author} {\bibinfo {author} {\bibfnamefont {C.}~\bibnamefont
  {Fang}}, \bibinfo {author} {\bibfnamefont {M.~J.}\ \bibnamefont {Gilbert}},
  \bibinfo {author} {\bibfnamefont {X.}~\bibnamefont {Dai}},\ and\ \bibinfo
  {author} {\bibfnamefont {B.~A.}\ \bibnamefont {Bernevig}},\ }\bibfield
  {title} {\bibinfo {title} {Multi-weyl topological semimetals stabilized by
  point group symmetry},\ }\href
  {https://doi.org/10.1103/PhysRevLett.108.266802} {\bibfield  {journal}
  {\bibinfo  {journal} {Phys. Rev. Lett.}\ }\textbf {\bibinfo {volume} {108}},\
  \bibinfo {pages} {266802} (\bibinfo {year} {2012})}\BibitemShut {NoStop}%
\bibitem [{\citenamefont {Tsirkin}\ \emph {et~al.}(2017)\citenamefont
  {Tsirkin}, \citenamefont {Souza},\ and\ \citenamefont
  {Vanderbilt}}]{2017_Multiweyl_PRB}%
  \BibitemOpen
  \bibfield  {author} {\bibinfo {author} {\bibfnamefont {S.~S.}\ \bibnamefont
  {Tsirkin}}, \bibinfo {author} {\bibfnamefont {I.}~\bibnamefont {Souza}},\
  and\ \bibinfo {author} {\bibfnamefont {D.}~\bibnamefont {Vanderbilt}},\
  }\bibfield  {title} {\bibinfo {title} {Composite weyl nodes stabilized by
  screw symmetry with and without time-reversal invariance},\ }\href
  {https://doi.org/10.1103/PhysRevB.96.045102} {\bibfield  {journal} {\bibinfo
  {journal} {Phys. Rev. B}\ }\textbf {\bibinfo {volume} {96}},\ \bibinfo
  {pages} {045102} (\bibinfo {year} {2017})}\BibitemShut {NoStop}%
\bibitem [{\citenamefont {El-Batanouny}\ and\ \citenamefont
  {Wooten}(2008)}]{el-batanouny_wooten_2008}%
  \BibitemOpen
  \bibfield  {author} {\bibinfo {author} {\bibfnamefont {M.}~\bibnamefont
  {El-Batanouny}}\ and\ \bibinfo {author} {\bibfnamefont {F.}~\bibnamefont
  {Wooten}},\ }\href {https://doi.org/10.1017/CBO9780511755736} {\emph
  {\bibinfo {title} {Symmetry and Condensed Matter Physics: A Computational
  Approach}}}\ (\bibinfo  {publisher} {Cambridge University Press},\ \bibinfo
  {year} {2008})\BibitemShut {NoStop}%
\bibitem [{\citenamefont {Sanchez}\ \emph {et~al.}(2019)\citenamefont
  {Sanchez}, \citenamefont {Belopolski}, \citenamefont {Cochran}, \citenamefont
  {Xu}, \citenamefont {Yin}, \citenamefont {Chang}, \citenamefont {Xie},
  \citenamefont {Manna}, \citenamefont {Süß}, \citenamefont {Huang},
  \citenamefont {Alidoust}, \citenamefont {Multer}, \citenamefont {Zhang},
  \citenamefont {Shumiya}, \citenamefont {Wang}, \citenamefont {Wang},
  \citenamefont {Chang}, \citenamefont {Felser}, \citenamefont {Xu},
  \citenamefont {Jia}, \citenamefont {Lin},\ and\ \citenamefont
  {Hasan}}]{sanchez_topological_2019}%
  \BibitemOpen
  \bibfield  {author} {\bibinfo {author} {\bibfnamefont {D.~S.}\ \bibnamefont
  {Sanchez}}, \bibinfo {author} {\bibfnamefont {I.}~\bibnamefont {Belopolski}},
  \bibinfo {author} {\bibfnamefont {T.~A.}\ \bibnamefont {Cochran}}, \bibinfo
  {author} {\bibfnamefont {X.}~\bibnamefont {Xu}}, \bibinfo {author}
  {\bibfnamefont {J.-X.}\ \bibnamefont {Yin}}, \bibinfo {author} {\bibfnamefont
  {G.}~\bibnamefont {Chang}}, \bibinfo {author} {\bibfnamefont
  {W.}~\bibnamefont {Xie}}, \bibinfo {author} {\bibfnamefont {K.}~\bibnamefont
  {Manna}}, \bibinfo {author} {\bibfnamefont {V.}~\bibnamefont {Süß}},
  \bibinfo {author} {\bibfnamefont {C.-Y.}\ \bibnamefont {Huang}}, \bibinfo
  {author} {\bibfnamefont {N.}~\bibnamefont {Alidoust}}, \bibinfo {author}
  {\bibfnamefont {D.}~\bibnamefont {Multer}}, \bibinfo {author} {\bibfnamefont
  {S.~S.}\ \bibnamefont {Zhang}}, \bibinfo {author} {\bibfnamefont
  {N.}~\bibnamefont {Shumiya}}, \bibinfo {author} {\bibfnamefont
  {X.}~\bibnamefont {Wang}}, \bibinfo {author} {\bibfnamefont {G.-Q.}\
  \bibnamefont {Wang}}, \bibinfo {author} {\bibfnamefont {T.-R.}\ \bibnamefont
  {Chang}}, \bibinfo {author} {\bibfnamefont {C.}~\bibnamefont {Felser}},
  \bibinfo {author} {\bibfnamefont {S.-Y.}\ \bibnamefont {Xu}}, \bibinfo
  {author} {\bibfnamefont {S.}~\bibnamefont {Jia}}, \bibinfo {author}
  {\bibfnamefont {H.}~\bibnamefont {Lin}},\ and\ \bibinfo {author}
  {\bibfnamefont {M.~Z.}\ \bibnamefont {Hasan}},\ }\bibfield  {title} {\bibinfo
  {title} {Topological chiral crystals with helicoid-arc quantum states},\
  }\href {https://doi.org/10.1038/s41586-019-1037-2} {\bibfield  {journal}
  {\bibinfo  {journal} {Nature}\ }\textbf {\bibinfo {volume} {567}},\ \bibinfo
  {pages} {500} (\bibinfo {year} {2019})}\BibitemShut {NoStop}%
\bibitem [{\citenamefont {Flicker}\ \emph {et~al.}(2018)\citenamefont
  {Flicker}, \citenamefont {de~Juan}, \citenamefont {Bradlyn}, \citenamefont
  {Morimoto}, \citenamefont {Vergniory},\ and\ \citenamefont
  {Grushin}}]{flicker_chiral_2018}%
  \BibitemOpen
  \bibfield  {author} {\bibinfo {author} {\bibfnamefont {F.}~\bibnamefont
  {Flicker}}, \bibinfo {author} {\bibfnamefont {F.}~\bibnamefont {de~Juan}},
  \bibinfo {author} {\bibfnamefont {B.}~\bibnamefont {Bradlyn}}, \bibinfo
  {author} {\bibfnamefont {T.}~\bibnamefont {Morimoto}}, \bibinfo {author}
  {\bibfnamefont {M.~G.}\ \bibnamefont {Vergniory}},\ and\ \bibinfo {author}
  {\bibfnamefont {A.~G.}\ \bibnamefont {Grushin}},\ }\bibfield  {title}
  {\bibinfo {title} {Chiral optical response of multifold fermions},\ }\href
  {https://doi.org/10.1103/PhysRevB.98.155145} {\bibfield  {journal} {\bibinfo
  {journal} {Phys. Rev. B}\ }\textbf {\bibinfo {volume} {98}},\ \bibinfo
  {pages} {155145} (\bibinfo {year} {2018})}\BibitemShut {NoStop}%
\bibitem [{\citenamefont {S\'anchez-Mart\'{\i}nez}\ \emph
  {et~al.}(2019)\citenamefont {S\'anchez-Mart\'{\i}nez}, \citenamefont
  {de~Juan},\ and\ \citenamefont {Grushin}}]{sanchez_linear2019}%
  \BibitemOpen
  \bibfield  {author} {\bibinfo {author} {\bibfnamefont {M.-A.}\ \bibnamefont
  {S\'anchez-Mart\'{\i}nez}}, \bibinfo {author} {\bibfnamefont
  {F.}~\bibnamefont {de~Juan}},\ and\ \bibinfo {author} {\bibfnamefont {A.~G.}\
  \bibnamefont {Grushin}},\ }\bibfield  {title} {\bibinfo {title} {Linear
  optical conductivity of chiral multifold fermions},\ }\href
  {https://doi.org/10.1103/PhysRevB.99.155145} {\bibfield  {journal} {\bibinfo
  {journal} {Phys. Rev. B}\ }\textbf {\bibinfo {volume} {99}},\ \bibinfo
  {pages} {155145} (\bibinfo {year} {2019})}\BibitemShut {NoStop}%
\bibitem [{\citenamefont {Jin}\ \emph {et~al.}(2017)\citenamefont {Jin},
  \citenamefont {Wang}, \citenamefont {Zhao}, \citenamefont {Du}, \citenamefont
  {Zheng}, \citenamefont {Gan}, \citenamefont {Liu}, \citenamefont {Xu},\ and\
  \citenamefont {Tong}}]{nodal_Nanoscale_2017}%
  \BibitemOpen
  \bibfield  {author} {\bibinfo {author} {\bibfnamefont {Y.-J.}\ \bibnamefont
  {Jin}}, \bibinfo {author} {\bibfnamefont {R.}~\bibnamefont {Wang}}, \bibinfo
  {author} {\bibfnamefont {J.-Z.}\ \bibnamefont {Zhao}}, \bibinfo {author}
  {\bibfnamefont {Y.-P.}\ \bibnamefont {Du}}, \bibinfo {author} {\bibfnamefont
  {C.-D.}\ \bibnamefont {Zheng}}, \bibinfo {author} {\bibfnamefont {L.-Y.}\
  \bibnamefont {Gan}}, \bibinfo {author} {\bibfnamefont {J.-F.}\ \bibnamefont
  {Liu}}, \bibinfo {author} {\bibfnamefont {H.}~\bibnamefont {Xu}},\ and\
  \bibinfo {author} {\bibfnamefont {S.~Y.}\ \bibnamefont {Tong}},\ }\bibfield
  {title} {\bibinfo {title} {The prediction of a family group of
  two-dimensional node-line semimetals},\ }\href
  {https://doi.org/10.1039/C7NR03520A} {\bibfield  {journal} {\bibinfo
  {journal} {Nanoscale}\ }\textbf {\bibinfo {volume} {9}},\ \bibinfo {pages}
  {13112} (\bibinfo {year} {2017})}\BibitemShut {NoStop}%
\bibitem [{\citenamefont {Ahn}\ \emph {et~al.}(2018)\citenamefont {Ahn},
  \citenamefont {Kim}, \citenamefont {Kim},\ and\ \citenamefont
  {Yang}}]{nodal_PRL_2018}%
  \BibitemOpen
  \bibfield  {author} {\bibinfo {author} {\bibfnamefont {J.}~\bibnamefont
  {Ahn}}, \bibinfo {author} {\bibfnamefont {D.}~\bibnamefont {Kim}}, \bibinfo
  {author} {\bibfnamefont {Y.}~\bibnamefont {Kim}},\ and\ \bibinfo {author}
  {\bibfnamefont {B.-J.}\ \bibnamefont {Yang}},\ }\bibfield  {title} {\bibinfo
  {title} {Band topology and linking structure of nodal line semimetals with
  ${Z}_{2}$ monopole charges},\ }\href
  {https://doi.org/10.1103/PhysRevLett.121.106403} {\bibfield  {journal}
  {\bibinfo  {journal} {Phys. Rev. Lett.}\ }\textbf {\bibinfo {volume} {121}},\
  \bibinfo {pages} {106403} (\bibinfo {year} {2018})}\BibitemShut {NoStop}%
\bibitem [{\citenamefont {Li}\ \emph {et~al.}(2018)\citenamefont {Li},
  \citenamefont {Liu}, \citenamefont {Wang}, \citenamefont {Yu}, \citenamefont
  {Guan}, \citenamefont {Sheng}, \citenamefont {Yao},\ and\ \citenamefont
  {Yang}}]{nodal_PRB_2018}%
  \BibitemOpen
  \bibfield  {author} {\bibinfo {author} {\bibfnamefont {S.}~\bibnamefont
  {Li}}, \bibinfo {author} {\bibfnamefont {Y.}~\bibnamefont {Liu}}, \bibinfo
  {author} {\bibfnamefont {S.-S.}\ \bibnamefont {Wang}}, \bibinfo {author}
  {\bibfnamefont {Z.-M.}\ \bibnamefont {Yu}}, \bibinfo {author} {\bibfnamefont
  {S.}~\bibnamefont {Guan}}, \bibinfo {author} {\bibfnamefont {X.-L.}\
  \bibnamefont {Sheng}}, \bibinfo {author} {\bibfnamefont {Y.}~\bibnamefont
  {Yao}},\ and\ \bibinfo {author} {\bibfnamefont {S.~A.}\ \bibnamefont
  {Yang}},\ }\bibfield  {title} {\bibinfo {title}
  {Nonsymmorphic-symmetry-protected hourglass dirac loop, nodal line, and dirac
  point in bulk and monolayer ${X}_{3}{\mathrm{site}}_{6}$ ($x$ = ta, nb)},\
  }\href {https://doi.org/10.1103/PhysRevB.97.045131} {\bibfield  {journal}
  {\bibinfo  {journal} {Phys. Rev. B}\ }\textbf {\bibinfo {volume} {97}},\
  \bibinfo {pages} {045131} (\bibinfo {year} {2018})}\BibitemShut {NoStop}%
\bibitem [{\citenamefont {Zhong}\ \emph {et~al.}(2019)\citenamefont {Zhong},
  \citenamefont {Wu}, \citenamefont {He}, \citenamefont {Ding}, \citenamefont
  {Liu}, \citenamefont {Li}, \citenamefont {Yang},\ and\ \citenamefont
  {Zhang}}]{nodal_Nanoscale_2019}%
  \BibitemOpen
  \bibfield  {author} {\bibinfo {author} {\bibfnamefont {C.}~\bibnamefont
  {Zhong}}, \bibinfo {author} {\bibfnamefont {W.}~\bibnamefont {Wu}}, \bibinfo
  {author} {\bibfnamefont {J.}~\bibnamefont {He}}, \bibinfo {author}
  {\bibfnamefont {G.}~\bibnamefont {Ding}}, \bibinfo {author} {\bibfnamefont
  {Y.}~\bibnamefont {Liu}}, \bibinfo {author} {\bibfnamefont {D.}~\bibnamefont
  {Li}}, \bibinfo {author} {\bibfnamefont {S.~A.}\ \bibnamefont {Yang}},\ and\
  \bibinfo {author} {\bibfnamefont {G.}~\bibnamefont {Zhang}},\ }\bibfield
  {title} {\bibinfo {title} {Two-dimensional honeycomb borophene oxide: strong
  anisotropy and nodal loop transformation},\ }\href
  {https://doi.org/10.1039/C8NR08729F} {\bibfield  {journal} {\bibinfo
  {journal} {Nanoscale}\ }\textbf {\bibinfo {volume} {11}},\ \bibinfo {pages}
  {2468} (\bibinfo {year} {2019})}\BibitemShut {NoStop}%
\bibitem [{\citenamefont {Wu}\ \emph {et~al.}(2019)\citenamefont {Wu},
  \citenamefont {Jiao}, \citenamefont {Li}, \citenamefont {Sheng},
  \citenamefont {Yu},\ and\ \citenamefont {Yang}}]{nodal_PRM_2019}%
  \BibitemOpen
  \bibfield  {author} {\bibinfo {author} {\bibfnamefont {W.}~\bibnamefont
  {Wu}}, \bibinfo {author} {\bibfnamefont {Y.}~\bibnamefont {Jiao}}, \bibinfo
  {author} {\bibfnamefont {S.}~\bibnamefont {Li}}, \bibinfo {author}
  {\bibfnamefont {X.-L.}\ \bibnamefont {Sheng}}, \bibinfo {author}
  {\bibfnamefont {Z.-M.}\ \bibnamefont {Yu}},\ and\ \bibinfo {author}
  {\bibfnamefont {S.~A.}\ \bibnamefont {Yang}},\ }\bibfield  {title} {\bibinfo
  {title} {Hourglass weyl loops in two dimensions: Theory and material
  realization in monolayer gatei family},\ }\href
  {https://doi.org/10.1103/PhysRevMaterials.3.054203} {\bibfield  {journal}
  {\bibinfo  {journal} {Phys. Rev. Materials}\ }\textbf {\bibinfo {volume}
  {3}},\ \bibinfo {pages} {054203} (\bibinfo {year} {2019})}\BibitemShut
  {NoStop}%
\bibitem [{\citenamefont {Bradlyn}\ \emph {et~al.}(2016)\citenamefont
  {Bradlyn}, \citenamefont {Cano}, \citenamefont {Wang}, \citenamefont
  {Vergniory}, \citenamefont {Felser}, \citenamefont {Cava},\ and\
  \citenamefont {Bernevig}}]{bradlyn_beyond_2016}%
  \BibitemOpen
  \bibfield  {author} {\bibinfo {author} {\bibfnamefont {B.}~\bibnamefont
  {Bradlyn}}, \bibinfo {author} {\bibfnamefont {J.}~\bibnamefont {Cano}},
  \bibinfo {author} {\bibfnamefont {Z.}~\bibnamefont {Wang}}, \bibinfo {author}
  {\bibfnamefont {M.~G.}\ \bibnamefont {Vergniory}}, \bibinfo {author}
  {\bibfnamefont {C.}~\bibnamefont {Felser}}, \bibinfo {author} {\bibfnamefont
  {R.~J.}\ \bibnamefont {Cava}},\ and\ \bibinfo {author} {\bibfnamefont
  {B.~A.}\ \bibnamefont {Bernevig}},\ }\bibfield  {title} {\bibinfo {title}
  {Beyond {Dirac} and {Weyl} fermions: {Unconventional} quasiparticles in
  conventional crystals},\ }\bibfield  {journal} {\bibinfo  {journal}
  {Science}\ }\textbf {\bibinfo {volume} {353}},\ \href
  {https://doi.org/10.1126/science.aaf5037} {10.1126/science.aaf5037} (\bibinfo
  {year} {2016})\BibitemShut {NoStop}%
\bibitem [{\citenamefont {Shekhar}\ \emph {et~al.}(2015)\citenamefont
  {Shekhar}, \citenamefont {Nayak}, \citenamefont {Sun}, \citenamefont
  {Schmidt}, \citenamefont {Nicklas}, \citenamefont {Leermakers}, \citenamefont
  {Zeitler}, \citenamefont {Skourski}, \citenamefont {Wosnitza}, \citenamefont
  {Liu}, \citenamefont {Chen}, \citenamefont {Schnelle}, \citenamefont
  {Borrmann}, \citenamefont {Grin}, \citenamefont {Felser},\ and\ \citenamefont
  {Yan}}]{shekhar_extremely_2015}%
  \BibitemOpen
  \bibfield  {author} {\bibinfo {author} {\bibfnamefont {C.}~\bibnamefont
  {Shekhar}}, \bibinfo {author} {\bibfnamefont {A.~K.}\ \bibnamefont {Nayak}},
  \bibinfo {author} {\bibfnamefont {Y.}~\bibnamefont {Sun}}, \bibinfo {author}
  {\bibfnamefont {M.}~\bibnamefont {Schmidt}}, \bibinfo {author} {\bibfnamefont
  {M.}~\bibnamefont {Nicklas}}, \bibinfo {author} {\bibfnamefont
  {I.}~\bibnamefont {Leermakers}}, \bibinfo {author} {\bibfnamefont
  {U.}~\bibnamefont {Zeitler}}, \bibinfo {author} {\bibfnamefont
  {Y.}~\bibnamefont {Skourski}}, \bibinfo {author} {\bibfnamefont
  {J.}~\bibnamefont {Wosnitza}}, \bibinfo {author} {\bibfnamefont
  {Z.}~\bibnamefont {Liu}}, \bibinfo {author} {\bibfnamefont {Y.}~\bibnamefont
  {Chen}}, \bibinfo {author} {\bibfnamefont {W.}~\bibnamefont {Schnelle}},
  \bibinfo {author} {\bibfnamefont {H.}~\bibnamefont {Borrmann}}, \bibinfo
  {author} {\bibfnamefont {Y.}~\bibnamefont {Grin}}, \bibinfo {author}
  {\bibfnamefont {C.}~\bibnamefont {Felser}},\ and\ \bibinfo {author}
  {\bibfnamefont {B.}~\bibnamefont {Yan}},\ }\bibfield  {title} {\bibinfo
  {title} {Extremely large magnetoresistance and ultrahigh mobility in the
  topological {Weyl} semimetal candidate {NbP}},\ }\bibfield  {journal}
  {\bibinfo  {journal} {Nature Physics}\ }\textbf {\bibinfo {volume} {11}},\
  \href {https://doi.org/10.1038/nphys3372} {10.1038/nphys3372} (\bibinfo
  {year} {2015})\BibitemShut {NoStop}%
\bibitem [{\citenamefont {Huang}\ \emph {et~al.}(2015)\citenamefont {Huang},
  \citenamefont {Zhao}, \citenamefont {Long}, \citenamefont {Wang},
  \citenamefont {Chen}, \citenamefont {Yang}, \citenamefont {Liang},
  \citenamefont {Xue}, \citenamefont {Weng}, \citenamefont {Fang},
  \citenamefont {Dai},\ and\ \citenamefont {Chen}}]{Huang_2015_PRX}%
  \BibitemOpen
  \bibfield  {author} {\bibinfo {author} {\bibfnamefont {X.}~\bibnamefont
  {Huang}}, \bibinfo {author} {\bibfnamefont {L.}~\bibnamefont {Zhao}},
  \bibinfo {author} {\bibfnamefont {Y.}~\bibnamefont {Long}}, \bibinfo {author}
  {\bibfnamefont {P.}~\bibnamefont {Wang}}, \bibinfo {author} {\bibfnamefont
  {D.}~\bibnamefont {Chen}}, \bibinfo {author} {\bibfnamefont {Z.}~\bibnamefont
  {Yang}}, \bibinfo {author} {\bibfnamefont {H.}~\bibnamefont {Liang}},
  \bibinfo {author} {\bibfnamefont {M.}~\bibnamefont {Xue}}, \bibinfo {author}
  {\bibfnamefont {H.}~\bibnamefont {Weng}}, \bibinfo {author} {\bibfnamefont
  {Z.}~\bibnamefont {Fang}}, \bibinfo {author} {\bibfnamefont {X.}~\bibnamefont
  {Dai}},\ and\ \bibinfo {author} {\bibfnamefont {G.}~\bibnamefont {Chen}},\
  }\bibfield  {title} {\bibinfo {title} {Observation of the
  chiral-anomaly-induced negative magnetoresistance in 3d weyl semimetal
  taas},\ }\href {https://doi.org/10.1103/PhysRevX.5.031023} {\bibfield
  {journal} {\bibinfo  {journal} {Phys. Rev. X}\ }\textbf {\bibinfo {volume}
  {5}},\ \bibinfo {pages} {031023} (\bibinfo {year} {2015})}\BibitemShut
  {NoStop}%
\bibitem [{\citenamefont {Wang}\ \emph {et~al.}(2017)\citenamefont {Wang},
  \citenamefont {Lin}, \citenamefont {Wang}, \citenamefont {Yu},\ and\
  \citenamefont {Liao}}]{Wang_2017_AdvPhysX}%
  \BibitemOpen
  \bibfield  {author} {\bibinfo {author} {\bibfnamefont {S.}~\bibnamefont
  {Wang}}, \bibinfo {author} {\bibfnamefont {B.-C.}\ \bibnamefont {Lin}},
  \bibinfo {author} {\bibfnamefont {A.-Q.}\ \bibnamefont {Wang}}, \bibinfo
  {author} {\bibfnamefont {D.-P.}\ \bibnamefont {Yu}},\ and\ \bibinfo {author}
  {\bibfnamefont {Z.-M.}\ \bibnamefont {Liao}},\ }\bibfield  {title} {\bibinfo
  {title} {Quantum transport in dirac and weyl semimetals: a review},\ }\href
  {https://doi.org/10.1080/23746149.2017.1327329} {\bibfield  {journal}
  {\bibinfo  {journal} {Advances in Physics: X}\ }\textbf {\bibinfo {volume}
  {2}},\ \bibinfo {pages} {518} (\bibinfo {year} {2017})},\ \Eprint
  {https://arxiv.org/abs/https://doi.org/10.1080/23746149.2017.1327329}
  {https://doi.org/10.1080/23746149.2017.1327329} \BibitemShut {NoStop}%
\bibitem [{\citenamefont {Tsirkin}\ \emph {et~al.}(2018)\citenamefont
  {Tsirkin}, \citenamefont {Puente},\ and\ \citenamefont
  {Souza}}]{Tsirkin_2018}%
  \BibitemOpen
  \bibfield  {author} {\bibinfo {author} {\bibfnamefont {S.~S.}\ \bibnamefont
  {Tsirkin}}, \bibinfo {author} {\bibfnamefont {P.~A.}\ \bibnamefont
  {Puente}},\ and\ \bibinfo {author} {\bibfnamefont {I.}~\bibnamefont
  {Souza}},\ }\bibfield  {title} {\bibinfo {title} {Gyrotropic effects in
  trigonal tellurium studied from first principles},\ }\href
  {https://doi.org/10.1103/PhysRevB.97.035158} {\bibfield  {journal} {\bibinfo
  {journal} {Phys. Rev. B}\ }\textbf {\bibinfo {volume} {97}},\ \bibinfo
  {pages} {035158} (\bibinfo {year} {2018})}\BibitemShut {NoStop}%
\bibitem [{\citenamefont {Ibañez-Azpiroz}\ \emph {et~al.}(2018)\citenamefont
  {Ibañez-Azpiroz}, \citenamefont {Tsirkin},\ and\ \citenamefont
  {Souza}}]{ibanez-azpiroz_ab_2018}%
  \BibitemOpen
  \bibfield  {author} {\bibinfo {author} {\bibfnamefont {J.}~\bibnamefont
  {Ibañez-Azpiroz}}, \bibinfo {author} {\bibfnamefont {S.~S.}\ \bibnamefont
  {Tsirkin}},\ and\ \bibinfo {author} {\bibfnamefont {I.}~\bibnamefont
  {Souza}},\ }\bibfield  {title} {\bibinfo {title} {\textit{{Ab} initio}
  calculation of the shift photocurrent by {Wannier} interpolation},\ }\href
  {https://doi.org/10.1103/PhysRevB.97.245143} {\bibfield  {journal} {\bibinfo
  {journal} {Phys. Rev. B}\ }\textbf {\bibinfo {volume} {97}},\ \bibinfo
  {pages} {245143} (\bibinfo {year} {2018})}\BibitemShut {NoStop}%
\bibitem [{\citenamefont {Pospischil}\ and\ \citenamefont
  {Mueller}(2016)}]{pospischil_optoelectronic_2016}%
  \BibitemOpen
  \bibfield  {author} {\bibinfo {author} {\bibfnamefont {A.}~\bibnamefont
  {Pospischil}}\ and\ \bibinfo {author} {\bibfnamefont {T.}~\bibnamefont
  {Mueller}},\ }\bibfield  {title} {\bibinfo {title} {Optoelectronic {Devices}
  {Based} on {Atomically} {Thin} {Transition} {Metal} {Dichalcogenides}},\
  }\bibfield  {journal} {\bibinfo  {journal} {Applied Sciences}\ }\textbf
  {\bibinfo {volume} {6}},\ \href {https://doi.org/10.3390/app6030078}
  {10.3390/app6030078} (\bibinfo {year} {2016}),\ \bibinfo {note} {number: 3
  keywords = {2D materials, optoelectronic devices, transition metal
  dichalcogenides}, pages = {78},}\BibitemShut {NoStop}%
\bibitem [{\citenamefont {Long}\ \emph {et~al.}(2020)\citenamefont {Long},
  \citenamefont {Sabatini}, \citenamefont {Saidaminov}, \citenamefont
  {Lakhwani}, \citenamefont {Rasmita}, \citenamefont {Liu}, \citenamefont
  {Sargent},\ and\ \citenamefont {Gao}}]{long_chiral-perovskite_2020}%
  \BibitemOpen
  \bibfield  {author} {\bibinfo {author} {\bibfnamefont {G.}~\bibnamefont
  {Long}}, \bibinfo {author} {\bibfnamefont {R.}~\bibnamefont {Sabatini}},
  \bibinfo {author} {\bibfnamefont {M.~I.}\ \bibnamefont {Saidaminov}},
  \bibinfo {author} {\bibfnamefont {G.}~\bibnamefont {Lakhwani}}, \bibinfo
  {author} {\bibfnamefont {A.}~\bibnamefont {Rasmita}}, \bibinfo {author}
  {\bibfnamefont {X.}~\bibnamefont {Liu}}, \bibinfo {author} {\bibfnamefont
  {E.~H.}\ \bibnamefont {Sargent}},\ and\ \bibinfo {author} {\bibfnamefont
  {W.}~\bibnamefont {Gao}},\ }\bibfield  {title} {\bibinfo {title}
  {Chiral-perovskite optoelectronics},\ }\bibfield  {journal} {\bibinfo
  {journal} {Nature Reviews Materials}\ }\textbf {\bibinfo {volume} {5}},\
  \href {https://doi.org/10.1038/s41578-020-0181-5} {10.1038/s41578-020-0181-5}
  (\bibinfo {year} {2020}),\ \bibinfo {note} {number: 6 pages =
  {423--439},}\BibitemShut {NoStop}%
\bibitem [{\citenamefont {Morell}\ \emph {et~al.}(2017)\citenamefont {Morell},
  \citenamefont {Chico},\ and\ \citenamefont {Brey}}]{Suarez_Chico_2017}%
  \BibitemOpen
  \bibfield  {author} {\bibinfo {author} {\bibfnamefont {E.~S.}\ \bibnamefont
  {Morell}}, \bibinfo {author} {\bibfnamefont {L.}~\bibnamefont {Chico}},\ and\
  \bibinfo {author} {\bibfnamefont {L.}~\bibnamefont {Brey}},\ }\bibfield
  {title} {\bibinfo {title} {Twisting dirac fermions: circular dichroism in
  bilayer graphene},\ }\href {https://doi.org/10.1088/2053-1583/aa7eb6}
  {\bibfield  {journal} {\bibinfo  {journal} {2D Materials}\ }\textbf {\bibinfo
  {volume} {4}},\ \bibinfo {pages} {035015} (\bibinfo {year}
  {2017})}\BibitemShut {NoStop}%
\bibitem [{\citenamefont {Addison}\ \emph {et~al.}(2019)\citenamefont
  {Addison}, \citenamefont {Park},\ and\ \citenamefont
  {Mele}}]{slip_twist_PRB_2019}%
  \BibitemOpen
  \bibfield  {author} {\bibinfo {author} {\bibfnamefont {Z.}~\bibnamefont
  {Addison}}, \bibinfo {author} {\bibfnamefont {J.}~\bibnamefont {Park}},\ and\
  \bibinfo {author} {\bibfnamefont {E.~J.}\ \bibnamefont {Mele}},\ }\bibfield
  {title} {\bibinfo {title} {Twist, slip, and circular dichroism in bilayer
  graphene},\ }\href {https://doi.org/10.1103/PhysRevB.100.125418} {\bibfield
  {journal} {\bibinfo  {journal} {Phys. Rev. B}\ }\textbf {\bibinfo {volume}
  {100}},\ \bibinfo {pages} {125418} (\bibinfo {year} {2019})}\BibitemShut
  {NoStop}%
\bibitem [{\citenamefont {Tan}\ \emph {et~al.}(2016)\citenamefont {Tan},
  \citenamefont {Zheng}, \citenamefont {Young}, \citenamefont {Wang},
  \citenamefont {Liu},\ and\ \citenamefont {Rappe}}]{tan_shift_2016}%
  \BibitemOpen
  \bibfield  {author} {\bibinfo {author} {\bibfnamefont {L.~Z.}\ \bibnamefont
  {Tan}}, \bibinfo {author} {\bibfnamefont {F.}~\bibnamefont {Zheng}}, \bibinfo
  {author} {\bibfnamefont {S.~M.}\ \bibnamefont {Young}}, \bibinfo {author}
  {\bibfnamefont {F.}~\bibnamefont {Wang}}, \bibinfo {author} {\bibfnamefont
  {S.}~\bibnamefont {Liu}},\ and\ \bibinfo {author} {\bibfnamefont {A.~M.}\
  \bibnamefont {Rappe}},\ }\bibfield  {title} {\bibinfo {title} {Shift current
  bulk photovoltaic effect in polar materials—hybrid and oxide perovskites
  and beyond},\ }\href {https://doi.org/10.1038/npjcompumats.2016.26}
  {\bibfield  {journal} {\bibinfo  {journal} {npj Comput Mater}\ }\textbf
  {\bibinfo {volume} {2}},\ \bibinfo {pages} {16026} (\bibinfo {year}
  {2016})}\BibitemShut {NoStop}%
\bibitem [{\citenamefont {Xu}\ \emph {et~al.}(2020)\citenamefont {Xu},
  \citenamefont {Zhang}, \citenamefont {Koepernik}, \citenamefont {Shi},
  \citenamefont {van~den Brink}, \citenamefont {Felser},\ and\ \citenamefont
  {Sun}}]{xu_comprehensive_2020}%
  \BibitemOpen
  \bibfield  {author} {\bibinfo {author} {\bibfnamefont {Q.}~\bibnamefont
  {Xu}}, \bibinfo {author} {\bibfnamefont {Y.}~\bibnamefont {Zhang}}, \bibinfo
  {author} {\bibfnamefont {K.}~\bibnamefont {Koepernik}}, \bibinfo {author}
  {\bibfnamefont {W.}~\bibnamefont {Shi}}, \bibinfo {author} {\bibfnamefont
  {J.}~\bibnamefont {van~den Brink}}, \bibinfo {author} {\bibfnamefont
  {C.}~\bibnamefont {Felser}},\ and\ \bibinfo {author} {\bibfnamefont
  {Y.}~\bibnamefont {Sun}},\ }\bibfield  {title} {\bibinfo {title}
  {Comprehensive scan for nonmagnetic {Weyl} semimetals with nonlinear optical
  response},\ }\href {https://doi.org/10.1038/s41524-020-0301-1} {\bibfield
  {journal} {\bibinfo  {journal} {npj Comput Mater}\ }\textbf {\bibinfo
  {volume} {6}},\ \bibinfo {pages} {32} (\bibinfo {year} {2020})}\BibitemShut
  {NoStop}%
\bibitem [{\citenamefont {Le}\ \emph {et~al.}(2020)\citenamefont {Le},
  \citenamefont {Zhang}, \citenamefont {Felser},\ and\ \citenamefont
  {Sun}}]{le_ab_2020}%
  \BibitemOpen
  \bibfield  {author} {\bibinfo {author} {\bibfnamefont {C.}~\bibnamefont
  {Le}}, \bibinfo {author} {\bibfnamefont {Y.}~\bibnamefont {Zhang}}, \bibinfo
  {author} {\bibfnamefont {C.}~\bibnamefont {Felser}},\ and\ \bibinfo {author}
  {\bibfnamefont {Y.}~\bibnamefont {Sun}},\ }\bibfield  {title} {\bibinfo
  {title} {\textit{{Ab} initio} study of quantized circular photogalvanic
  effect in chiral multifold semimetals},\ }\href
  {https://doi.org/10.1103/PhysRevB.102.121111} {\bibfield  {journal} {\bibinfo
   {journal} {Phys. Rev. B}\ }\textbf {\bibinfo {volume} {102}},\ \bibinfo
  {pages} {121111} (\bibinfo {year} {2020})}\BibitemShut {NoStop}%
\bibitem [{\citenamefont {Ni}\ \emph {et~al.}(2020)\citenamefont {Ni},
  \citenamefont {Xu}, \citenamefont {Sanchez-Martinez}, \citenamefont {Zhang},
  \citenamefont {Manna}, \citenamefont {Bernhard}, \citenamefont {Venderbos},
  \citenamefont {de~Juan}, \citenamefont {Felser}, \citenamefont {Grushin},\
  and\ \citenamefont {Wu}}]{ni_linear_2020}%
  \BibitemOpen
  \bibfield  {author} {\bibinfo {author} {\bibfnamefont {Z.}~\bibnamefont
  {Ni}}, \bibinfo {author} {\bibfnamefont {B.}~\bibnamefont {Xu}}, \bibinfo
  {author} {\bibfnamefont {M.~A.}\ \bibnamefont {Sanchez-Martinez}}, \bibinfo
  {author} {\bibfnamefont {Y.}~\bibnamefont {Zhang}}, \bibinfo {author}
  {\bibfnamefont {K.}~\bibnamefont {Manna}}, \bibinfo {author} {\bibfnamefont
  {C.}~\bibnamefont {Bernhard}}, \bibinfo {author} {\bibfnamefont {J.~W.~F.}\
  \bibnamefont {Venderbos}}, \bibinfo {author} {\bibfnamefont {F.}~\bibnamefont
  {de~Juan}}, \bibinfo {author} {\bibfnamefont {C.}~\bibnamefont {Felser}},
  \bibinfo {author} {\bibfnamefont {A.~G.}\ \bibnamefont {Grushin}},\ and\
  \bibinfo {author} {\bibfnamefont {L.}~\bibnamefont {Wu}},\ }\bibfield
  {title} {\bibinfo {title} {Linear and nonlinear optical responses in the
  chiral multifold semimetal {RhSi}},\ }\href {http://arxiv.org/abs/2005.13473}
  {\bibfield  {journal} {\bibinfo  {journal} {arXiv:2005.13473 [cond-mat,
  physics:physics]}\ } (\bibinfo {year} {2020})},\ \bibinfo {note} {arXiv:
  2005.13473}\BibitemShut {NoStop}%
\bibitem [{\citenamefont {Cook}\ \emph {et~al.}(2017)\citenamefont {Cook},
  \citenamefont {M.~Fregoso}, \citenamefont {de~Juan}, \citenamefont {Coh},\
  and\ \citenamefont {Moore}}]{cook_design_2017}%
  \BibitemOpen
  \bibfield  {author} {\bibinfo {author} {\bibfnamefont {A.~M.}\ \bibnamefont
  {Cook}}, \bibinfo {author} {\bibfnamefont {B.}~\bibnamefont {M.~Fregoso}},
  \bibinfo {author} {\bibfnamefont {F.}~\bibnamefont {de~Juan}}, \bibinfo
  {author} {\bibfnamefont {S.}~\bibnamefont {Coh}},\ and\ \bibinfo {author}
  {\bibfnamefont {J.~E.}\ \bibnamefont {Moore}},\ }\bibfield  {title} {\bibinfo
  {title} {Design principles for shift current photovoltaics},\ }\href
  {https://doi.org/10.1038/ncomms14176} {\bibfield  {journal} {\bibinfo
  {journal} {Nat Commun}\ }\textbf {\bibinfo {volume} {8}},\ \bibinfo {pages}
  {14176} (\bibinfo {year} {2017})}\BibitemShut {NoStop}%
\bibitem [{\citenamefont {Gallego}\ \emph {et~al.}(2019)\citenamefont
  {Gallego}, \citenamefont {Etxebarria}, \citenamefont {Elcoro}, \citenamefont
  {Tasci},\ and\ \citenamefont {Perez-Mato}}]{Gallego_tensor}%
  \BibitemOpen
  \bibfield  {author} {\bibinfo {author} {\bibfnamefont {S.~V.}\ \bibnamefont
  {Gallego}}, \bibinfo {author} {\bibfnamefont {J.}~\bibnamefont {Etxebarria}},
  \bibinfo {author} {\bibfnamefont {L.}~\bibnamefont {Elcoro}}, \bibinfo
  {author} {\bibfnamefont {E.~S.}\ \bibnamefont {Tasci}},\ and\ \bibinfo
  {author} {\bibfnamefont {J.~M.}\ \bibnamefont {Perez-Mato}},\ }\bibfield
  {title} {\bibinfo {title} {{Automatic calculation of symmetry-adapted tensors
  in magnetic and non-magnetic materials: a new tool of the Bilbao
  Crystallographic Server}},\ }\href
  {https://doi.org/10.1107/S2053273319001748} {\bibfield  {journal} {\bibinfo
  {journal} {Acta Crystallographica Section A}\ }\textbf {\bibinfo {volume}
  {75}},\ \bibinfo {pages} {438} (\bibinfo {year} {2019})}\BibitemShut
  {NoStop}%
\bibitem [{\citenamefont {Rangel}\ \emph {et~al.}(2017)\citenamefont {Rangel},
  \citenamefont {Fregoso}, \citenamefont {Mendoza}, \citenamefont {Morimoto},
  \citenamefont {Moore},\ and\ \citenamefont {Neaton}}]{Rangel_GeS}%
  \BibitemOpen
  \bibfield  {author} {\bibinfo {author} {\bibfnamefont {T.}~\bibnamefont
  {Rangel}}, \bibinfo {author} {\bibfnamefont {B.~M.}\ \bibnamefont {Fregoso}},
  \bibinfo {author} {\bibfnamefont {B.~S.}\ \bibnamefont {Mendoza}}, \bibinfo
  {author} {\bibfnamefont {T.}~\bibnamefont {Morimoto}}, \bibinfo {author}
  {\bibfnamefont {J.~E.}\ \bibnamefont {Moore}},\ and\ \bibinfo {author}
  {\bibfnamefont {J.~B.}\ \bibnamefont {Neaton}},\ }\bibfield  {title}
  {\bibinfo {title} {Large bulk photovoltaic effect and spontaneous
  polarization of single-layer monochalcogenides},\ }\href
  {https://doi.org/10.1103/PhysRevLett.119.067402} {\bibfield  {journal}
  {\bibinfo  {journal} {Phys. Rev. Lett.}\ }\textbf {\bibinfo {volume} {119}},\
  \bibinfo {pages} {067402} (\bibinfo {year} {2017})}\BibitemShut {NoStop}%
\bibitem [{\citenamefont {Zhang}\ \emph {et~al.}(2018)\citenamefont {Zhang},
  \citenamefont {Ishizuka}, \citenamefont {van~den Brink}, \citenamefont
  {Felser}, \citenamefont {Yan},\ and\ \citenamefont
  {Nagaosa}}]{zhang_photogalvanic_2018}%
  \BibitemOpen
  \bibfield  {author} {\bibinfo {author} {\bibfnamefont {Y.}~\bibnamefont
  {Zhang}}, \bibinfo {author} {\bibfnamefont {H.}~\bibnamefont {Ishizuka}},
  \bibinfo {author} {\bibfnamefont {J.}~\bibnamefont {van~den Brink}}, \bibinfo
  {author} {\bibfnamefont {C.}~\bibnamefont {Felser}}, \bibinfo {author}
  {\bibfnamefont {B.}~\bibnamefont {Yan}},\ and\ \bibinfo {author}
  {\bibfnamefont {N.}~\bibnamefont {Nagaosa}},\ }\bibfield  {title} {\bibinfo
  {title} {Photogalvanic effect in {Weyl} semimetals from first principles},\
  }\href {https://doi.org/10.1103/PhysRevB.97.241118} {\bibfield  {journal}
  {\bibinfo  {journal} {Phys. Rev. B}\ }\textbf {\bibinfo {volume} {97}},\
  \bibinfo {pages} {241118} (\bibinfo {year} {2018})}\BibitemShut {NoStop}%
\bibitem [{\citenamefont {Zhang}\ \emph {et~al.}(2019)\citenamefont {Zhang},
  \citenamefont {de~Juan}, \citenamefont {Grushin}, \citenamefont {Felser},\
  and\ \citenamefont {Sun}}]{zhang_strong_2019}%
  \BibitemOpen
  \bibfield  {author} {\bibinfo {author} {\bibfnamefont {Y.}~\bibnamefont
  {Zhang}}, \bibinfo {author} {\bibfnamefont {F.}~\bibnamefont {de~Juan}},
  \bibinfo {author} {\bibfnamefont {A.~G.}\ \bibnamefont {Grushin}}, \bibinfo
  {author} {\bibfnamefont {C.}~\bibnamefont {Felser}},\ and\ \bibinfo {author}
  {\bibfnamefont {Y.}~\bibnamefont {Sun}},\ }\bibfield  {title} {\bibinfo
  {title} {Strong bulk photovoltaic effect in chiral crystals in the visible
  spectrum},\ }\href {https://doi.org/10.1103/PhysRevB.100.245206} {\bibfield
  {journal} {\bibinfo  {journal} {Phys. Rev. B}\ }\textbf {\bibinfo {volume}
  {100}},\ \bibinfo {pages} {245206} (\bibinfo {year} {2019})}\BibitemShut
  {NoStop}%
\bibitem [{\citenamefont {Young}\ and\ \citenamefont
  {Rappe}(2012)}]{young_first_2012}%
  \BibitemOpen
  \bibfield  {author} {\bibinfo {author} {\bibfnamefont {S.~M.}\ \bibnamefont
  {Young}}\ and\ \bibinfo {author} {\bibfnamefont {A.~M.}\ \bibnamefont
  {Rappe}},\ }\bibfield  {title} {\bibinfo {title} {First {Principles}
  {Calculation} of the {Shift} {Current} {Photovoltaic} {Effect} in
  {Ferroelectrics}},\ }\href {https://doi.org/10.1103/PhysRevLett.109.116601}
  {\bibfield  {journal} {\bibinfo  {journal} {Phys. Rev. Lett.}\ }\textbf
  {\bibinfo {volume} {109}},\ \bibinfo {pages} {116601} (\bibinfo {year}
  {2012})}\BibitemShut {NoStop}%
\bibitem [{\citenamefont {Young}\ \emph {et~al.}(2012)\citenamefont {Young},
  \citenamefont {Zheng},\ and\ \citenamefont
  {Rappe}}]{young_first-principles_2012}%
  \BibitemOpen
  \bibfield  {author} {\bibinfo {author} {\bibfnamefont {S.~M.}\ \bibnamefont
  {Young}}, \bibinfo {author} {\bibfnamefont {F.}~\bibnamefont {Zheng}},\ and\
  \bibinfo {author} {\bibfnamefont {A.~M.}\ \bibnamefont {Rappe}},\ }\bibfield
  {title} {\bibinfo {title} {First-{Principles} {Calculation} of the {Bulk}
  {Photovoltaic} {Effect} in {Bismuth} {Ferrite}},\ }\href
  {https://doi.org/10.1103/PhysRevLett.109.236601} {\bibfield  {journal}
  {\bibinfo  {journal} {Phys. Rev. Lett.}\ }\textbf {\bibinfo {volume} {109}},\
  \bibinfo {pages} {236601} (\bibinfo {year} {2012})}\BibitemShut {NoStop}%
\bibitem [{\citenamefont {Zenkevich}\ \emph {et~al.}(2014)\citenamefont
  {Zenkevich}, \citenamefont {Matveyev}, \citenamefont {Maksimova},
  \citenamefont {Gaynutdinov}, \citenamefont {Tolstikhina},\ and\ \citenamefont
  {Fridkin}}]{zenkevich_giant_2014}%
  \BibitemOpen
  \bibfield  {author} {\bibinfo {author} {\bibfnamefont {A.}~\bibnamefont
  {Zenkevich}}, \bibinfo {author} {\bibfnamefont {Y.}~\bibnamefont {Matveyev}},
  \bibinfo {author} {\bibfnamefont {K.}~\bibnamefont {Maksimova}}, \bibinfo
  {author} {\bibfnamefont {R.}~\bibnamefont {Gaynutdinov}}, \bibinfo {author}
  {\bibfnamefont {A.}~\bibnamefont {Tolstikhina}},\ and\ \bibinfo {author}
  {\bibfnamefont {V.}~\bibnamefont {Fridkin}},\ }\bibfield  {title} {\bibinfo
  {title} {Giant bulk photovoltaic effect in thin ferroelectric {BaTiO} 3
  films},\ }\href {https://doi.org/10.1103/PhysRevB.90.161409} {\bibfield
  {journal} {\bibinfo  {journal} {Phys. Rev. B}\ }\textbf {\bibinfo {volume}
  {90}},\ \bibinfo {pages} {161409} (\bibinfo {year} {2014})}\BibitemShut
  {NoStop}%
\bibitem [{\citenamefont {Bir}\ and\ \citenamefont
  {Pikus}(1974)}]{bir1974symmetry}%
  \BibitemOpen
  \bibfield  {author} {\bibinfo {author} {\bibfnamefont {G.}~\bibnamefont
  {Bir}}\ and\ \bibinfo {author} {\bibfnamefont {G.}~\bibnamefont {Pikus}},\
  }\href@noop {} {\emph {\bibinfo {title} {Symmetry and Strain-induced Effects
  in Semiconductors}}},\ A Halsted Press book\ (\bibinfo  {publisher} {Wiley},\
  \bibinfo {year} {1974})\BibitemShut {NoStop}%
\end{thebibliography}%


\begin{thebibliography}{10}

\bibitem{QESPRESSO2020}
Paolo Giannozzi, Oscar Baseggio, Pietro Bonfà, Davide Brunato, Roberto Car,
  Ivan Carnimeo, Carlo Cavazzoni, Stefano de~Gironcoli, Pietro Delugas,
  Fabrizio Ferrari~Ruffino, Andrea Ferretti, Nicola Marzari, Iurii Timrov,
  Andrea Urru, and Stefano Baroni.
\newblock Quantum espresso toward the exascale.
\newblock {\em The Journal of Chemical Physics}, 152(15):154105, 2020.

\bibitem{GPAW_paper2010}
JE~Enkovaara, Carsten Rostgaard, J~J{\o}rgen Mortensen, Jingzhe Chen,
  M~Du{\l}ak, Lara Ferrighi, Jeppe Gavnholt, Christian Glinsvad, V~Haikola,
  HA~Hansen, et~al.
\newblock Electronic structure calculations with gpaw: a real-space
  implementation of the projector augmented-wave method.
\newblock {\em Journal of Physics: Condensed Matter}, 22(25):253202, 2010.

\bibitem{Real-spacegrid_2005}
Jens~J{\o}rgen Mortensen, Lars~Bruno Hansen, and Karsten~Wedel Jacobsen.
\newblock Real-space grid implementation of the projector augmented wave
  method.
\newblock {\em Physical Review B}, 71(3):035109, 2005.

\bibitem{ASE1_2002}
S.~R. Bahn and K.~W. Jacobsen.
\newblock An object-oriented scripting interface to a legacy electronic
  structure code.
\newblock {\em Comput. Sci. Eng.}, 4(3):56--66, MAY-JUN 2002.

\bibitem{ASE2_2017}
Ask~Hjorth Larsen, Jens~Jørgen Mortensen, Jakob Blomqvist, Ivano~E Castelli,
  Rune Christensen, Marcin Dułak, Jesper Friis, Michael~N Groves, Bjørk
  Hammer, Cory Hargus, Eric~D Hermes, Paul~C Jennings, Peter~Bjerre Jensen,
  James Kermode, John~R Kitchin, Esben~Leonhard Kolsbjerg, Joseph Kubal,
  Kristen Kaasbjerg, Steen Lysgaard, Jón~Bergmann Maronsson, Tristan Maxson,
  Thomas Olsen, Lars Pastewka, Andrew Peterson, Carsten Rostgaard, Jakob
  Schiøtz, Ole Schütt, Mikkel Strange, Kristian~S Thygesen, Tejs Vegge, Lasse
  Vilhelmsen, Michael Walter, Zhenhua Zeng, and Karsten~W Jacobsen.
\newblock The atomic simulation environment—a python library for working with
  atoms.
\newblock {\em Journal of Physics: Condensed Matter}, 29(27):273002, 2017.

\bibitem{PBE_1996}
John~P Perdew, Kieron Burke, and Matthias Ernzerhof.
\newblock Generalized gradient approximation made simple.
\newblock {\em Physical review letters}, 77(18):3865, 1996.

\bibitem{HSE06_2003}
Jochen Heyd, Gustavo~E Scuseria, and Matthias Ernzerhof.
\newblock Hybrid functionals based on a screened coulomb potential.
\newblock {\em The Journal of chemical physics}, 118(18):8207--8215, 2003.

\bibitem{Phonon_LABS_2009}
A.~H. Larsen, M.~Vanin, J.~J. Mortensen, K.~S. Thygesen, and K.~W. Jacobsen.
\newblock Localized atomic basis set in the projector augmented wave method.
\newblock {\em Phys. Rev. B}, 80(19):195112, 2009.

\bibitem{spglib_phyton_2018}
Atsushi {Togo} and Isao {Tanaka}.
\newblock {$\texttt{Spglib}$: a software library for crystal symmetry search}.
\newblock {\em arXiv e-prints}, page arXiv:1808.01590, August 2018.

\bibitem{FindSymm_2005}
Harold~T. Stokes and Dorian~M. Hatch.
\newblock {{\it FINDSYM}: program for identifying the space-group symmetry of a
  crystal}.
\newblock {\em Journal of Applied Crystallography}, 38(1):237--238, Feb 2005.

\end{thebibliography}

\end{document}


\author{\text{Sergio Bravo}$^\dagger$}
\author{\text{M. Pacheco}$^\mathsection$}
\author{V. Nuñez}%
\affil{Departamento de F\' isica, Universidad T\'ecnica Federico Santa Mar\' ia, Valpara\' iso, Chile}

\author{J.D. Correa}%
\affil{Facultad de Ciencias  B\'asicas, Universidad de Medell\' \i n, 
Medell\' \i n, Colombia}

\author{Leonor Chico$^\star$}%
\affil{Departamento de Física de Materiales, 
Facultad de Ciencias Físicas, 
Universidad Complutense de Madrid, 
28040 Madrid, Spain}
\affil[ ]{ }
\affil[$\dagger$]{sergio.bravo.14@sansano.usm.cl}
\affil[$\mathsection$]{monica.pacheco@usm.cl}
\affil[$\star$]{leochico@ucm.es}
\date{}

\maketitle

\bigskip

Outline: The following supplementary material is divided in two parts. Part A presents
extended information for the electronic, dynamical and optical calculations for all 
the materials presented in the main article. Part B deals briefly with the extension of
the space group characterization for multilayer pentagonal structures, with two 
and three layers. 

\bigskip

\subsection*{Part A: Additional information for pentagonal materials}

We present further details for the first-principles
results related to the novel pentagonal materials proposed in the main article.
In particular, the geometric parameters and
the band structure with GGA (PBE) and hybrid functionals are reported along
with additional spin textures for each material. Also, the phonon band 
structure is given for the novel materials first proposed in this work. 
Finally, the optical responses 
not included in the 
main article are presented here. 

\subsubsection*{Geometric properties of novel pentagonal materials}

\begin{figure}[H]
    \includegraphics[width=1.1\textwidth,center]{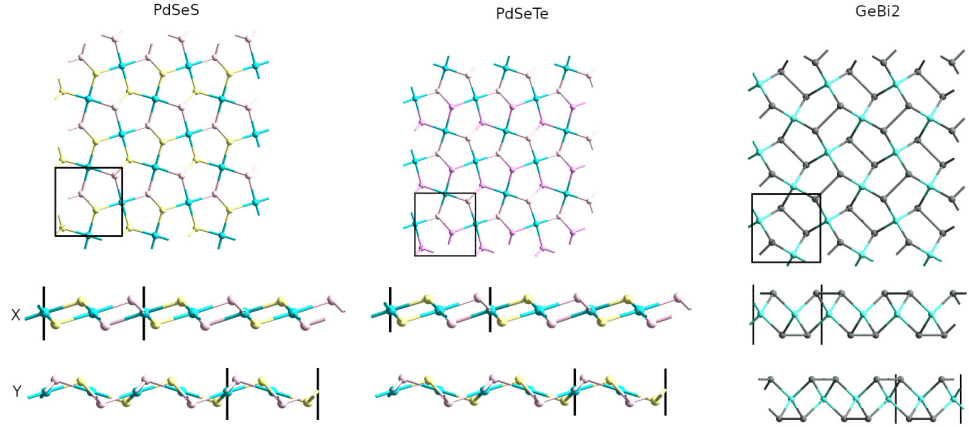}
    \caption{Lattice 
structure for the pentagonal materials presented in the main article.}
\label{FS1_GPAW_latt}
\end{figure}

The structural parameters were calculated with two first-principles packages: 
QUANTUM ESPRESSO (QE) \cite{QESPRESSO2020} and GPAW \cite{GPAW_paper2010}. 
The reason for this is the possibility to 
calculate Wannier-interpolated bands with QE, and on the other side, to 
compute hybrid and phonon band structures with GPAW. We have checked 
carefully that both codes give the same electronic band structures for all 
the materials reported. Although relaxed geometric parameters could differ between 
the two codes, the difference is under the expected bounds related to the 
different numerical procedures that are applied. The GPAW relaxed lattice
structures for PdSeTe PdSeS and GeBi$_2$ are depicted in Fig. \ref{FS1_GPAW_latt}. 
The geometric parameters for both QE and GPAW are presented in Table \ref{Table_GPAW}
and Table \ref{Table_QE}, respectively.      

\subsubsection*{Computational details for GPAW calculations}

\begin{figure}[H]
    \includegraphics[width=0.6\textwidth,center]{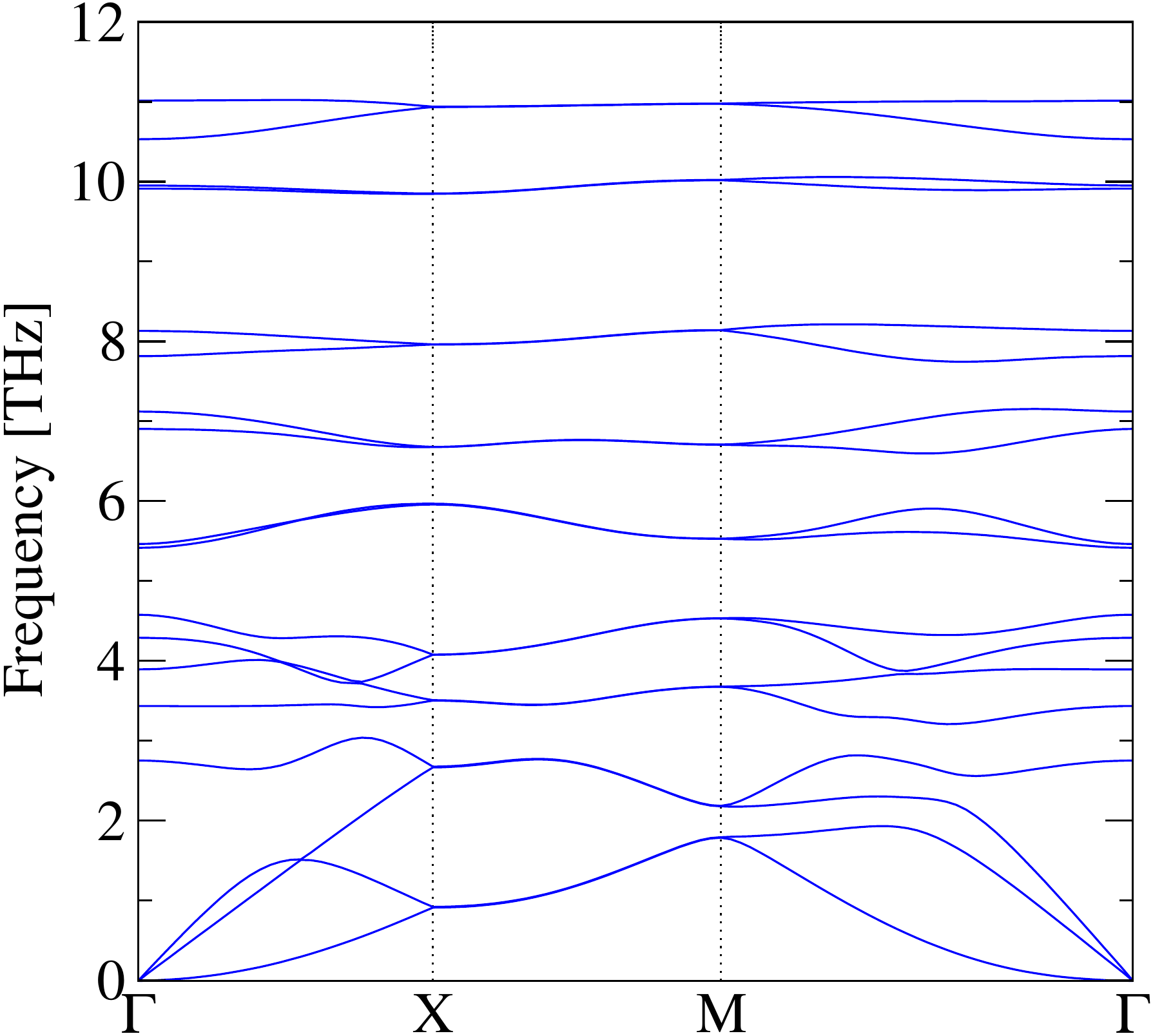}
    \caption{Phonon band structure for PdSeS from GPAW.}
\label{PdSeS_phonons}
\end{figure}

\begin{figure}[H]
    \includegraphics[width=0.6\textwidth,center]{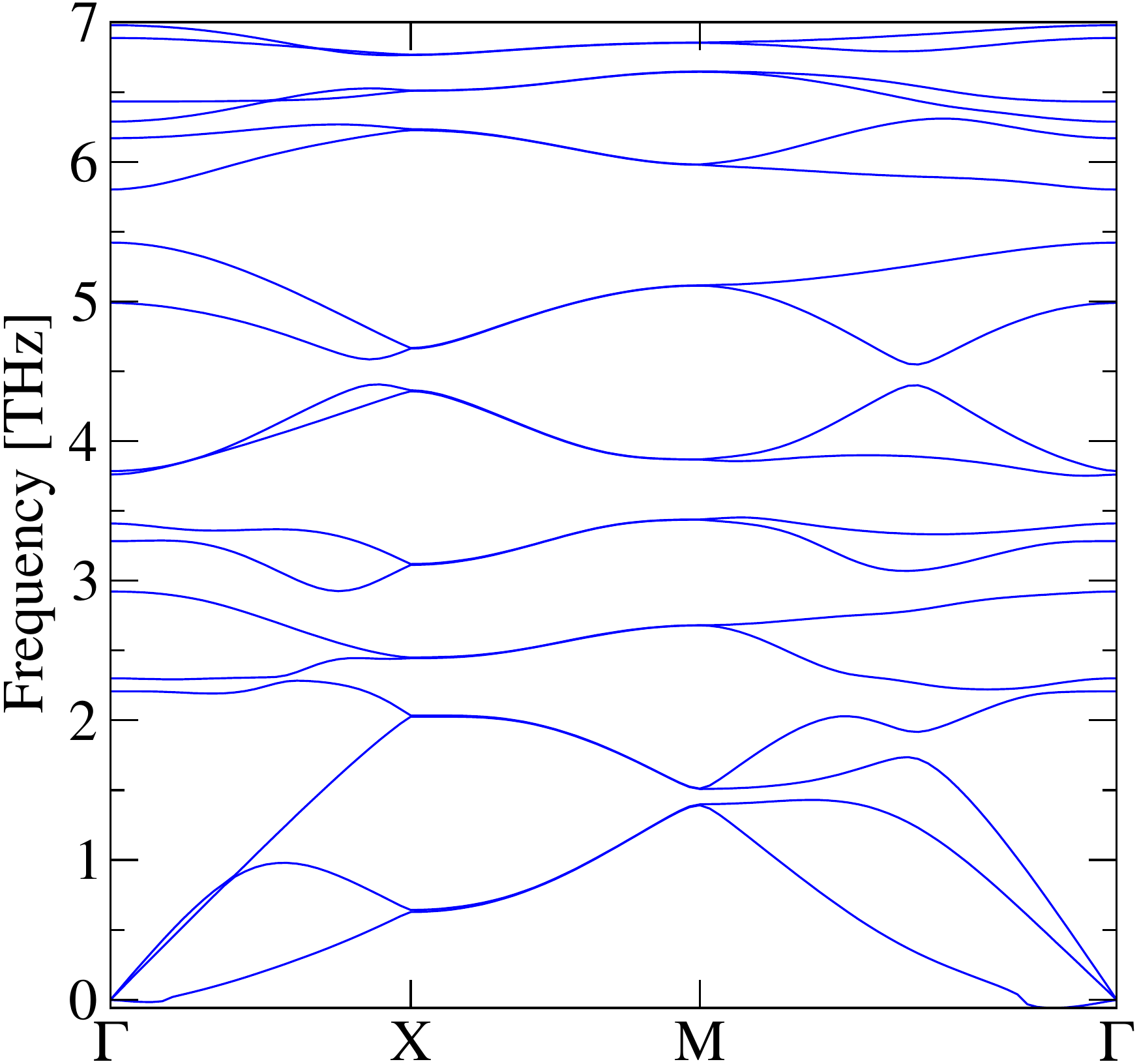}
    \caption{Phonon band structure for PdSeTe from GPAW.}
\label{PdSeTe_phonons}
\end{figure}

\begin{figure}[H]
    \includegraphics[width=0.6\textwidth,center]{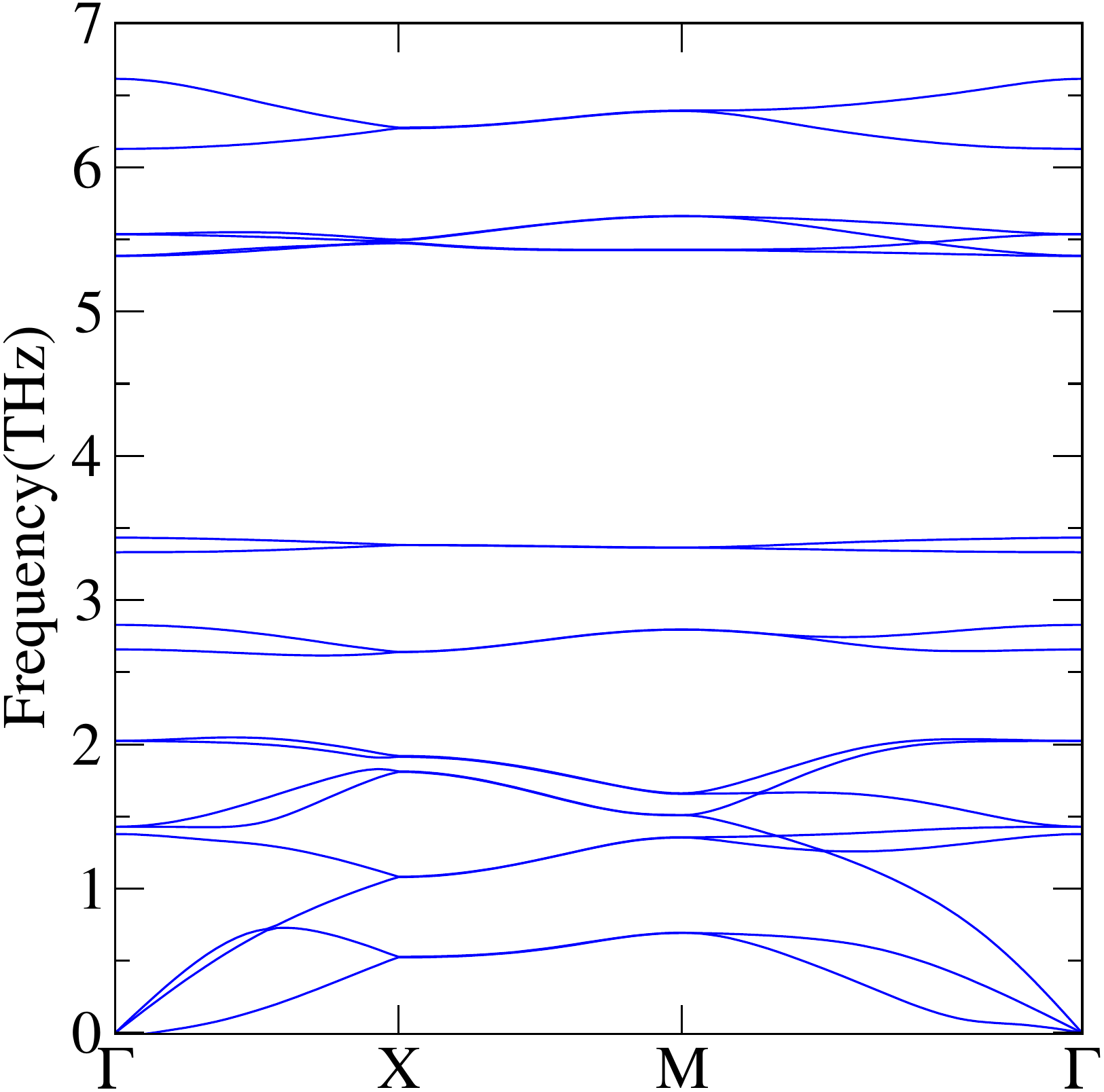}
    \caption{Phonon band structure for GeBi$_2$ from GPAW.}
\label{GeBi2_phonons}
\end{figure}

For the electronic structure calculations the real-space projector
augmented wavefunction method was employed, \cite{Real-spacegrid_2005}
together with the atomic simulation environment (ASE) \cite{ASE1_2002,ASE2_2017}.
In order to check the QE results, the first exchange-correlation functional 
used was a PBE functional \cite{PBE_1996}. 
There is a very good agreement between the outcomes of both codes. 
This gives way to perform further computations with additional features. 
Thus, additional band structure calculations with a hybrid functional in 
the form of the HSE06 implementation \cite{HSE06_2003} 
were obtained for PdSeS, PdSeTe and GeBi$_2$ as presented in
Fig. \ref{PdSeS_Hybrid}, Fig. \ref{PdSeTe_hybrid} and Fig. \ref{GeBi2_hybrid}, respectively. 
The energy cutoff was set to $850$ eV and the Brillouin zone was
sampled via a $10\times10\times1$ Monkhorst-Pack k-grid.
A vacuum space of 15 \AA\ in the direction normal to the monolayer plane
was used and the unit cell was relaxed until the atomic forces were
less than 0.01 eV/atom. In addition to these electronic band calculations, 
we obtained the phonon dispersions. For this we employed a
LCAO mode computation \cite{Phonon_LABS_2009} with localized double-$\zeta$ and 
single-polarized atomic orbitals, with a $5\times 5\times 1$ supercell. 
Phonon bands for PdSeS, PdSeTe and GeBi$_2$ are depicted in 
Fig. \ref{PdSeS_phonons}, Fig. \ref{PdSeTe_phonons} and \ref{GeBi2_phonons}, 
respectively. Results show that all the materials presented are dynamically stable. 

\begin{table}[ht]
\centering{}%
\begin{tabular}{ccccccc}
\hline 
\textbf{Material} & \textbf{a}(\AA) & \textbf{b}(\AA) & \textbf{d}(\AA) & \multicolumn{3}{c}{\textbf{Bonds}(\AA)}\tabularnewline
\hline 
\multirow{2}{*}{PdSeS} & \multirow{2}{*}{5.619} & \multirow{2}{*}{5.754} & \multirow{2}{*}{1.469} & \textbf{Pd-S} & \textbf{Pd-Se} & \textbf{Se-S}\tabularnewline
\cline{5-7} \cline{6-7} \cline{7-7} 
 &  &  &  & 2.351 & 2.445 & 2.278\tabularnewline
\multirow{2}{*}{PdSeTe} & \multirow{2}{*}{5.939} & \multirow{2}{*}{6.168} & \multirow{2}{*}{1.663} & \textbf{Pd-Te} & \textbf{Pd-Se} & \textbf{Se-Te}\tabularnewline
\cline{5-7} \cline{6-7} \cline{7-7} 
 &  &  &  & 2.595 & 2.486 & 2.604\tabularnewline
\multirow{2}{*}{GeBi$_{2}$} & \multirow{2}{*}{5.831} & \multirow{2}{*}{5.831} & \multirow{2}{*}{3.558} &  & \textbf{Ge-Bi} & \tabularnewline
\cline{5-7} \cline{6-7} \cline{7-7} 
 &  &  &  &  & 2.775 & \tabularnewline
\hline 
\end{tabular}\caption{Structural parameters for PdSeS, PdSeTe and GeBi$_2$ extracted from GPAW. 
 $\boldsymbol{a}$ and $\boldsymbol{b}$ are the lattice vectors in the $x$ and $y$ directions,
 respectively; $\boldsymbol{d}$ is the distance between the bottom and top atoms. The bond length 
 is given in the last column.}
\label{Table_GPAW}
\end{table}

\begin{table}[ht]
\centering{}%
\begin{tabular}{ccccccc}
\hline 
\textbf{Material} & \textbf{a}(\AA) & \textbf{b}(\AA) & \textbf{d}(\AA) & \multicolumn{3}{c}{\textbf{Bonds}(\AA)}\tabularnewline
\hline 
\multirow{2}{*}{PdSeS} & \multirow{2}{*}{5.616} & \multirow{2}{*}{5.752} & \multirow{2}{*}{1.468} & \textbf{Pd-S} & \textbf{Pd-Se} & \textbf{Se-S}\tabularnewline
\cline{5-7} \cline{6-7} \cline{7-7} 
 &  &  &  & 2.350 & 2.444 & 2.277\tabularnewline
\multirow{2}{*}{PdSeTe} & \multirow{2}{*}{5.951} & \multirow{2}{*}{6.183} & \multirow{2}{*}{1.664} & \textbf{Pd-Te} & \textbf{Pd-Se} & \textbf{Se-Te}\tabularnewline
\cline{5-7} \cline{6-7} \cline{7-7} 
 &  &  &  & 2.610 & 2.481 & 2.608\tabularnewline
\multirow{2}{*}{GeBi$_{2}$} & \multirow{2}{*}{5.849} & \multirow{2}{*}{5.849} & \multirow{2}{*}{3.500} &   \textbf{Ge-Bi} & \textbf{Bi-Bi}\tabularnewline
\cline{5-7} \cline{6-7} \cline{7-7} 
 &  &  &  & 2.776 & 3.016 & \tabularnewline
\hline 
\end{tabular}\caption{Structural parameters for PdSeS, PdSeTe and GeBi$_2$ extracted from QE. 
 $\boldsymbol{a}$ and $\boldsymbol{b}$ are the lattice vectors in the $x$ and $y$ directions,
 respectively; $\boldsymbol{d}$ is the distance between the bottom and top atoms. The bond length 
 is given in the last column.}
\label{Table_QE}
\end{table}

\subsubsection*{PdSeS information}

\begin{figure}[H]
    \includegraphics[width=1.0\textwidth,center]{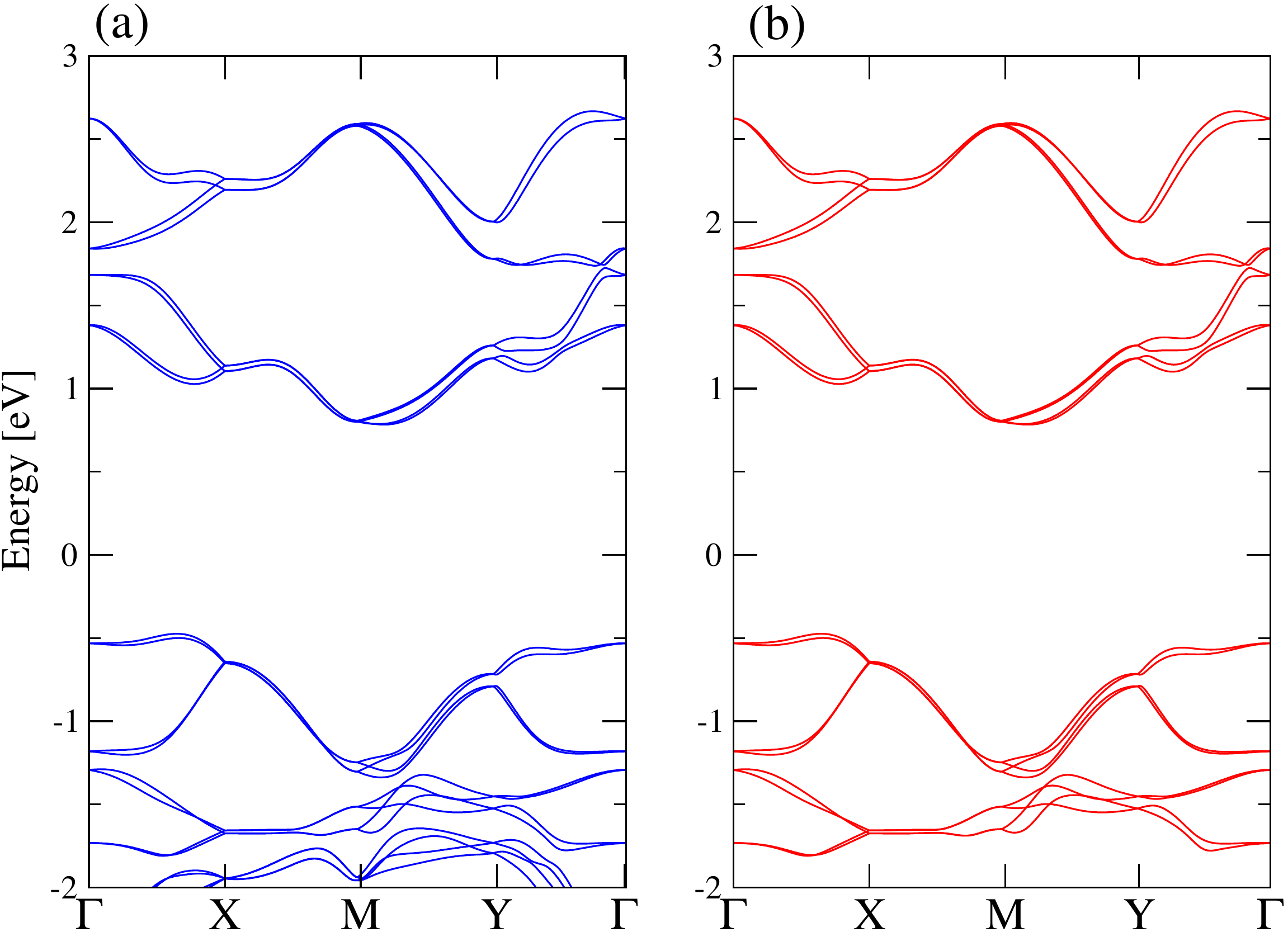}
    \caption{Electronic band structure for PdSeS from a) QE and b) Wannier interpolation.}
\label{PdSeS_bandsQE_w90}
\end{figure}

\begin{figure}[H]
    \includegraphics[width=1.1\textwidth,center]{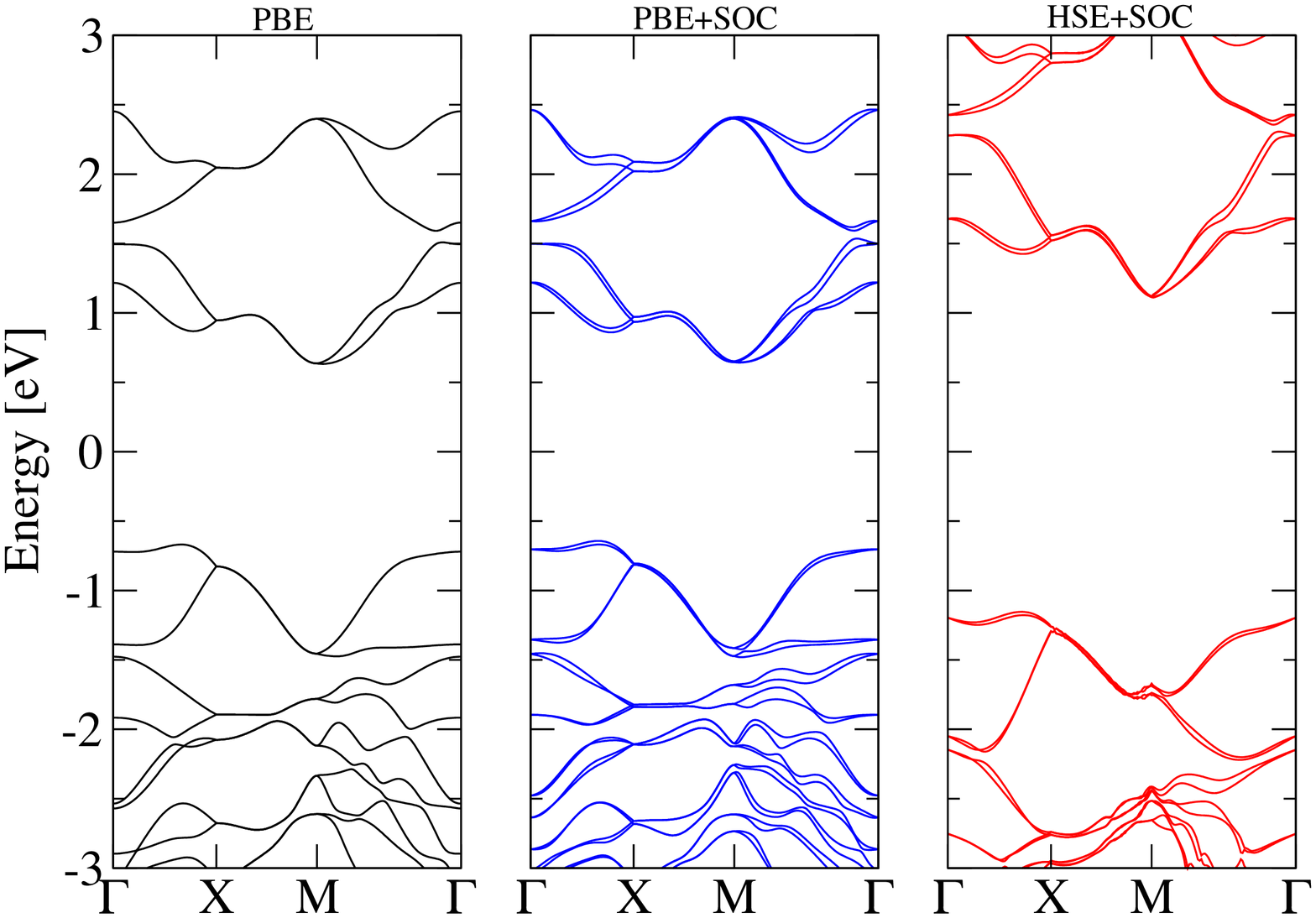}
    \caption{Electronic band structure for PdSeS obtained with GPAW. Including calculations 
    without SOC, with SOC and the PBE functional and with the hybrid functional with SOC.
    The results agree very well with those obtained from QE in Fig.  \ref{PdSeS_bandsQE_w90}.}
\label{PdSeS_Hybrid}
\end{figure}

\begin{figure}[H]
    \includegraphics[width=1.2\textwidth,center]{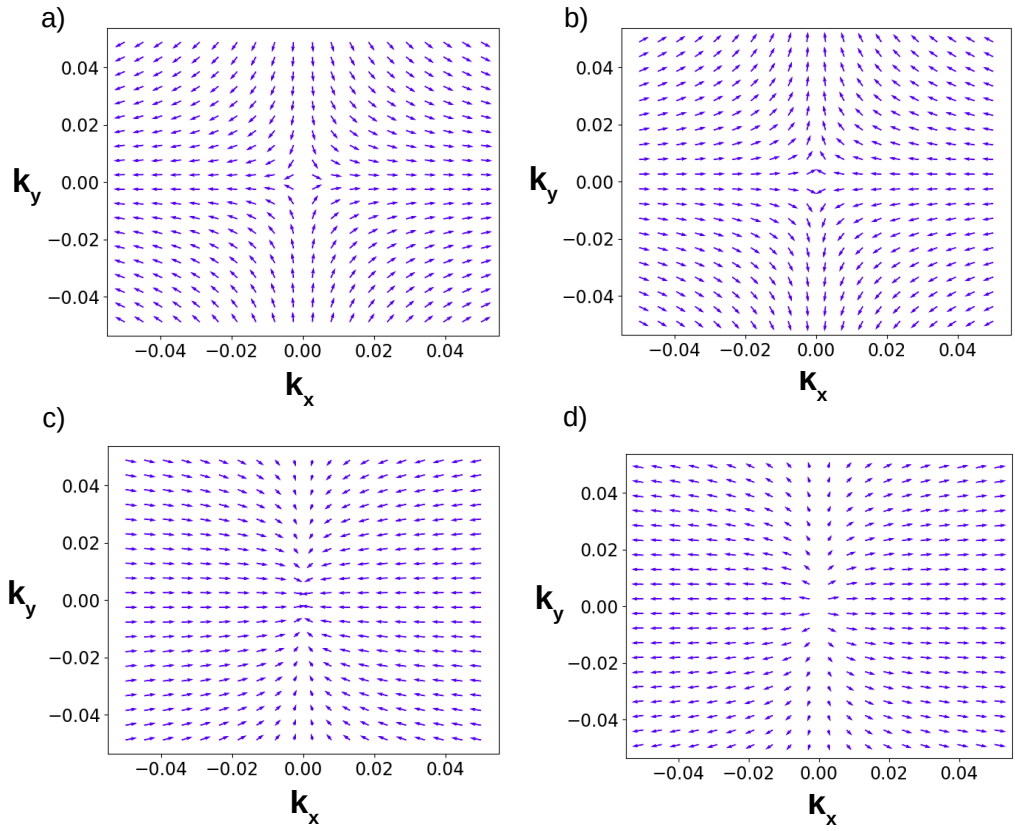}
    \caption{Spin texture near the $\Gamma$ point for a) top b) second c) third and 
    d) fourth valence bands for PdSeS.}
\label{PdSeS_spint}
\end{figure}

\begin{figure}[H]
    \includegraphics[width=1.\textwidth,center]{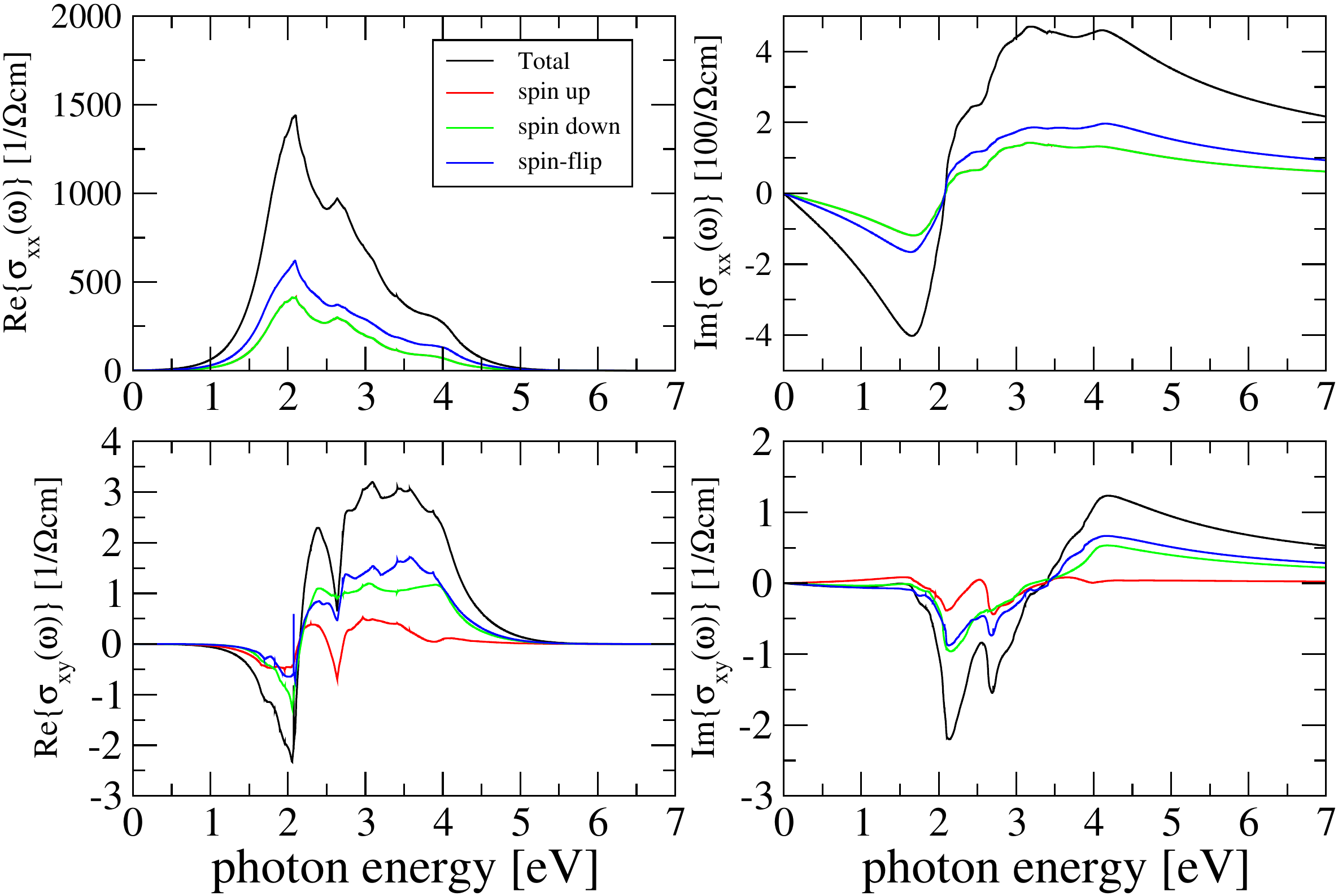}
    \caption{Optical conductivity with spin-resolved components along the spin $x$ direction for PdSeS.
    Other directions give similar results.}
\label{PdSeS_kubo}
\end{figure}

 \begin{figure}[H]
    \includegraphics[width=1.\textwidth,center]{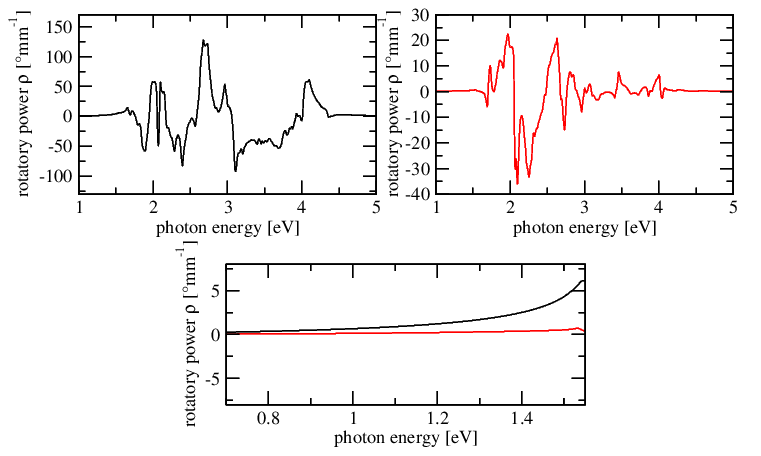}
    \caption{Natural optical activity in terms of the rotatory power $\rho$.}
\label{PdSeS_NOA}
\end{figure}

\begin{figure}[H]
    \includegraphics[width=1.\textwidth,center]{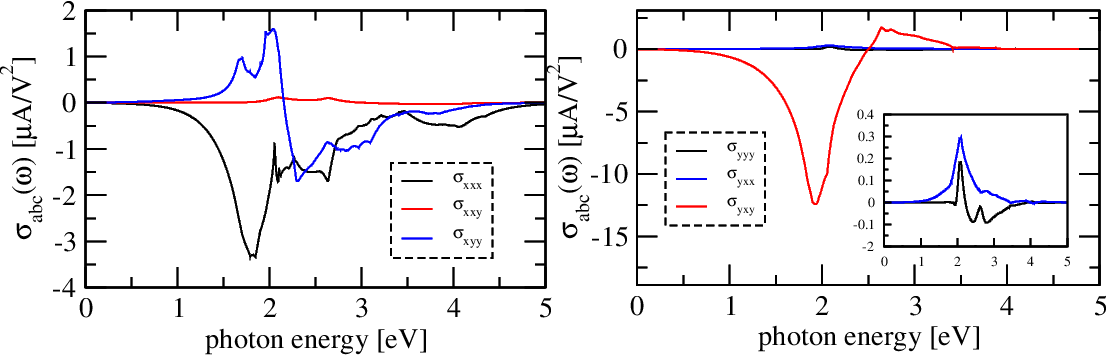}
    \caption{Nonzero components of the shift current tensor $\sigma_{abc}$.}
\label{PdSeS_shiftc}
\end{figure}

\subsubsection*{PdSeTe additional information}

\begin{figure}[H]
    \includegraphics[width=1.\textwidth,center]{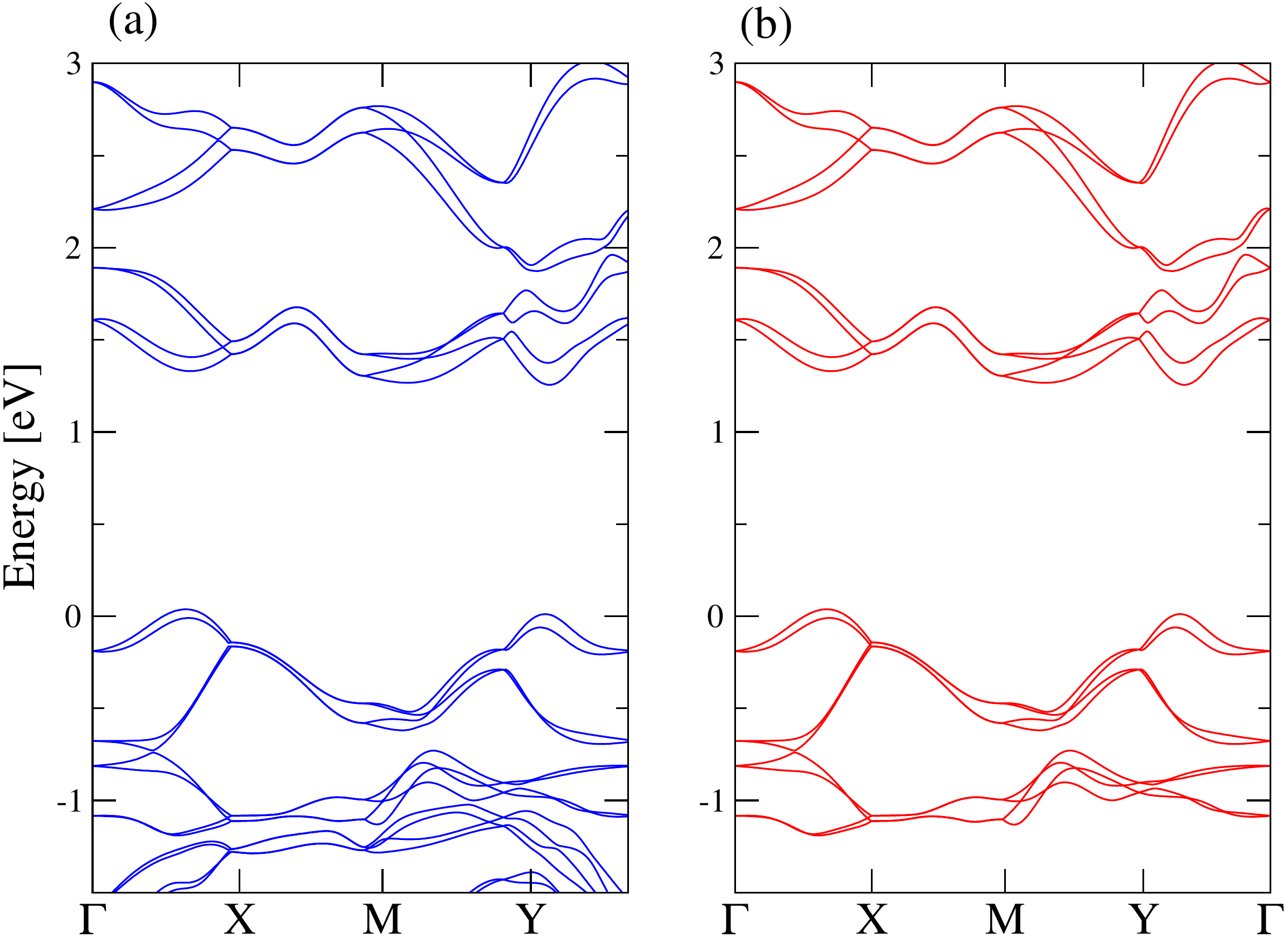}
    \caption{Electronic band structure for PdSeTe from a) QE and b) Wannier interpolation. 
    The band structure from QE has been repeated here for the sake of  comparison.}
\label{PdSeTe_QE_w90}
\end{figure}

\begin{figure}[H]
    \includegraphics[width=1.\textwidth,center]{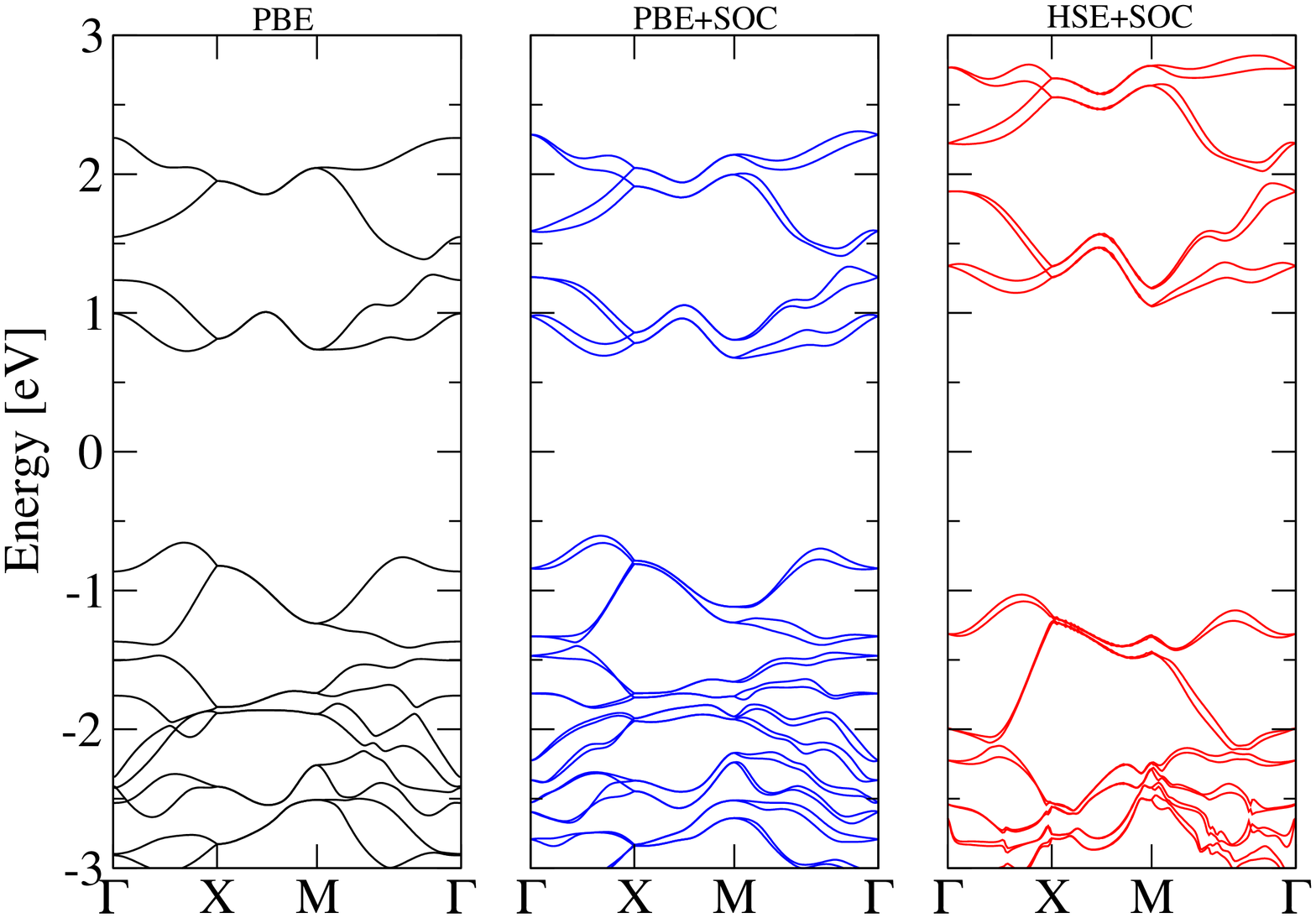}
    \caption{Electronic band structure for PdSeS obtained with GPAW. Including calculations 
    without SOC, with SOC and the PBE functional and with the hybrid functional with SOC.
    The results agree very well with those obtained from QE in Fig. \ref{PdSeTe_QE_w90}.}
\label{PdSeTe_hybrid}
\end{figure}

\begin{figure}[H]
    \includegraphics[width=1.1\textwidth,center]{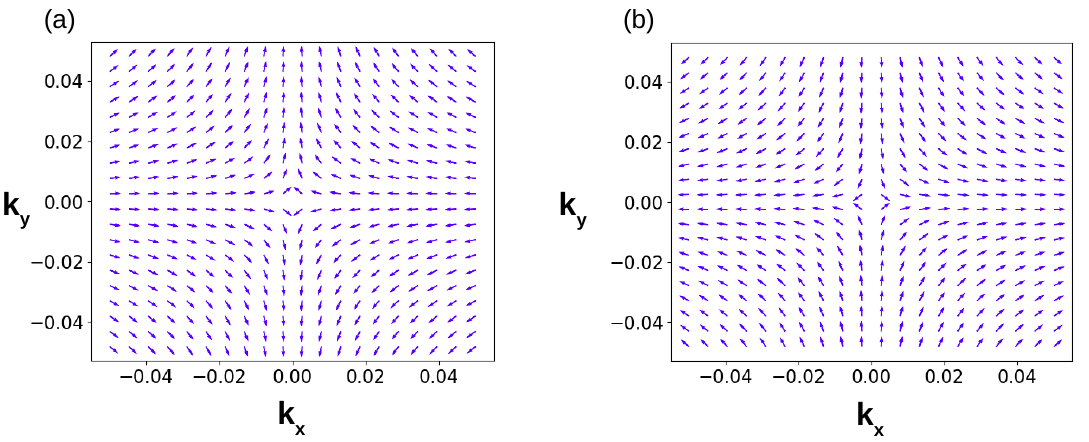}
    \caption{Spin texture for a) the top valence band and b) the second valence band for PdSeTe.}
\label{PdSeTe_spint}
\end{figure}

\subsubsection*{InP$_5$ additional information}

\begin{figure}[H]
    \includegraphics[width=1.\textwidth,center]{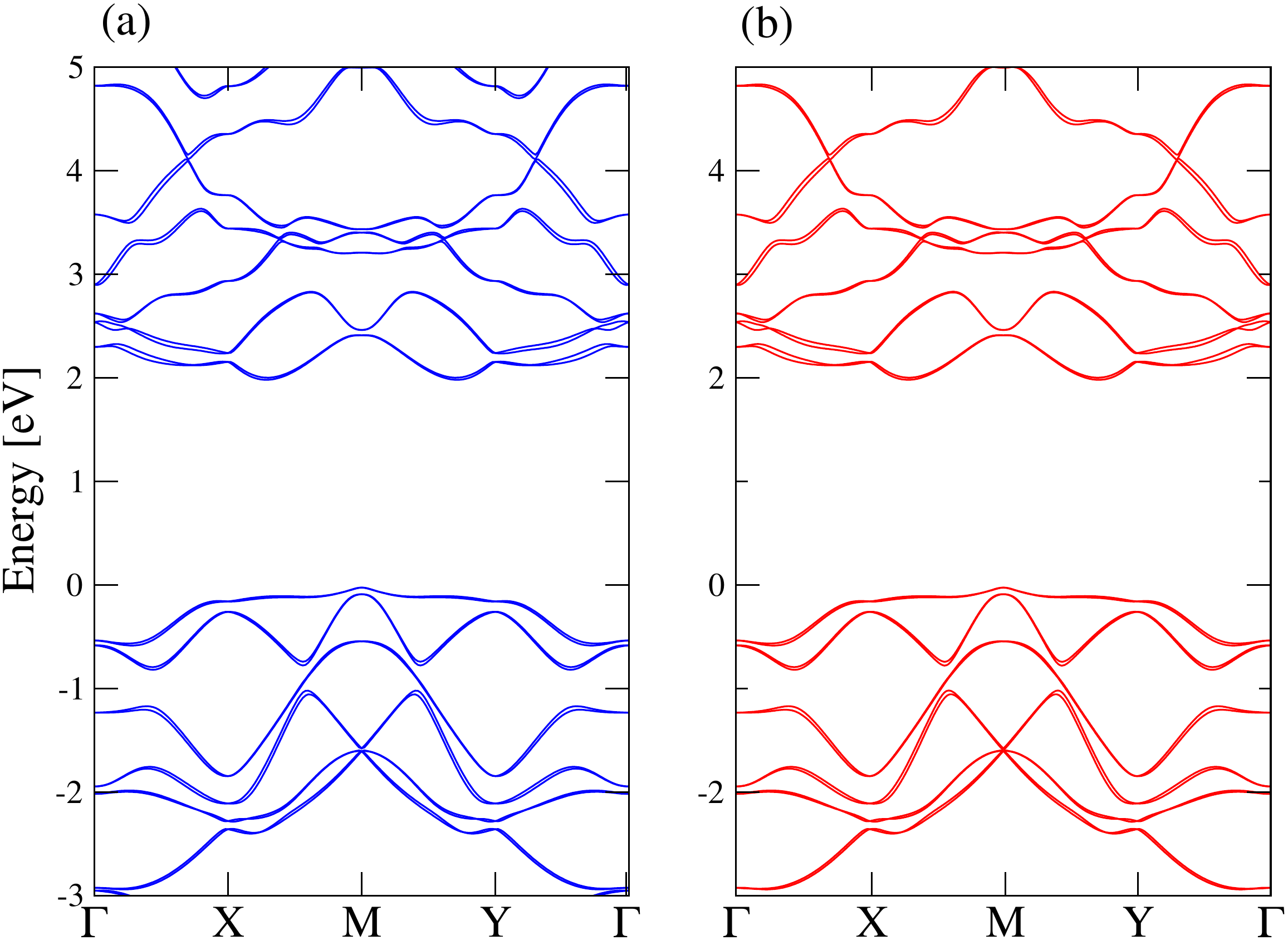}
    \caption{Electronic band structure for InP$_5$ from a) QE and b) Wannier interpolation. 
    The band structure from QE has been repeated here to facilitate comparison. This material 
    has been reported elsewhere; therefore, no further calculations are provided for phonons and 
    band structure.}
\label{InP5_QE_w90}
\end{figure}

\begin{figure}[H]
    \includegraphics[width=1.1\textwidth,center]{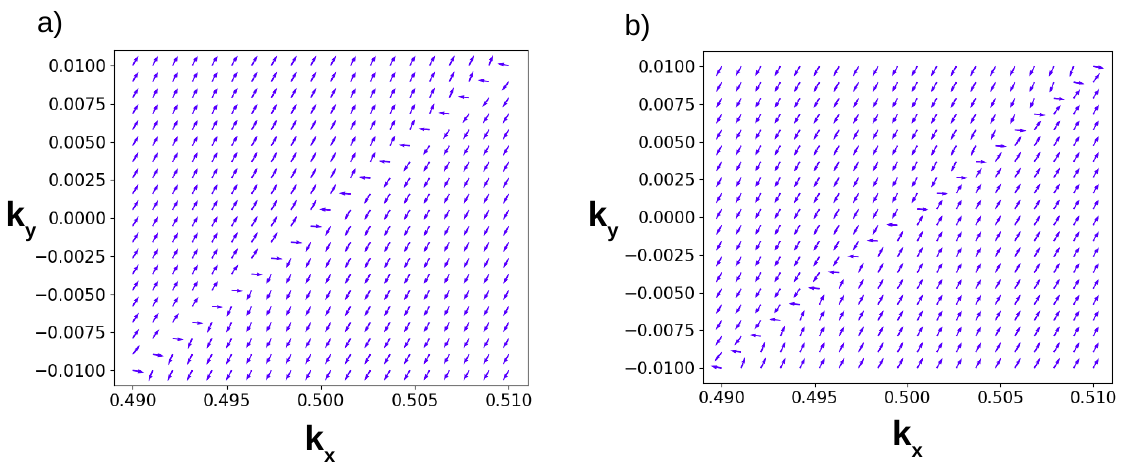}
    \caption{Spin texture near the X point for a) the top valence band and b) the second uppermost
    valence band for InP$_5$. }
\label{InP5_spint}
\end{figure}

\begin{figure}[H]
    \includegraphics[width=1.1\textwidth,center]{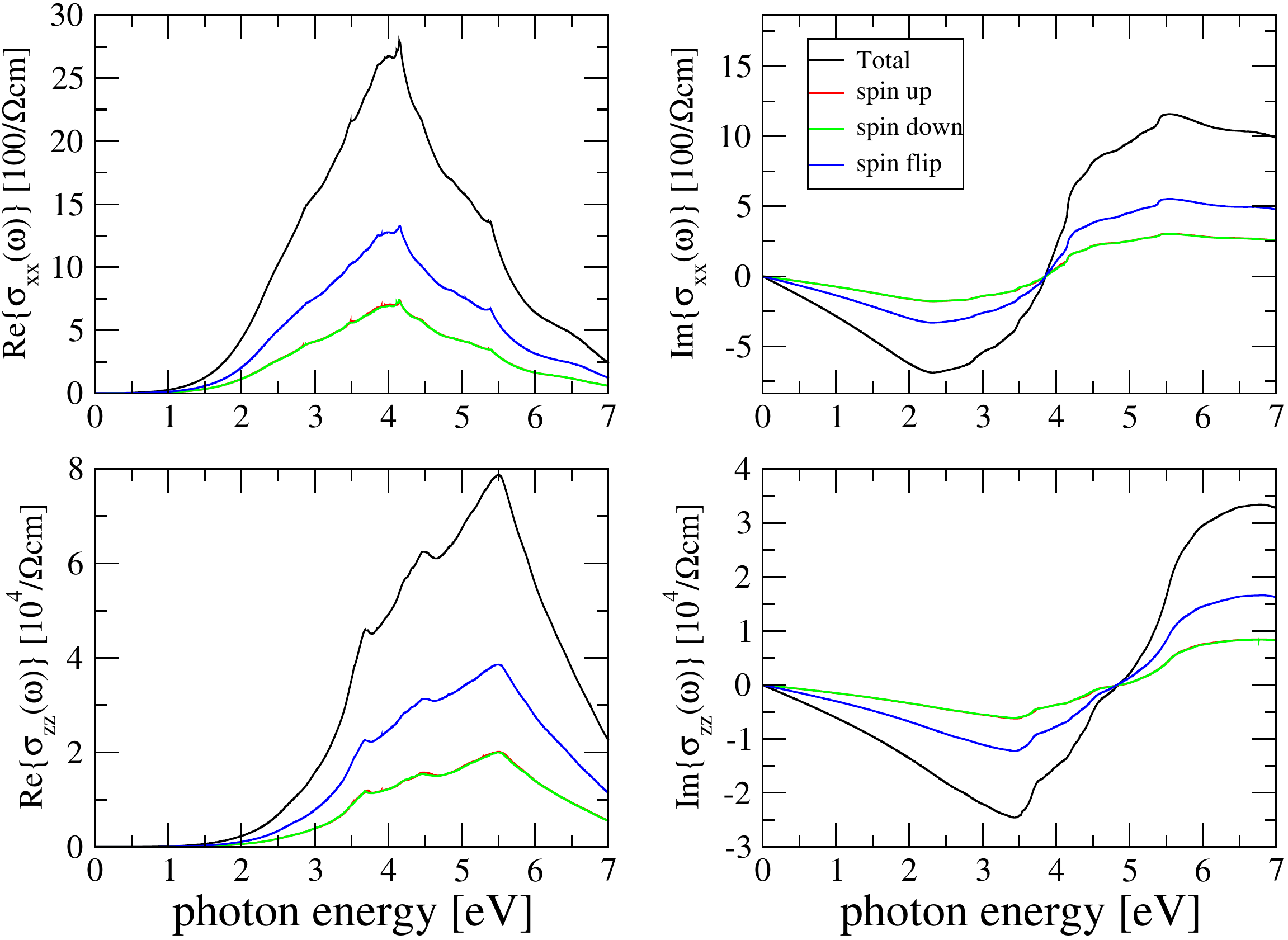}
    \caption{Optical conductivity with spin-resolved components along the spin $z$ direction
    for InP$_5$. Other directions give similar results.}
\label{InP5_kubo}
\end{figure}

 \begin{figure}[H]
    \includegraphics[width=1.\textwidth,center]{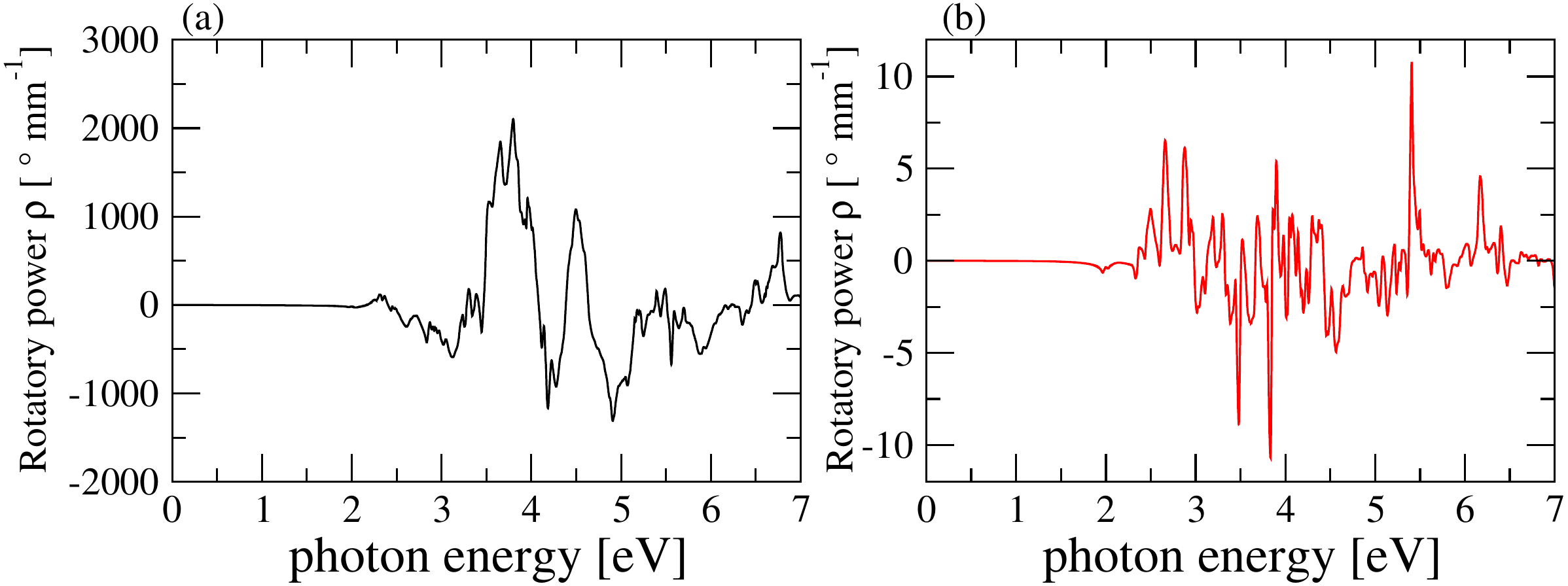}
    \caption{Natural optical activity in terms of the rotatory power $\rho$ for InP$_5$.}
\label{InP5_NOA}
\end{figure}

\subsubsection*{GeBi$_2$ additional information}

\begin{figure}[H]
    \includegraphics[width=1.2\textwidth,center]{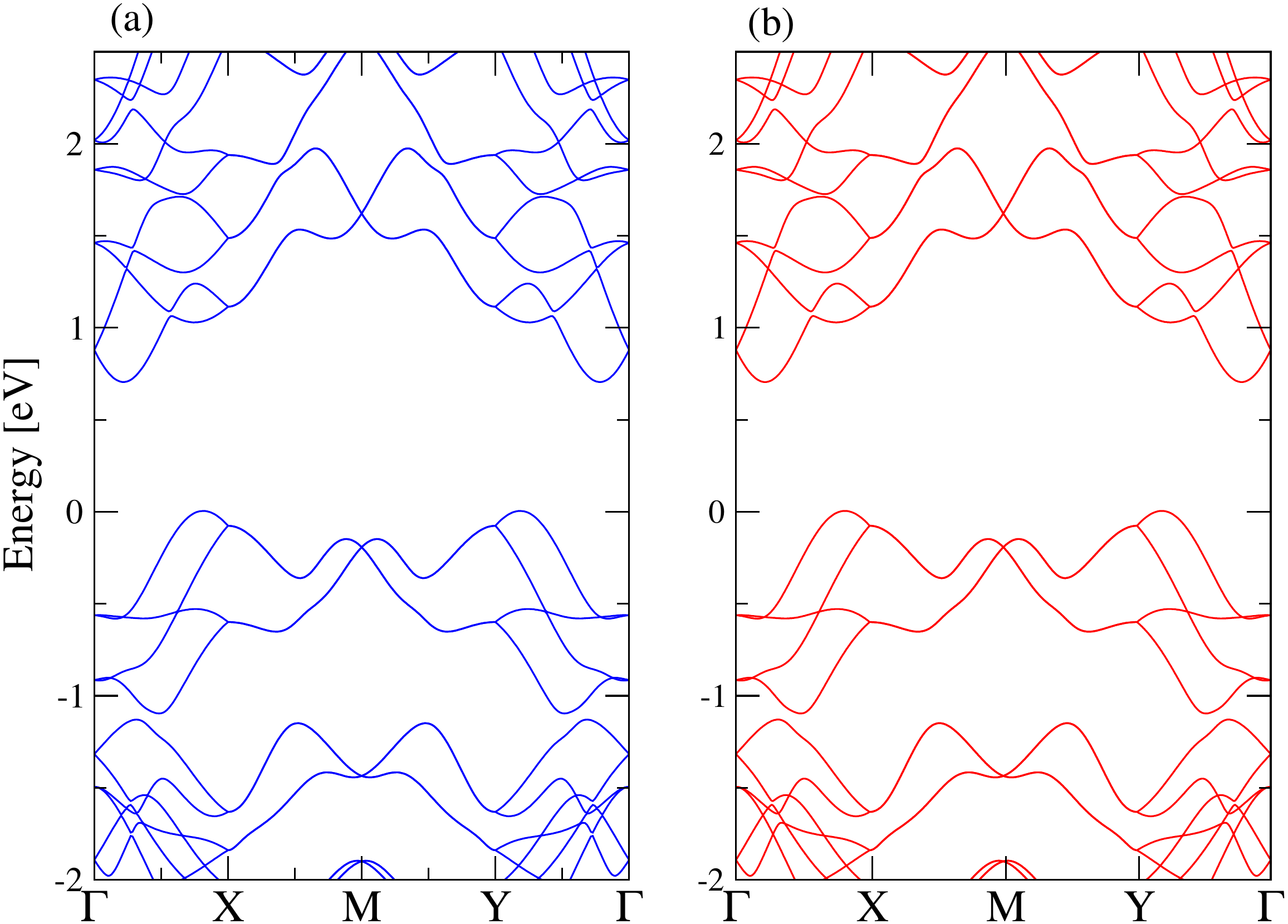}
    \caption{Electronic band structure for GeBi$_2$ from (a) QE and (b) Wannier interpolation. 
    The band structure from QE has been repeated here to facilitate comparison.}
\label{GeBi2_QE_w90}
\end{figure}

\begin{figure}[H]
    \includegraphics[width=1.1\textwidth,center]{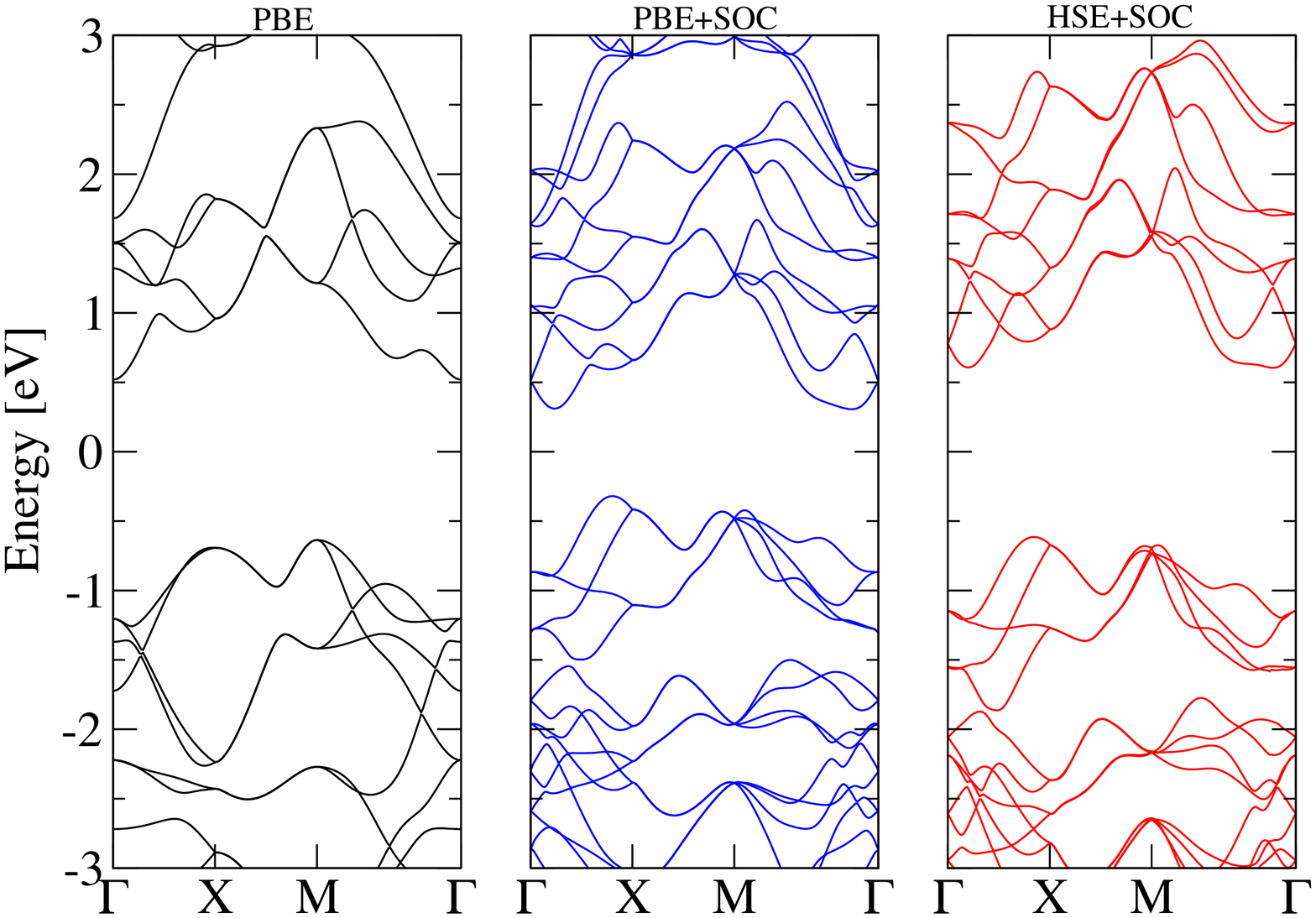}
    \caption{Electronic band structure for GeBi$_2$ obtained with GPAW. Including calculations 
    without SOC, with SOC and the PBE functional and with the hybrid functional with SOC.
    The results agree very well with those obtained from QE in Fig.  \ref{GeBi2_QE_w90}.}
\label{GeBi2_hybrid}
\end{figure}

\begin{figure}[H]
    \includegraphics[width=1.1\textwidth,center]{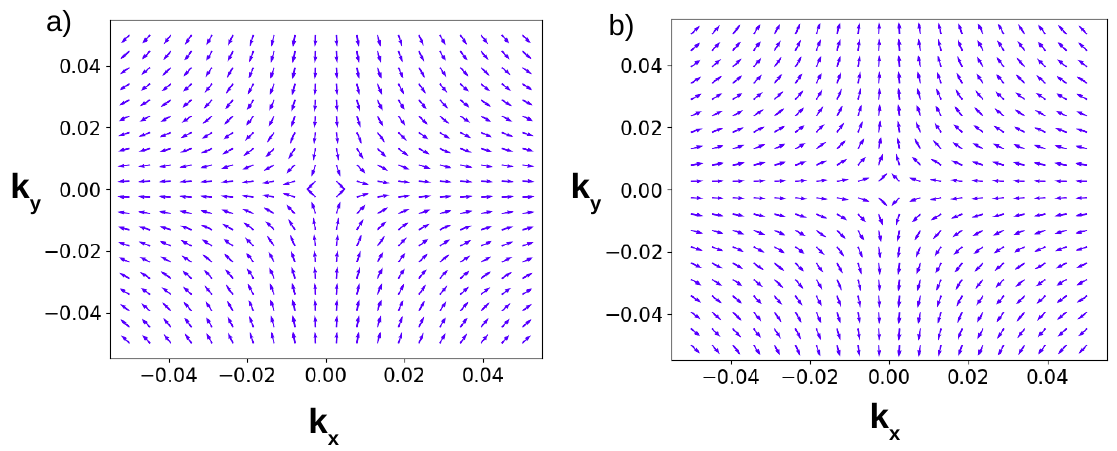}
    \caption{Spin texture in the $\Gamma$ point vicinity for a) bottom conduction band 
    b) second conduction band for GeBi$_2$.}
\label{GeBi2_spint}
\end{figure}

\begin{figure}[H]
    \includegraphics[width=1.1\textwidth,center]{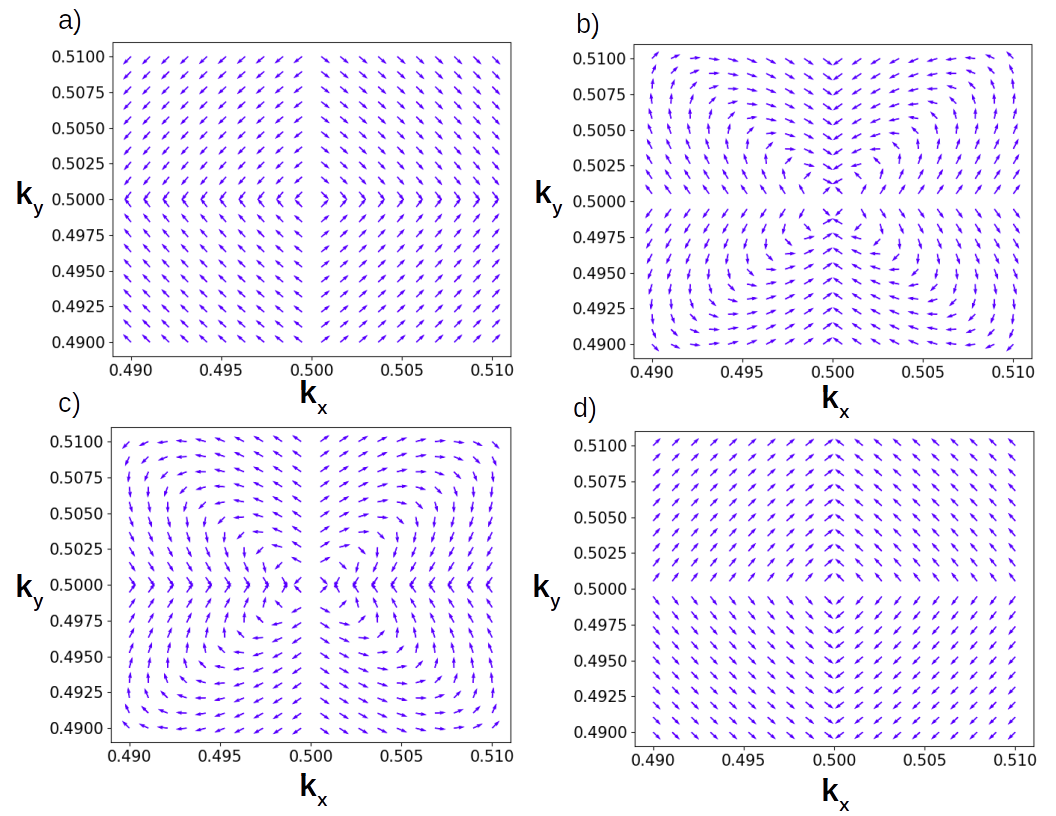}
    \caption{Spin texture in the neighborhood of the M point for a) top valence band, 
    b) second valence band, c) third valence band and d) fourth valence band 
    for GeBi$_2$. Note that the spin (band) partners are not contiguous for the a) and d) case. }
\label{GeBi2_spint_M}
\end{figure}

\begin{figure}[H]
    \includegraphics[width=1.1\textwidth,center]{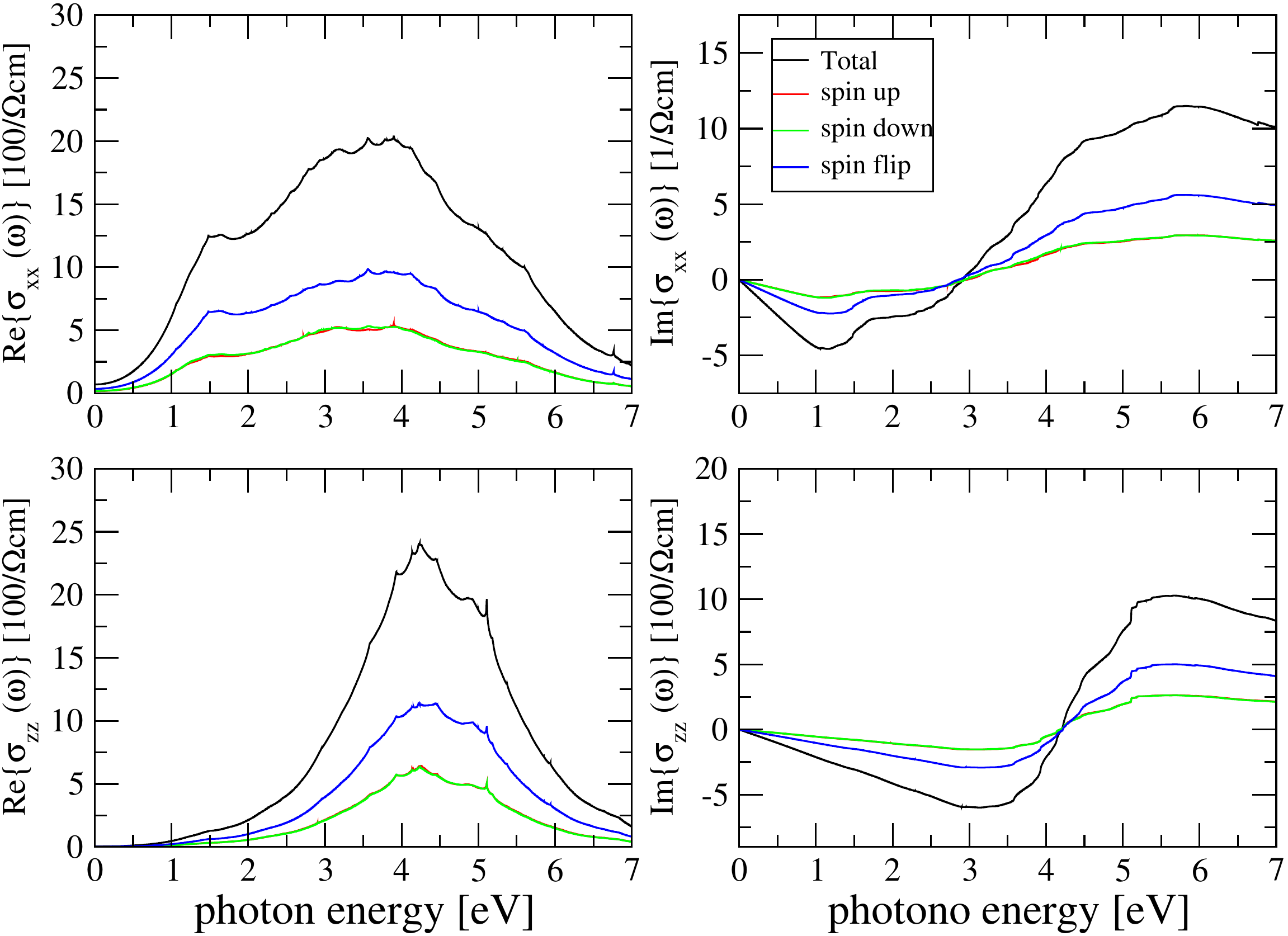}
    \caption{Optical conductivity with spin-resolved components along the spin $z$ direction
    for GeBi$_2$. Other directions give similar results.}
\label{GeBi2_kubo}
\end{figure}

\begin{figure}[H]
    \includegraphics[width=1.\textwidth,center]{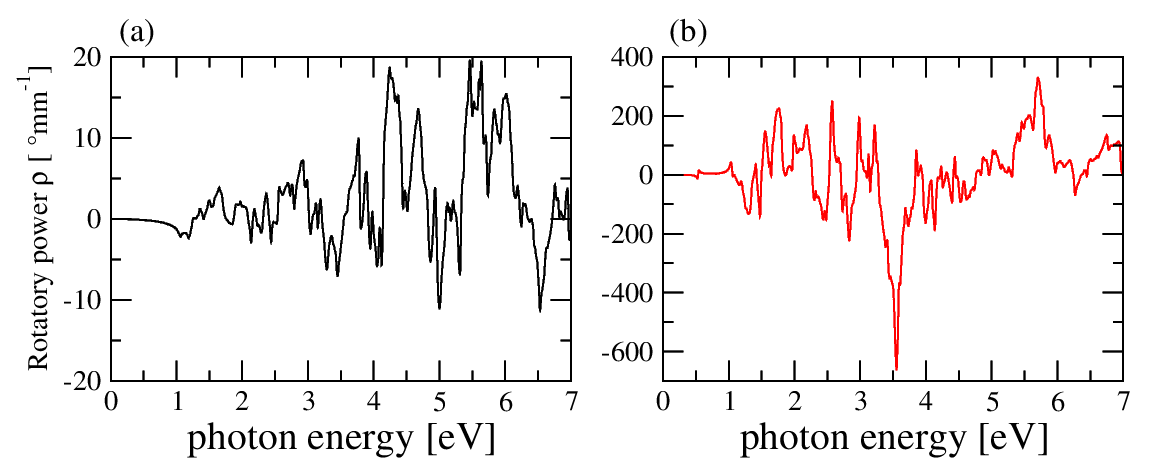}
    \caption{Natural optical activity in terms of the rotatory power $\rho$ for GeBi$_2$.}
\label{GeBi2_NOA}
\end{figure}

\begin{figure}[H]
    \includegraphics[width=.8\textwidth,center]{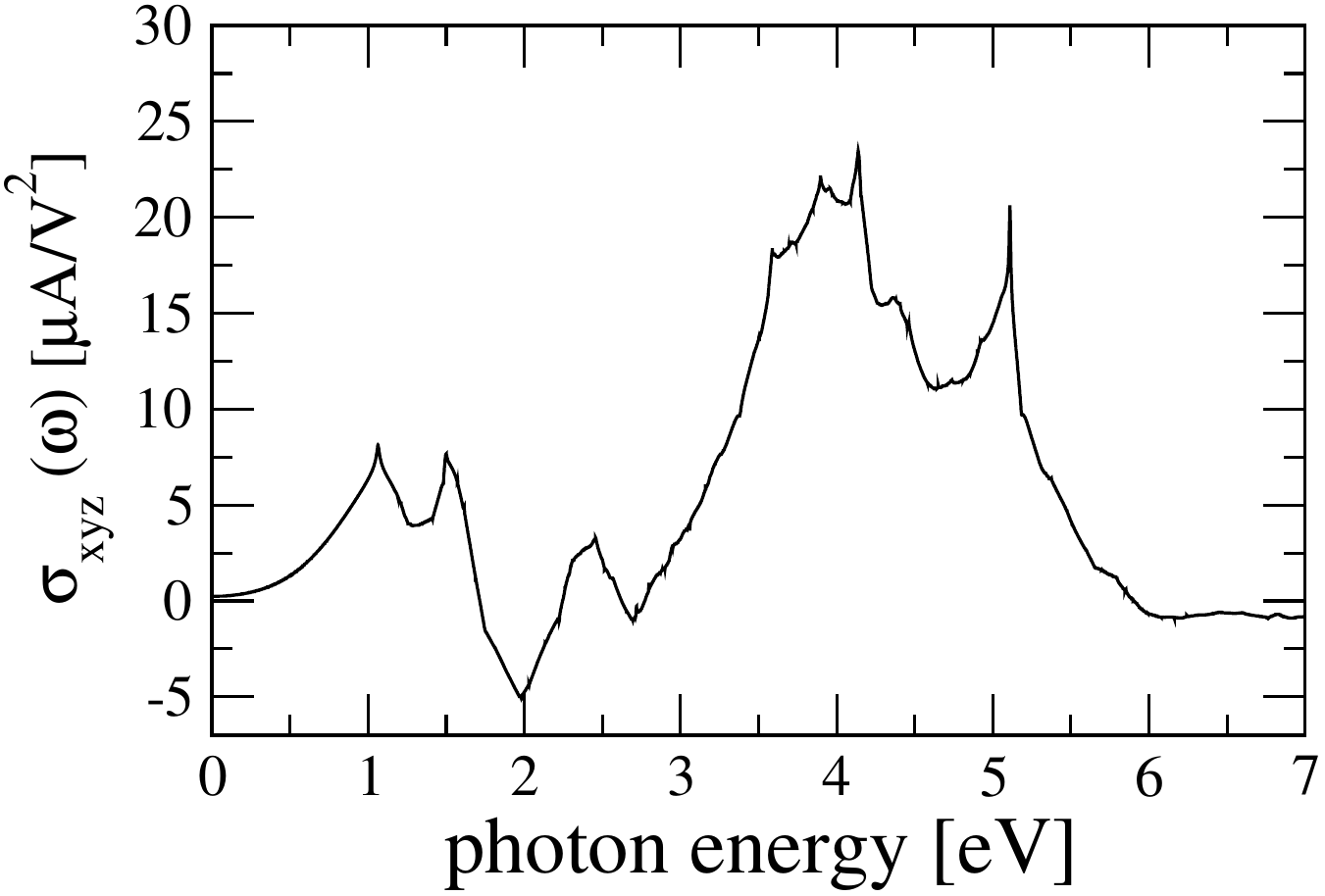}
    \caption{The nonzero component of the shift current tensor $\sigma_{abc}$ for GeBi$_2$.}
\label{GeBi2_shiftc}
\end{figure}

\subsection*{Part B: Multilayer space groups information}

\subsubsection*{Bilayer space groups}

The most straightforward way to build multilayer structures 
is using the so-called slip (translational) stacking. In the bilayer case presented 
here, two monolayers with the same space group are positioned with a relative 
displacement described by the vector ($t_1$,$t_2$,$t_3$). We consider the 
following relative translation vectors for the analysis (in units of the 
lattice vectors) \\

\begin{compactitem}
\item$\tau_{1}=(0,0,t_{3})$
\item$\tau_{2,1}=(\frac{1}{2},0,t_{3})$ and $\tau_{2,2}=(0,\frac{1}{2},t_{3})$.
\item$\tau_{3}=(\frac{1}{2},\frac{1}{2},t_{3})$.
\item$\tau_{4,1}=(\frac{1}{4},0,t_{3})$ and $\tau_{4,2}=(0,\frac{1}{4},t_{3})$.
\item$\tau_{5}=(\frac{1}{4},\frac{1}{4},t_{3})$. 
\item$\tau_{6,1}=(\frac{1}{2},\frac{1}{4},t_{3})$ and $\tau_{6,2}=(\frac{1}{4},\frac{1}{2},t_{3})$.
\item$\tau_{7,1}=(t_{1},0,t_{3})$ and $\tau_{7,1}=(0,t_{2},t_{3})$. \end{compactitem}
\bigskip

For each of these vectors we computed the space group that describes the structure. 
The calculations were done with the aid of the \textit{spglib} python library 
\cite{spglib_phyton_2018} and checked with the \textit{FindSym} code
\cite{FindSymm_2005}. The results of the computation of the space group
for each particular stacking are summarized in Table \ref{Tab_2L}, with the exception 
of vector $\tau_{1}=(0,0,t_{3})$ which yields the same space group as the composing 
monolayers. 

\begin{table}[H]
\centering
\begin{tabular}{lll}
\hline
\hline
\textbf{Space group of} & \textbf{Translation vectors} & \textbf{Bilayer space group}\\
\textbf{the monolayer} & & \\
\hline
\hline
$P4/mbm$ (\#127*) & $\tau_{2,1}$ and $\tau_{2,2}$  & $Pcca$ (\#54)  \\
 & $\tau_{3}$  & $P4/nbm$ (\#125)  \\
 & $\tau_{4,1}$ and $\tau_{4,2}$  & $P2_{1}/c$ (\#14)  \\
 & $\tau_{5}$  & $C2/m$ (\#12)  \\
 & $\tau_{6,1}$ and $\tau_{6,2}$  & $P2/c$ (\#13)  \\
 & $\tau_{7,1}$ and $\tau_{7,2}$  & $P2_{1}/c$ (\#14)  \\
 $P\overline{4}2_{1}m$ (\#113) & $\tau_{2,1}$ and $\tau_{2,2}$  & $P222_{1}$ (\#17)  \\
 & $\tau_{3}$  & $P\overline{4}2m$ (\#111)  \\
 & $\tau_{4,1}$ and $\tau_{4,2}$  & $P2_{1}$ (\#4)  \\
 & $\tau_{5}$  & $Cm$ (\#8)  \\
 & $\tau_{6,1}$ and $\tau_{6,2}$  & $P2$ (\#3)  \\
 & $\tau_{7,1}$ and $\tau_{7,2}$  & $P2_{1}$ (\#4)  \\
$P4bm$ (\#100) & $\tau_{2,1}$ and $\tau_{2,2}$  & $Pba2$ (\#32)  \\
 & $\tau_{3}$  & $P4bm$ (\#100)  \\
 & $\tau_{4,1}$ and $\tau_{4,2}$  & $Pc$ (\#7)  \\
 & $\tau_{5}$  & $Cm$ (\#8)  \\
 & $\tau_{6,1}$ and $\tau_{6,2}$  & $Pc$ (\#7)  \\
 & $\tau_{7,1}$ and $\tau_{7,2}$  & $Pc$ (\#7)  \\ 
$P42_{1}2$ (\#90) & $\tau_{2,1}$ and $\tau_{2,2}$  & $P222_{1}$ (\#17)  \\
 & $\tau_{3}$  & $P422$ (\#89)  \\
 & $\tau_{4,1}$ and $\tau_{4,2}$  & $P2_{1}$ (\#4)  \\
 & $\tau_{5}$  & $C2$ (\#5)  \\
 & $\tau_{6,1}$ and $\tau_{6,2}$  & $P2$ (\#3)  \\
 & $\tau_{7,1}$ and $\tau_{7,2}$  & $P2_{1}$ (\#4)  \\ 
$Pbam$ (\#55) & $\tau_{2,1}$ and $\tau_{2,2}$  & $Pcca$ (\#54)  \\
 & $\tau_{3}$  & $Pban$ (\#50)  \\
 & $\tau_{4,1}$ and $\tau_{4,2}$  & $P2_{1}/c$ (\#14)  \\
 & $\tau_{5}$  & $P\overline{1}$ (\#2)  \\
 & $\tau_{6,1}$ and $\tau_{6,2}$  & $P2/c$ (\#13)  \\
 & $\tau_{7,1}$ and $\tau_{7,2}$  & $P2_{1}/c$ (\#14)  \\ 
 $Pba2$ (\#32) & $\tau_{2,1}$ and $\tau_{2,2}$  & $Pba2$ (\#32)  \\
 & $\tau_{3}$  & $Pba2$ (\#32)  \\
 & $\tau_{4,1}$ and $\tau_{4,2}$  & $Pc$ (\#7)  \\
 & $\tau_{5}$  & $P1$ (\#1)  \\
 & $\tau_{6,1}$ and $\tau_{6,2}$  & $Pc$ (\#7)  \\
 & $\tau_{7,1}$ and $\tau_{7,2}$  & $Pc$ (\#7)  \\ 
 $P2_{1}2_{1}2$ (\#18) & $\tau_{2,1}$ and $\tau_{2,2}$  & $P222_{1}$ (\#17)  \\
 & $\tau_{3}$  & $P2$ (\#3)  \\
 & $\tau_{4,1}$ and $\tau_{4,2}$  & $P2_1$ (\#4)  \\
 & $\tau_{5}$  & $P1$ (\#1)  \\
 & $\tau_{6,1}$ and $\tau_{6,2}$  & $P2$ (\#3)  \\
 & $\tau_{7,1}$ and $\tau_{7,2}$  & $P2_{1}$ (\#4)  \\ 
$P2_{1}/c$ (\#14) & $\tau_{2,1}$ and $\tau_{2,2}$  & $P2/c$ (\#17) and $P2_{1}/c$ (\#14)\\
 & $\tau_{3}$  & $P2/c$ (\#13)  \\
 & $\tau_{4,1}$ and $\tau_{4,2}$  & $P1$ (\#1) and $P2_{1}/c$ (\#14) \\
 & $\tau_{5}$ & $C2$ (\#5)  \\
 & $\tau_{6,1}$ and $\tau_{6,2}$  & $P2$ (\#3)  \\
 & $\tau_{7,1}$ and $\tau_{7,2}$  & $P1$ (\#1)  \\ 
\hline
\hline
\end{tabular}
\caption{\label{Tab_2L}Space groups for bilayer pentagonal structures 
based on the monolayers space groups. *The SG \#117 gives the same 
groups as the same WP coordinates are used for both groups.}
\end{table}

\subsubsection*{Frequency tables for the case with $n = 3$ layers}

\begin{figure}[H]
    \includegraphics[width=.9\textwidth,center]{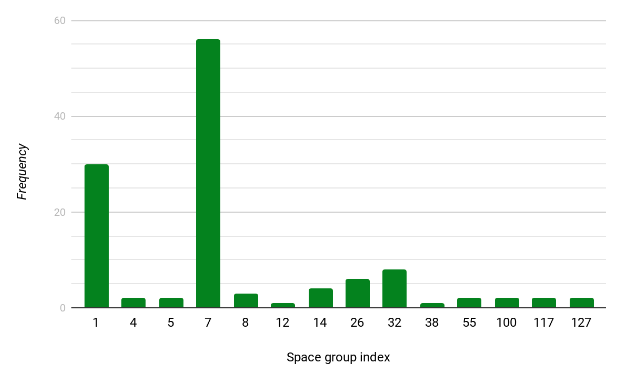}
    \caption{Frequency graph for the possible space groups that can be formed by translational
    stacking of SG \#127 monoloyers.}
\label{Freq_127}
\end{figure}

\begin{figure}[H]
    \includegraphics[width=.9\textwidth,center]{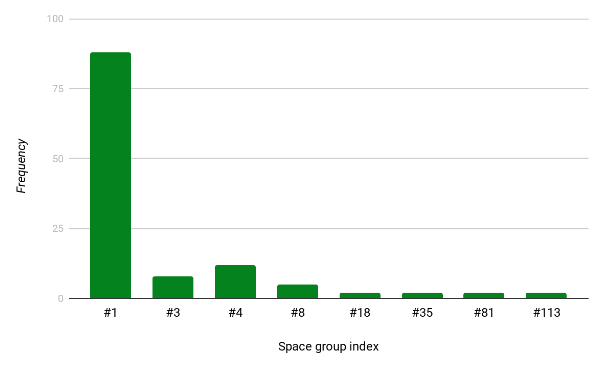}
    \caption{Frequency graph for the possible space groups that can be formed by translational
    stacking of SG \#113 monoloyers.}
\label{Freq_113}
\end{figure}

When we add a third layer to the system, the stacking combinations increase substantially, implying that the enumeration of the space groups for these configurations becomes
cumbersome and fairly impractical. Evidently this extends to $n>3$.
Despite that, we can do a general analysis for some space groups chosen as examples. We select SG \#127
and SG \#113, whose multilayers could be of interest due to great number of reported 
materials belonging to these two space groups.
We present in Fig. \ref{Freq_127} and Fig. \ref{Freq_113} a summary 
for parent SG \#127 and SG \#113, respectively, in the form of two frequency charts 
that show the most recurring space groups arising from the stackings defined above.

It can be noted in Fig. \ref{Freq_127} that for the space group \#127, the most frequent 
three-layer group is \#7. This is in contrast with what is is generally expected; 
that the \#1 space group dominates the frequency count, as happens for example with
space group \#113 in Fig. \ref{Freq_113}.  

The above procedure can be easily continued for $n > 3$ but in general the trivial
space group $P1$, will have an even greater incidence.

\bibliographystyle{unsrt}
\bibliography{supplement}